\documentclass[authoryear]{elsarticle} 

\usepackage{fullpage}
\journal{Applied Ocean Research (Accepted March 2025)}

\usepackage{enumitem}

\usepackage{upgreek}
\usepackage{mathtools}
\usepackage{rotating}
\usepackage{amsmath}
\usepackage{amssymb}
\usepackage{bbm}
\usepackage{subcaption}
\usepackage{graphicx}
\usepackage{url}
\usepackage[linesnumbered]{algorithm2e}
\usepackage{placeins}
\usepackage{xcolor}

\definecolor{MyChange}{rgb}{0,0,0}

\def\bSig\mathbf{\Sigma}

\newcommand {\un}[1]{\boldsymbol{#1}}
\newcommand {\pbi}{\begin{itemize}}
\newcommand {\pei}{\end{itemize}}

\newcommand {\pbc}{\begin{center}}
\newcommand {\pec}{\end{center}}
\newcommand {\pbe}{\begin{eqnarray*}}
\newcommand {\pee}{\end{eqnarray*}}
\newcommand {\pben}{\begin{eqnarray}}
\newcommand {\peen}{\end{eqnarray}}
\let\hat=\widehat

\newcommand{\ed}[1]{\textcolor{black}{#1}}
\usepackage {tikz}
\usetikzlibrary{arrows}
\begin{document}
	
\begin{frontmatter}
		
\title{Calibration of medium-range metocean forecasts for the North Sea}
\author[lancs]{Conor Murphy}
\author[shelluk]{Ross Towe}
\author[lancs]{Philip Jonathan\corref{cor1}}
\address[lancs]{School of Mathematical Sciences, Lancaster University LA1 4YF, United Kingdom.}		
\address[shelluk]{Shell Information Technology International Ltd., London SE1 7NA, United Kingdom.}
\cortext[cor1]{Corresponding author {\tt p.jonathan@lancaster.ac.uk}}		

\begin{abstract}
We assess the value of calibrating forecast models for significant wave height $H_S$, wind speed $W$ and mean spectral wave period $T_m$ for forecast horizons between zero and 168 hours from a commercial forecast provider, to improve forecast performance for a location in the central North Sea. We consider two straightforward calibration models, linear regression (LR) and non-homogeneous Gaussian regression (NHGR), incorporating deterministic, control and ensemble mean forecast covariates. We show that relatively simple calibration models (with at most three covariates) provide good calibration and that addition of further covariates cannot be justified. Optimal calibration models (for the forecast mean of a physical quantity) always make use of the deterministic forecast and ensemble mean forecast for the same quantity, together with a covariate associated with a different physical quantity. The selection of optimal covariates is performed independently per forecast horizon, and the set of optimal covariates shows a large degree of consistency across forecast horizons. As a result, it is possible to specify a consistent model to calibrate a given physical quantity, incorporating a common set of three covariates for all horizons. For NHGR models of a given physical quantity, the ensemble forecast standard deviation for that quantity is skilful in predicting forecast error standard deviation, strikingly so for $H_S$. We show that the consistent LR and NHGR calibration models facilitate reduction in forecast bias to near zero for all of $H_S$, $W$ and $T_m$, and that there is little difference between LR and NHGR calibration for the mean. Both LR and NHGR models facilitate reduction in forecast error standard deviation relative to naive adoption of the (uncalibrated) deterministic forecast, with NHGR providing somewhat better performance. Distributions of standardised residuals from NHGR are generally more similar to a standard Gaussian than those from LR.
\end{abstract}

\begin{keyword}
metocean, forecast, calibration, weather, linear regression, non-homogeneous Gaussian regression.
\end{keyword}

\end{frontmatter}

\section{Introduction} \label{Sct:Int}
%
Safe execution of offshore activities requires the forecasting of environmental time series to improve decision making, for e.g. platform evacuation in severe weather, wind farm maintenance scheduling, and on- and off-loading from floating LNG facilities. Good forecast performance for a range of environmental conditions is particularly important, as is reliable quantification of forecast uncertainty; in principle, we are interested in forecasting the full joint spatio-temporal distribution of metocean sea state variables ({incorporating} sea state significant wave height $H_S$, mean wave period $T_M$ and wind speed $W$) well, including marginal and joint extremes.

Weather-forecasting organisations now routinely provide metocean forecasts to inform offshore activities. Usually, for use at specific locations, it is possible to calibrate the original forecasts further, so that the calibrated forecast exhibits smaller bias and uncertainty than the original forecast. Moreover, modern forecasts tend to come in the form of a combination of different components, including e.g. a deterministic forecast, a control forecast and an ensemble of forecasts representing a range of possible future metocean temporal trajectories; all of these are available to calibrate the original forecast. 

There is a large literature on forecast calibration. \cite{GntEA05} presents a calibration method for probabilistic forecasts, based on multiple linear regression, to address both forecast bias and under-dispersion. \cite{PnsEA09} discusses the generation of statistical scenarios of short-term wind generation that accounts for  interdependence of prediction errors and predictive distributions of wind power production. \cite{Gnt11} provides a framwework for estimation and evaluation of point forecasts. \cite{Gnt14} discusses calibration of medium-range weather forecasts, presenting alternative strategies including Bayesian model averaging (BMA), non-homogeneous Gaussian regression (NHGR) and empirical copula coupling (ECC). The latter incorporates marginal calibration of raw ensembles together with a reordering to retain rank correlation structure across variates. \cite{SchEA13} discusses ECC for uncertainty quantification in computer models. \cite{GntKtz14} provides a review of probabilistic forecasting. \cite{BssEA17} and \cite{Swn20} review the challenges facing forecasters in electric power and renewable energy, emphasising the advantages of probabilistic methods. \cite{GlbEA21} proposes boosted semi-parametric models for probabilistic forecasts which outperform those estimated via maximum likelihood.  \cite{HnrEA21} proposes probabilistic post-processing of multivariate forecasts, incorporating moving averages and covariance matrix regularization, allowing for non-stationary, non-isotropic and negative correlations in the forecasting error. \cite{Mer21} proposes a multivariate probabilistic ensemble model to forecast solar irradiance. \cite{BjrEA21} provides an introduction to multivariate probabilistic forecast evaluation. \cite{GaoEA22} consider probabilistic forecasting of Arctic Sea ice thickness. \cite{AstEA23} emphasises that proper scoring rules (see e.g. \citealt{Wnk69}, \citealt{GntEA07}) are necessary for the evaluation of probabilistic forecasts, and illustrates their performance in an operational maritime engineering context. Proper scoring rules evaluate forecasting performance in terms of forecast sharpness and calibration, such that a model score is optimized when the reported forecast distribution is equal to the true predictive distribution given information (from e.g. data). \cite{AllEA23} considers the adoption of transformed kernel scores to emphasise the importance of forecasting events with high impact. \cite{AdnEA23} uses BMA in conjunction with various machine learning methods to provide short-term probabilistic forecasting of $H_S$. \cite{CrqTrg24} discusses direct forecasting of exceedance probability for $H_S$. \cite{HhlEA24} examines the use of permutation-invariant neural networks for ensemble-based forecasting, that treat forecast ensembles as a set of unordered member forecasts, learning link functions that are by design invariant to permutations of the member ordering. \cite{TrlPpc24} provides a recent review of predictive uncertainty estimation using machine learning.

\subsection*{Objectives and outline}
{The typical practising metocean engineer uses the simplest tools (e.g. linear regression) for calibration, rarely exploiting the richness of data provided by modern forecast models. The aspiration of the current paper is to demonstrate that there is material benefit from exploiting the output of modern forecasts more fully within pragmatic forecast calibration procedures for realistic offshore application. Specifically we do not claim that the calibration procedures considered are the best currently available, but we do contend that they provide useful tools with which the metocean engineer can reasonably expect to improve the performance of their forecast calibrations.}

The objective of the current work is to evaluate the performance of simple methods to calibrate forecasts for $H_S$, $T_M$ and $W$, based on forecasts from a commercial forecast provider and offshore measurements for a central North Sea (CNS) location. Specifically, we evaluate the relative performance of different marginal linear regression (LR) and non-homogeneous Gaussian regression (NHGR) calibration models for a given measured metocean variable (one of $H_S$, $T_M$ and $W$) in terms of deterministic, control and ensemble forecasts for the variable, and also potentially forecasts for other variables. 

The layout of the article is as follows. Section~\ref{Sct:MtvApp} describes the motivating CNS application, and Section~\ref{Sct:Mth} outlines the calibration methodologies employed. \ed{Section~\ref{Sct:Rsl} provides results from the calibration studies, including assessment of both within-sample and out-of-sample performance. Section~\ref{Sct:DscCnc} provides discussion and conclusions. Online Supplementary Material (SM) provides supporting figures.}

\FloatBarrier
\section{Motivating application}  \label{Sct:MtvApp}
%
We consider the calibration of forecast data, from weather forecast provider StormGeo, for a location in the central North Sea. At this location, we have access to in-situ measured data for sea state $H_S$ and mean spectral wave period $T_m$ (measured using a downward-looking Saab Rex wave radar) and wind speed $W$ (measured using a Gill ultrasonic wind sensor) for the interval 17 May 2022 to 6 September 2023. $W$ is sampled at 10 minute intervals, whereas $H_S$ and $T_m$ estimates are provided every 30 minutes. These data were averaged to provide hourly values.

Forecast data for the same period was issued every 6 hours (from mid-night), providing forecasts for horizons at three-hourly resolution (for forecast horizons $\le$ 72 hours) and at six-hourly resolution for longer forecast horizons $\le$ 168 hours. At each issue time, the forecast data consist of a deterministic forecast, a control forecast and an ensembles forecast with 50 exchangeable members (see Section~\ref{Sct:Mth:ExcEns}) for all of $H_S$, $W$ and $T_m$. These will be referred to henceforth as ``forecast components'' for definiteness. 

\begin{figure}[!ht]
	\centering
	\begin{subfigure}{1\textwidth}
		\centering
		\includegraphics[width=1\textwidth]{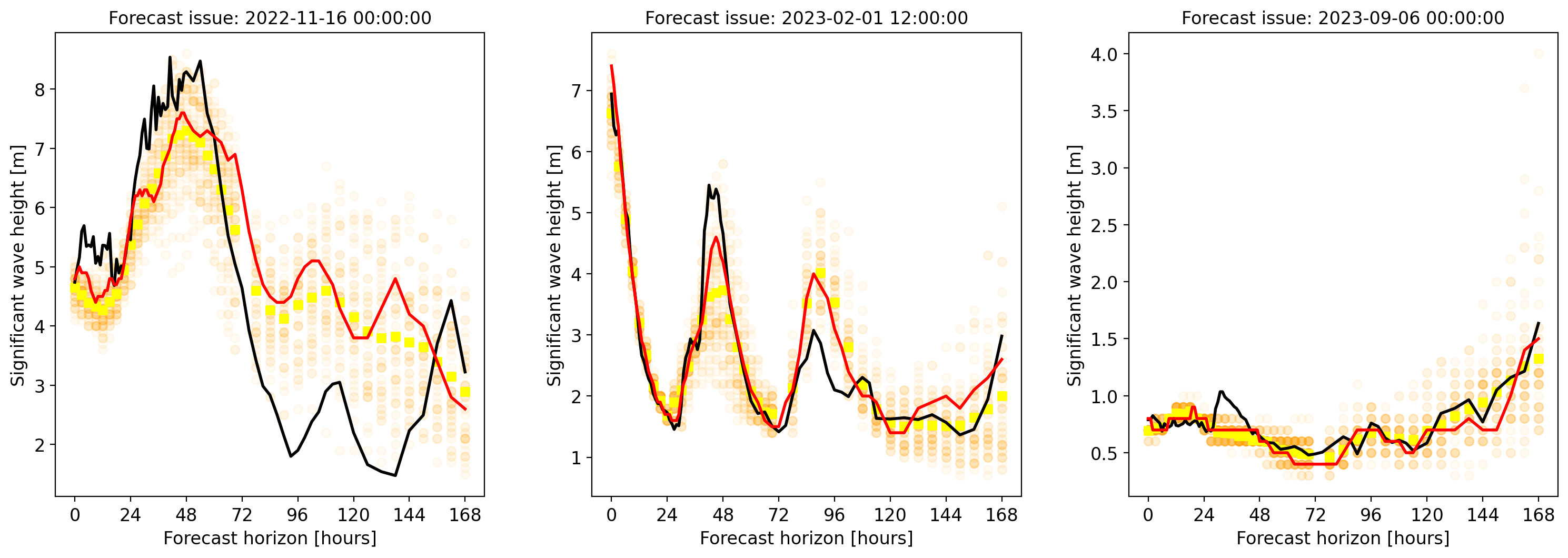}
		\caption{Significant wave height $H_S$.}
	\end{subfigure}
	\begin{subfigure}{1\textwidth}
		\centering
		\includegraphics[width=1\textwidth]{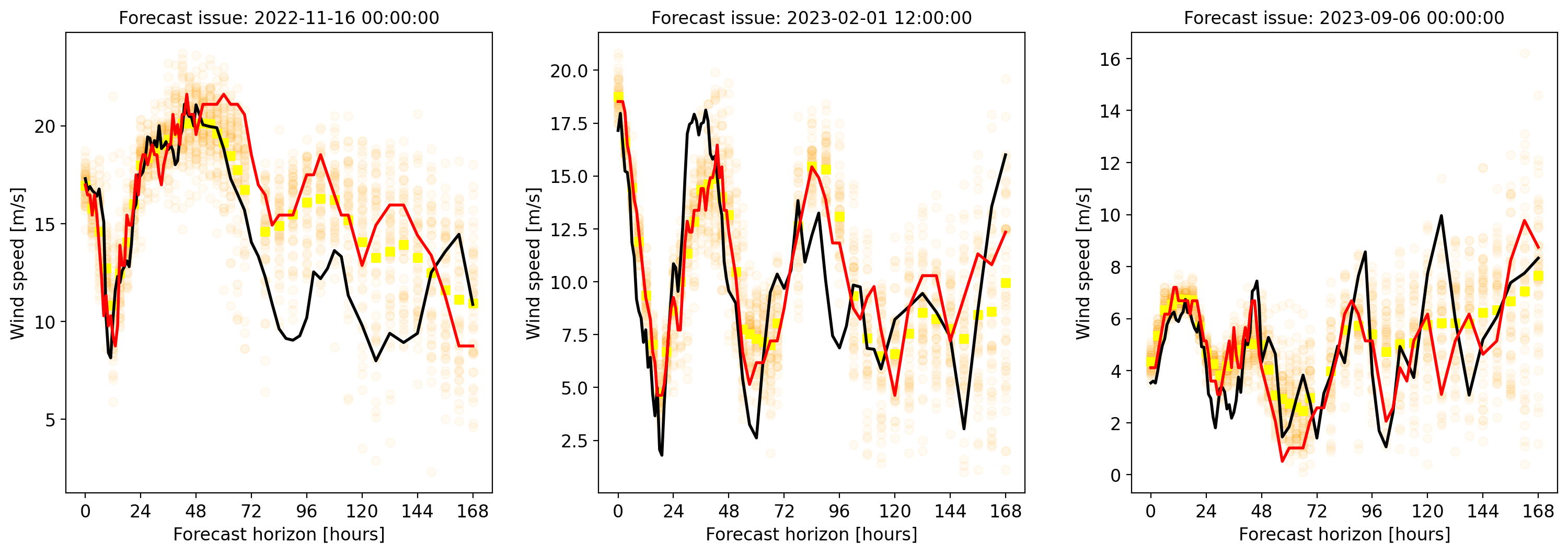}
		\caption{Wind speed $W$.}
	\end{subfigure}
	\begin{subfigure}{1\textwidth}
		\centering
		\includegraphics[width=1\textwidth]{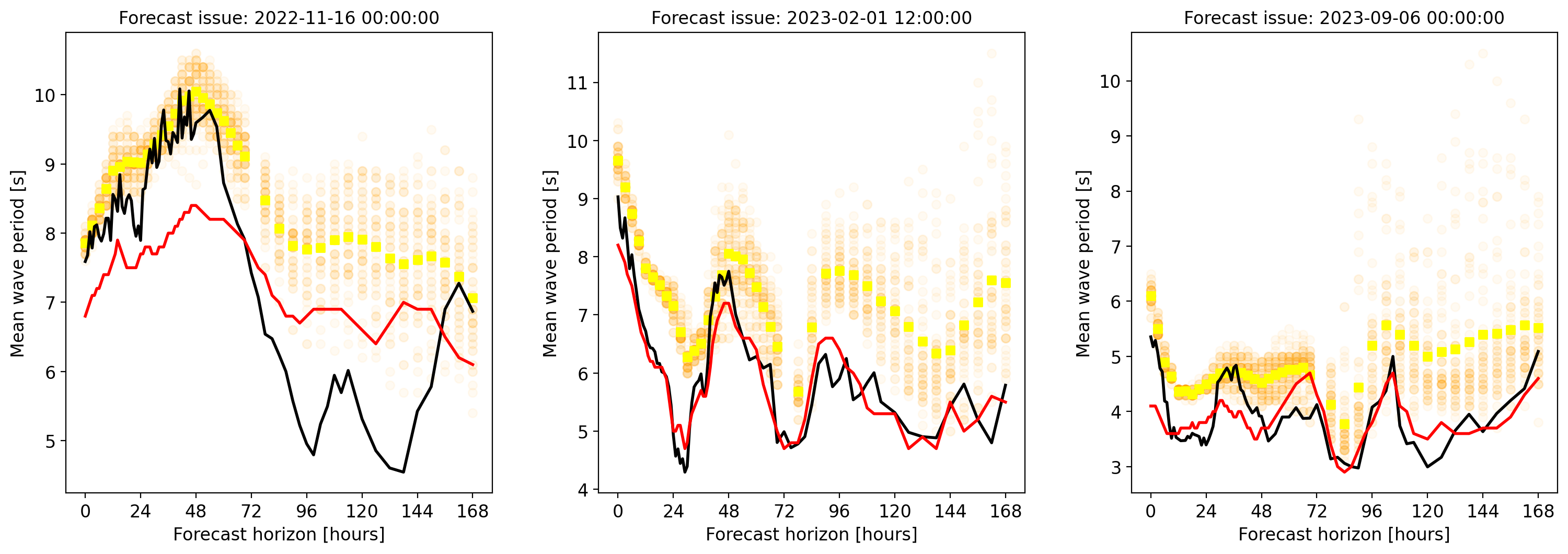}
		\caption{Mean wave period $T_M$.}
	\end{subfigure}
	\caption{Forecasts and (future) reality for forecast components at three illustrative forecast times. Columns show forecasts issued on 16 November 2022 (left), 1 February 2023 (centre) and 6 September 2023 (right) with forecast horizons $\in [1,168]$ hours for (a) significant wave height $H_S$, (b) wind speed $W$ and (c) mean wave period $T_M$. Title of each column gives the forecast issue time. (Future) reality illustrated using black line. Deterministic forecast illustrated using a red line. Forecast ensemble members shown as orange circles, and the ensemble mean as yellow circles. Note that the future reality and deterministic forecast lines are shown as continuous for ease of inspection, whereas in fact they are discrete as described in the text.}
	\label{Fgr:EDA:1}
\end{figure}

Figure~\ref{Fgr:EDA:1} provides illustrations of the data for three forecast issue times as described in the figure caption. For $H_S$ and $W$, the correspondence between reality and forecast components is generally good. For $T_m$, there is evidence that ensemble forecasts are somewhat larger than both reality and the deterministic forecast.

\begin{figure}[!ht]
	\centering
	\begin{subfigure}{1\textwidth}
		\centering
		\includegraphics[width=1\textwidth]{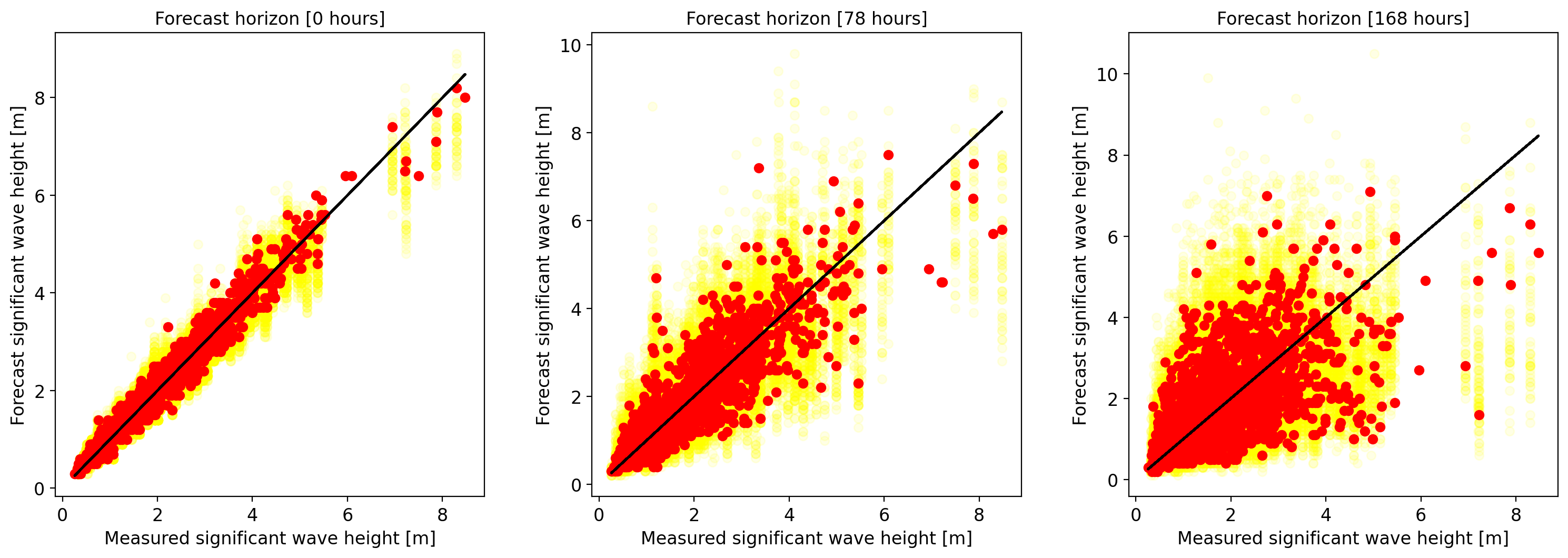}
		\caption{Significant wave height $H_S$.}
	\end{subfigure}
	\begin{subfigure}{1\textwidth}
		\centering
		\includegraphics[width=1\textwidth]{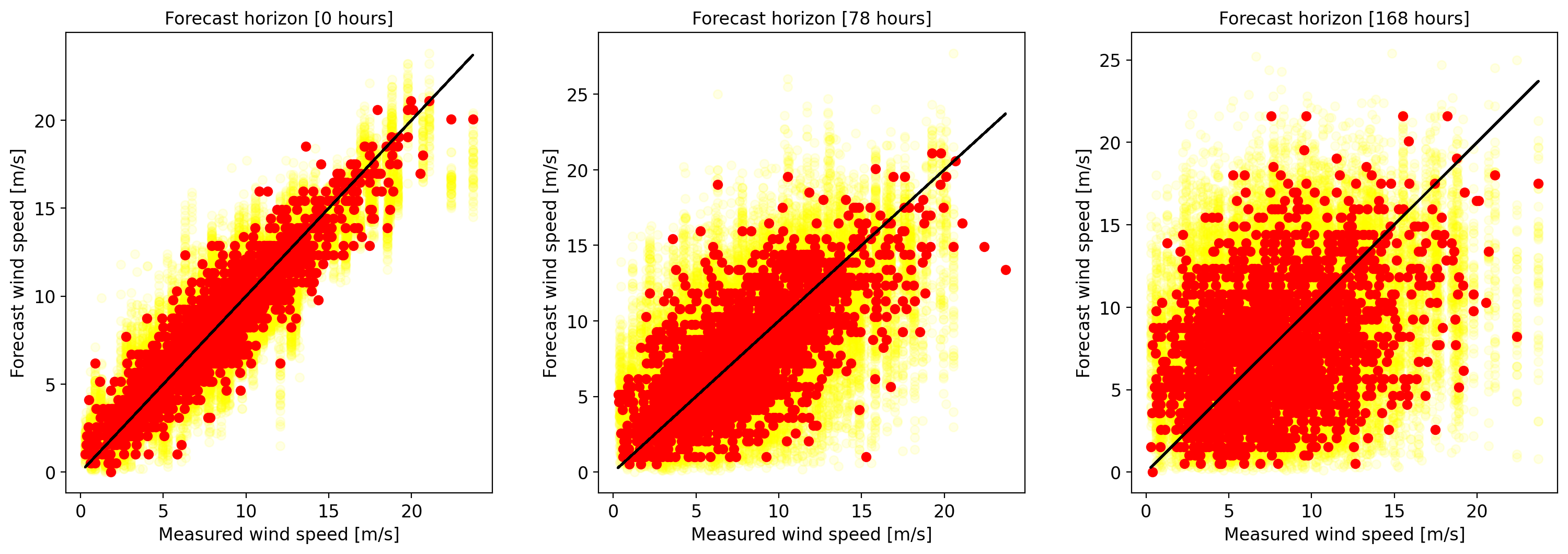}
		\caption{Wind speed $W$.}
	\end{subfigure}
	\begin{subfigure}{1\textwidth}
		\centering
		\includegraphics[width=1\textwidth]{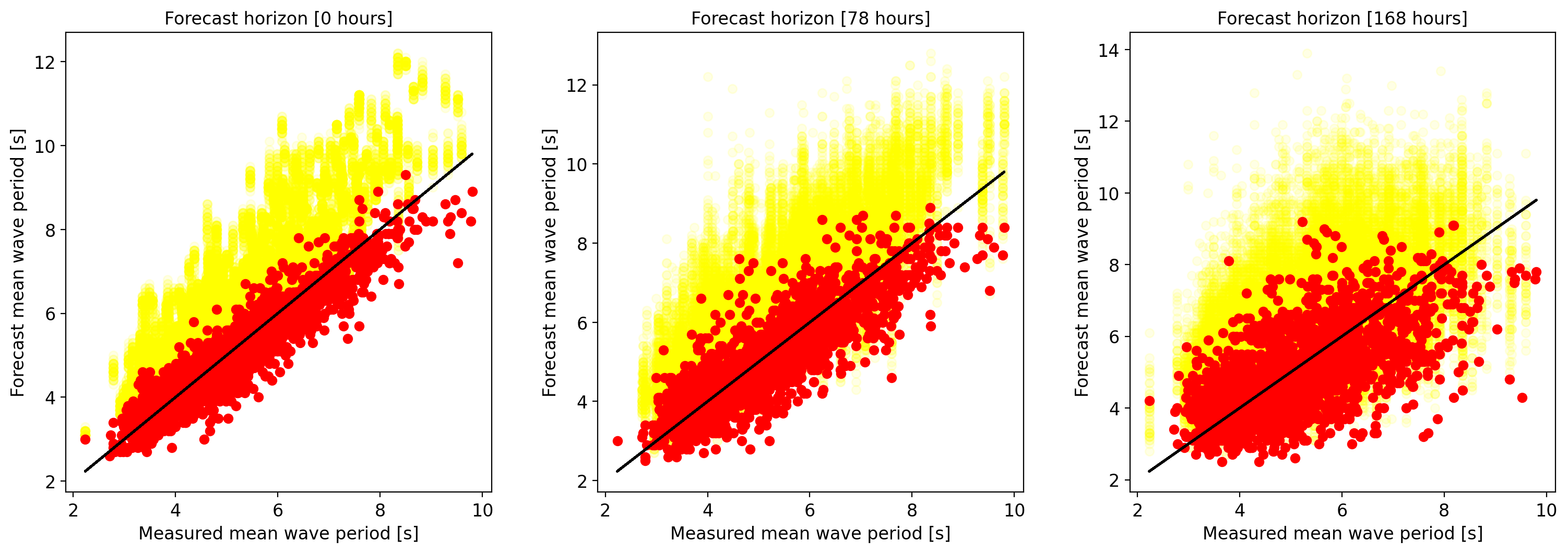}
		\caption{Mean wave period $T_M$.}
	\end{subfigure}
	\caption{Scatter plots of forecast values (y) on measured (x) corresponding to three forecast horizons (columns) for (a) significant wave height $H_S$, (b) wind speed $W$ and (c) mean wave period $T_M$. Title of each panel gives the forecast horizon. Deterministic forecasts shown in red. Individual ensemble member forecasts shown in yellow.}
	\label{Fgr:EDA:2}
\end{figure}

Figure~\ref{Fgr:EDA:2} gives scatter plots of the deterministic forecast on reality, and individual ensemble member forecasts on reality, for three forecast horizons, for the full period of observation. As would be expected, the now-cast prediction (forecast horizon = 0 hours) provides the least scatter about the $y=x$ line of agreement; scatter about this line increases with forecast horizon for all variables. For $H_S$ and $W$, the clouds of red and yellow points are centred on the line of agreement, indicating that the bias in these forecasts is low. The bias of the deterministic forecast for $T_m$ is also relatively low, but considerable bias for the ensemble forecast is present, suggesting that a simple offset (or linear regression calibration) model would improve forecast performance. There is some evidence also, for all variables, that forecasts of the highest values may be biased low (see e.g. the forecast of $H_S$ at forecast horizons of 78 and 168 hours.) The figure therefore suggests that the characteristics of the calibration model should vary with both the value of the actual response, and the forecast horizon.

\begin{figure}[!ht]
	\centering
	\begin{subfigure}{0.9\textwidth}
		\centering
		\includegraphics[width=1\textwidth]{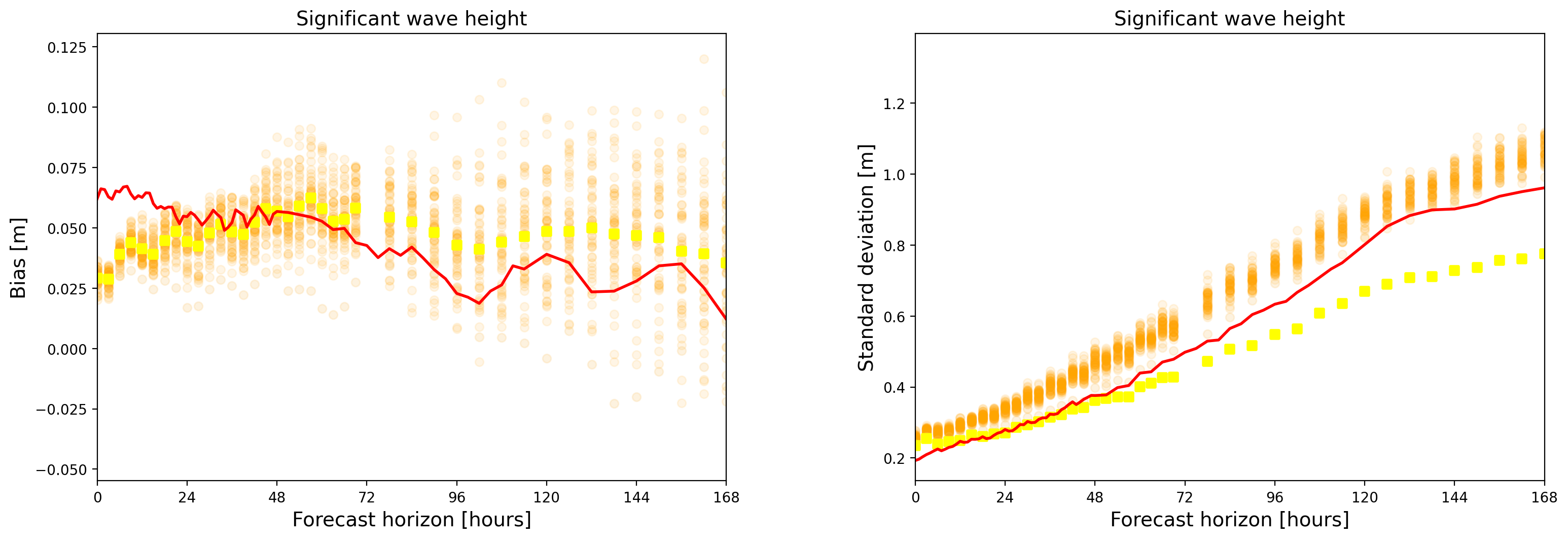}
		\caption{Significant wave height $H_S$.}
	\end{subfigure}
	\begin{subfigure}{0.9\textwidth}
		\centering
		\includegraphics[width=1\textwidth]{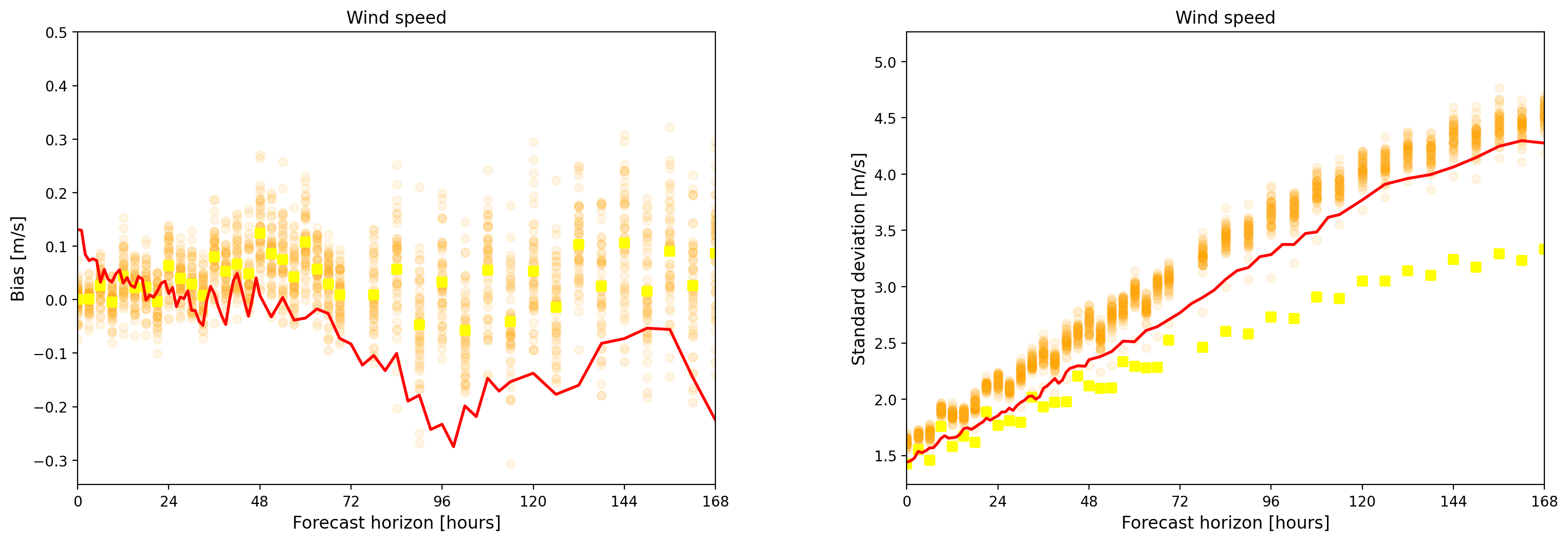}
		\caption{Wind speed $W$.}
	\end{subfigure}
	\begin{subfigure}{0.9\textwidth}
		\centering
		\includegraphics[width=1\textwidth]{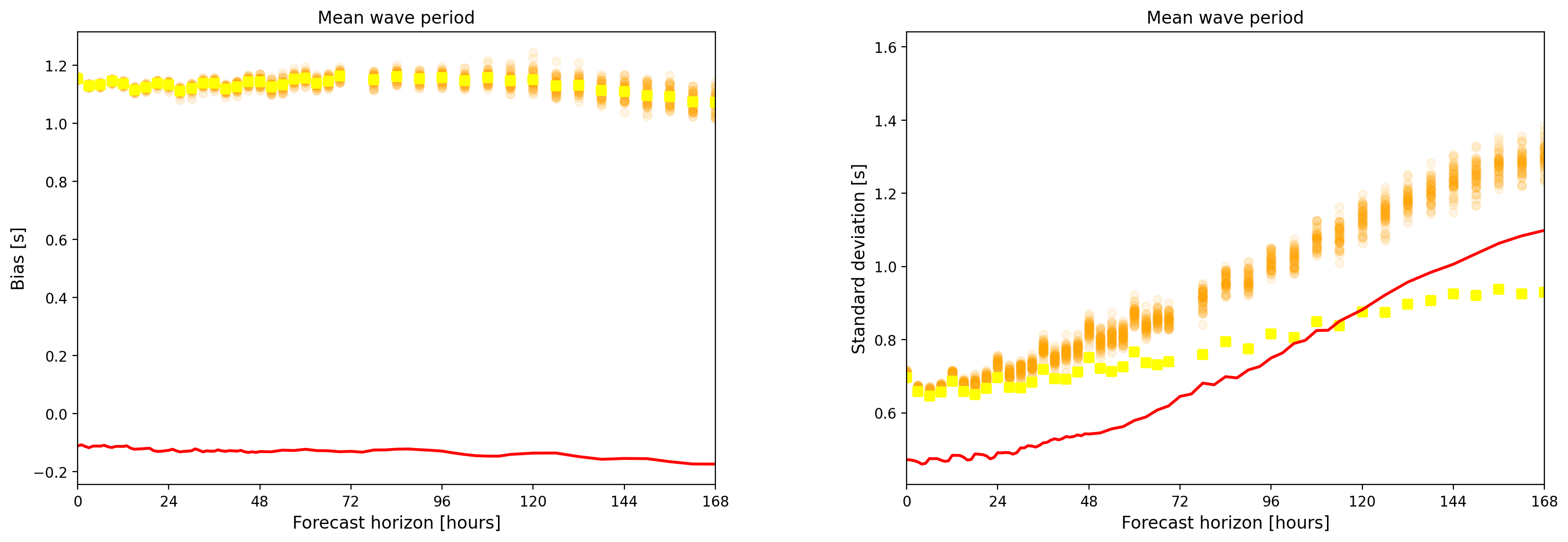}
		\caption{Mean wave period $T_M$.}
	\end{subfigure}
	\caption{Bias (left) and standard deviation (right) of forecast error as a function of horizon for (a) significant wave height $H_S$, (b) wind speed $W$ and (c) mean wave period $T_M$. Bias for deterministic forecast (red line), individual ensemble members (orange circles) and ensemble mean (yellow discs). Note that deterministic forecast bias line is shown as continuous for ease of inspection, whereas in fact it is only available at a discrete set of forecast horizons.}
	\label{Fgr:EDA:3}
\end{figure}

Figure~\ref{Fgr:EDA:3} {quantifies these findings, by} summarising the bias and standard deviation of forecast components for the full period of observation. There is a small positive bias in the deterministic and ensemble mean forecast for $H_S$ of around 0.05 m for all forecast horizons. For $W$, the absolute mean ensemble bias is $\le$ 0.1 m/s over all horizons, whereas bias for the deterministic forecast shows a more systematic trend with absolute mean ensemble bias is $\le$ 0.3 m/s everywhere. For $T_m$, the deterministic forecast shows a typical bias of around -0.1 s everywhere, whereas as the ensemble bias is relatively large at around 1.1 s everywhere. In terms of forecast standard deviation, the expected trend of increasing forecast error with horizon is observed for all variables. At longest horizons, the ensemble mean is seen to provide lower forecast error than the deterministic forecast for all variables; this effect is particularly strong for $H_S$ and $W$. For $T_m$, the deterministic forecast yields lowest standard deviations at short horizons. Again as would be expected, the forecast error from individual (exchangeable) forecast ensemble members is poorer than from the ensemble mean.

\section{Methodology}  \label{Sct:Mth}
%
In this section we introduce two straightforward methods used to establish calibration models, which provide best estimates of (future) reality (i.e. of each of measured $H_S$, $W$ and $T_m$ independently) as a function of forecast components (i.e. the deterministic, control and ensemble forecasts) for the same set of metocean variables. The first approach is simple linear regression (LR), outlined in Section~\ref{Sct:Mth:LR}; the second is an extension of linear regression to accommodate non-stationary error variance, namely non-homogeneous Gaussian regression (NHGR), outlined in Section~\ref{Sct:Mth:NHGR}. The performance of calibration models with different levels of complexity is evaluated as a function of forecast horizon by comparison of values for the Akaike Information Criterion (AIC), as explained in Section~\ref{Sct:Mth:AIC}. Further, for ease of interpretation of regression output, we choose to standardise the covariates in a particular manner, as discussed in Section~\ref{Sct:Mth:Stn}. We stress that these choices of calibration models are meant to represent the simplest approaches the metocean engineer might consider for practical application. Results of the modelling exercise are then presented in Section~\ref{Sct:Rsl}. First, we outline the manner in which we accommodate the exchangeable nature of the ensemble members in the calibration models.

\subsection{Accommodating exchangeable ensemble members in a calibration model} \label{Sct:Mth:ExcEns}
A forecast ensemble of 50 members is available for each of $H_S$, $W$ and $T_m$ for each combination of forecast issue time $t$ hours and forecast horizon $\tau$ hours, with $t,\tau \in \mathbb{Z}_{\ge0}$. These ensemble members are exchangeable, in the sense that for any two forecast issue times $t$ and $t'$, the association between the values $E_j(t)$ and $E_{j'}(t')$ for ensemble members $E_j$ and $E_{j'}$ is no different when $j=j'$ to when $j \neq j'$, $j,j'=1,2,...,50$. Hence it makes no sense to include individual ensemble members directly as covariates in the calibration model. However, we are free to use summary statistics of the 50-member ensemble, such as the ensemble mean $M_E(\tau;t)$ and the ensemble standard deviation $S_E(\tau;t)$, the values of which are unaffected by permutations of the ensemble members (see e.g. \citealt{Gnt14}). For this reason, in the LR and NHGR models below, the only ensemble covariates we consider are the ensemble mean $M_E(\tau;t)$ (in the LR and NHGR mean), and the ensemble standard deviation $S_E(\tau;t)$ (for the NHGR error variance). 

\subsection{Linear regression (LR)} \label{Sct:Mth:LR}
We estimate a linear regression model for a directly-measured metocean variable $Y \ge 0$ (i.e. one of $H_S$, $W$ and $T_m$) at time $t+\tau$ based on forecast components for the \emph{same} metocean quantity; i.e. the deterministic, control and ensemble mean forecasts for that quantity, written $\un{X}(\tau;t)$ $=(X_1(\tau;t), X_2(\tau;t), ..., X_j(\tau;t), ..., X_{n_X}(\tau;t))$, and forecast components $\un{Z}(\tau;t)$ $=(Z_1(\tau;t), Z_2(\tau;t), ..., Z_j(\tau;t), ..., Z_{n_Z}(\tau;t))$ for \emph{other} metocean quantities at forecast issue time $t$ and forecast horizon $\tau$. With $X_j(\tau;t), Z_{k}(\tau;t) \ge 0$ $\forall j, k = 1,2,3, ...$, we write 
\begin{eqnarray}
	Y(t+\tau|\un{X}(\tau;t)=\un{x}, \un{Z}(\tau;t)=\un{z}) \sim N(a(\tau) + \sum_{j=1}^{n_X} b_j(\tau) x_j + \sum_{k=1}^{n_Z} c_k(\tau) z_k,s^2)
	\label{Eqn:LR}
\end{eqnarray}
for real-valued parameters $a(\tau)$, $b_j(\tau)$ and $c_j(\tau)$, for values $\un{x}$ $=(x_1, x_2, ..., x_j, ..., x_{n_X})$ of covariate vector $\un{X}(\tau;t)$, and values $\un{z}$ $=(z_1, z_2, ..., z_j, ..., z_{n_Z})$ of covariate vector $\un{Z}(\tau;t)$. The regression error variance $s^2$ (and standard deviation $s>0$) is assumed fixed for all forecast horizons $\tau$. We estimate the parameters $a$, $b$ and $c$ using maximum likelihood (or equivalently least squares). Once the parameter estimates are available, we estimate $s^2$ as the ratio of the residual sum of squares for the fitted model, and the number of degrees of freedom in the model.

The covariate vector $\un{X}(\tau;t)$ always consists of one or more of the deterministic, control and ensemble mean forecasts for forecast horizon $\tau$ issues at time $t$, for the metocean quantity being predicted. Thus the components of $\un{X}(\tau;t)$ in a model for measured $H_S$ will consist of one or more of the $H_S$ forecast components only; analogously, the covariate vector $\un{Z}(\tau;t)$ will consist of one or more of the forecast components for $W$ and $T_m$ only. We will refer to ``type-$X$'' and ``type-$Z$'' covariates in subsequent sections for clarity and brevity of description.

\subsection{Non-homogeneous Gaussian regression (NHGR)} \label{Sct:Mth:NHGR}
Using the notation of Section~\ref{Sct:Mth:LR}, non-homogeneous Gaussian regression (NHGR) can be seen as an extension of the linear regression in Equation~\ref{Eqn:LR} to non-stationary error variance. Thus, in the current work, the NHGR model form is
\begin{eqnarray}
	Y(t+\tau|\un{X}(\tau;t)=\un{x}, \un{Z}(\tau;t)=\un{z}, S_E(\tau;t)=s_E) \sim N(a(\tau) + \sum_{j=1}^{n_X} b_j(\tau) x_j + \sum_{k=1}^{n_Z} c_k(\tau) z_k, (d(\tau) + e(\tau) s_E)^2)
	\label{Eqn:NHGR}
\end{eqnarray}
That is, the NHGR forecast error standard deviation $d(\tau) + e(\tau) s_E>0$ is assumed to be a linear function of the ensemble forecast standard deviation $s_E>0$; thus, when the ensemble forecast variability is large (e.g. for longer forecast horizons), a larger NHGR forecast error variance is implied (for the same choice of model parameters $d(\tau)$ and $e(\tau)$). {As for LR, NHGR parameters $a$, $b$, $c$, $d$ and $e$ are estimated using maximum likelihood estimation}.

\subsection{Identifying best-performing models using AIC} \label{Sct:Mth:AIC}
We assess the performance of models using AIC, defined as $2p+2\hat{\ell}$, where $p$ is the number of parameters in the model ($p=n_X + n_Z+1$ for LR, and $p=n_X + n_Z + 3$ for NHGR), and $\hat{\ell}$ is the value of the negative log likelihood of the sample for the model evaluated at its minimum with respect to model parameters; that is, $\hat{\ell}$ is simply the sum of negative log Gaussian densities of the sample points at the maximum likelihood solution. Better performing models provide lower values for AIC (see \citealt{Akk74}, \citealt{EmlEA14}). For all responses, and for both LR and NHGR methodologies, we find that including a total of at most three covariates (i.e. $n_X + n_Z \le 3$ in Equations~\ref{Eqn:LR} and \ref{Eqn:NHGR}) reduces AIC sufficiently that addition of further covariates cannot be justified.

\subsection{Standardisation of covariates} \label{Sct:Mth:Stn}
Standardisation of covariates often facilitates more intuitive interpretation of regression analysis. For this reason, in the current work, covariates for forecast components $Z_k(\tau;t)$ ($k=1,2,...,n_Z$) representing different (``type-$Z$'') physical quantities to the response $Y(t+\tau)$ are standardised prior to model fitting so that their sample variance is equal to that of the response. Using this standardisation, the estimated regression coefficient $c_k(\tau)$ indicates exactly the fraction of the response variance explained by covariate $Z_k(\tau;t)$; e.g. $c_k(\tau)=1$ indicates that $Z_k(\tau;t)$ explains exactly 100\% of the response variance. Specifically, to achieve standardisation, we simply substitute the expression 
\begin{eqnarray}
	z_k^*=\frac{\sigma_Y}{\sigma_{Z_k}} (z_k - {\mu_{Z_k}})
	\label{Eqn:Stn}
\end{eqnarray}
for $z_k$ in Equations~\ref{Eqn:LR} and ~\ref{Eqn:NHGR}, where for $\sigma_Y>0$ and $\sigma_{Z_k}>0$ we use sample estimates for the standard deviation of $Y(t+\tau)$ and $Z_k(\tau;t)$, and for $\mu_{Z_k}$ we use the sample mean of $Z_k(\tau;t)$. However, covariates $X_j(\tau;t)$ ($j=1,2,...,n_X$) representing the same (``type-$X$'') physical quantity as the response are not standardised. In this case, an estimate for parameter $b_j>1$ would indicate that the variance of the corresponding forecast variable is smaller than that of the response. We stress that standardisation of covariates does not affect predictions made using the LR and NHGR models, nor the choice of optimal models; it is merely a convenient transformation to aid interpretation of regression coefficients. (For comparison, the {Supplementary Material} provides plots of relative contributions of covariates to the regression in which both response and all covariates are standardised to zero mean and unit standard deviation; see Figures~SM1 and SM3).

\FloatBarrier
\section{Results}  \label{Sct:Rsl}
%
For each metocean response variable $Y(t+\tau)$ (i.e. measured $H_S$, $W$ and $T_m$), we estimate LR and NHGR calibration models independently, using covariates $X(\tau;t)$ (i.e. one or more of the deterministic, ensemble mean and control forecasts for the same metocean variable) and covariates $Z(\tau;t)$ (i.e. one or more of the deterministic, ensemble mean and control forecasts for different metocean variables). The estimated LR and NHGR models are then illustrated in more detail in Sections~\ref{Sct:Rsl:LR} and \ref{Sct:Rsl:NHGR}. Supporting plots are provided in on-line Supplementary Material (SM). Where possible, uncertainties in parameters and predictions are quantified using 95\% confidence and Gaussian forecast uncertainty intervals available in closed form. Otherwise bootstrap resampling is used to estimate corresponding 95\% uncertainty bands (estimated from repeated model fitting to bootstrap resamples of the original sample for model fitting). \ed{The characteristics of the original uncalibrated deterministic forecast, and those of LR- and NHGR-calibrated forecasts, are assessed within-sample in Section 4.3. In Section 4.4, we assess the performance of the three forecasts using data for two subsequent time periods, not used for calibration model estimation.}

\FloatBarrier
\subsection{Linear regression}  \label{Sct:Rsl:LR}
\begin{figure}[!ht]
	\centering
	\begin{subfigure}{1\textwidth}
		\centering
		\includegraphics[width=1\textwidth]{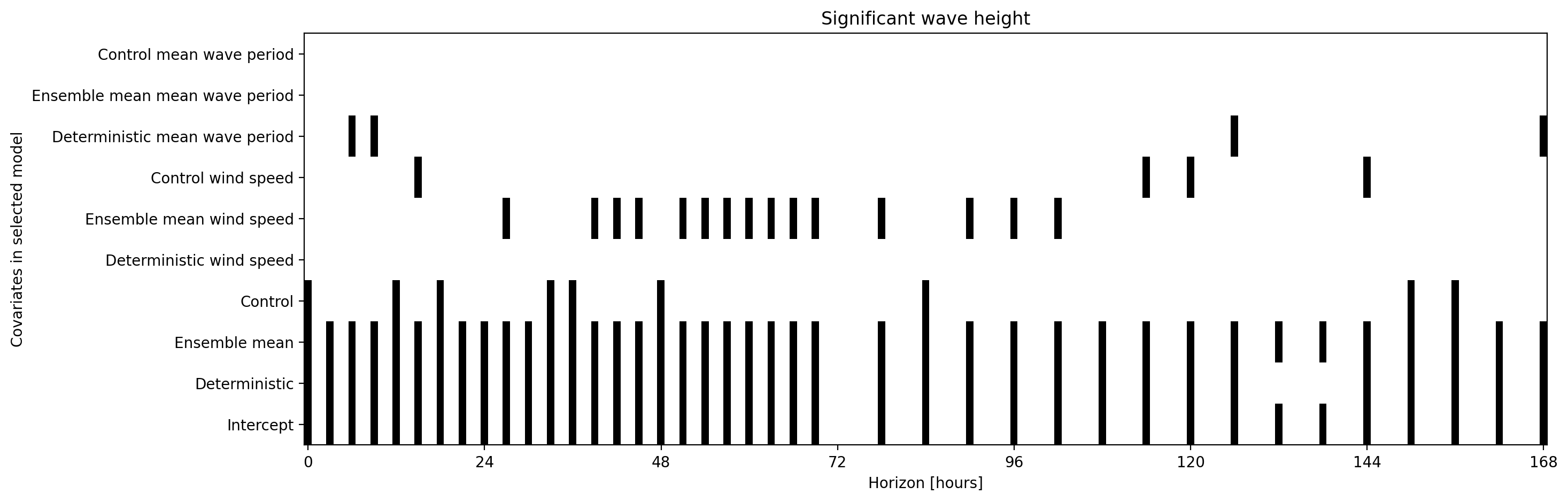}
		\caption{Significant wave height $H_S$.}
	\end{subfigure}
	\begin{subfigure}{1\textwidth}
		\centering
		\includegraphics[width=1\textwidth]{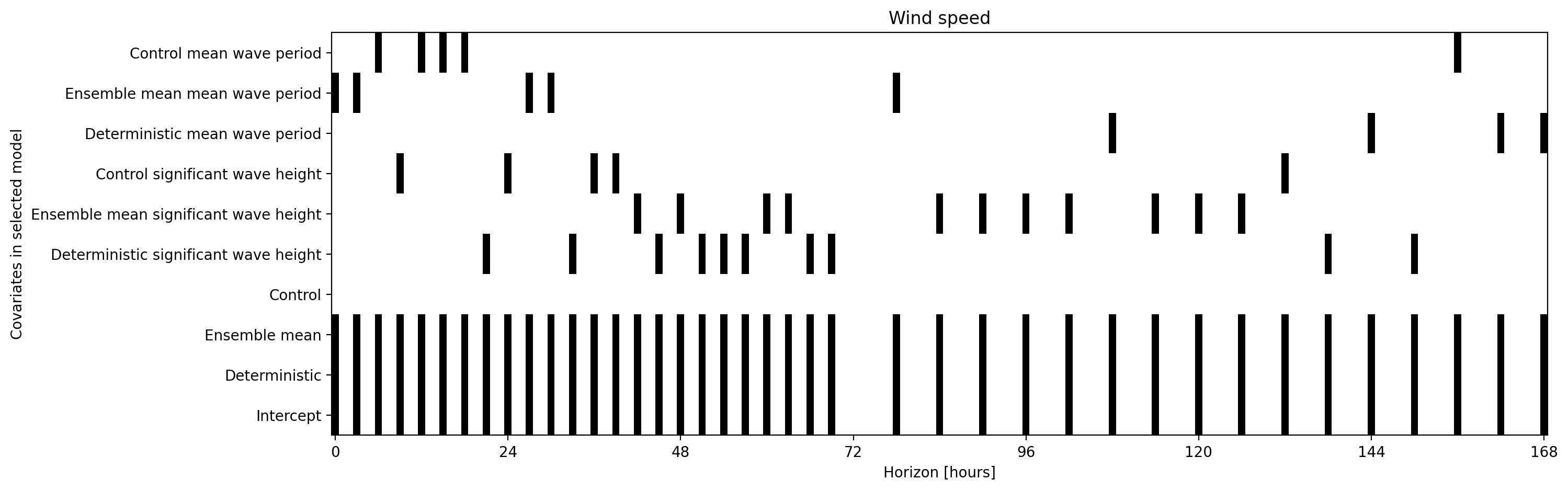}
		\caption{Wind speed $W$.}
	\end{subfigure}
	\begin{subfigure}{1\textwidth}
		\centering
		\includegraphics[width=1\textwidth]{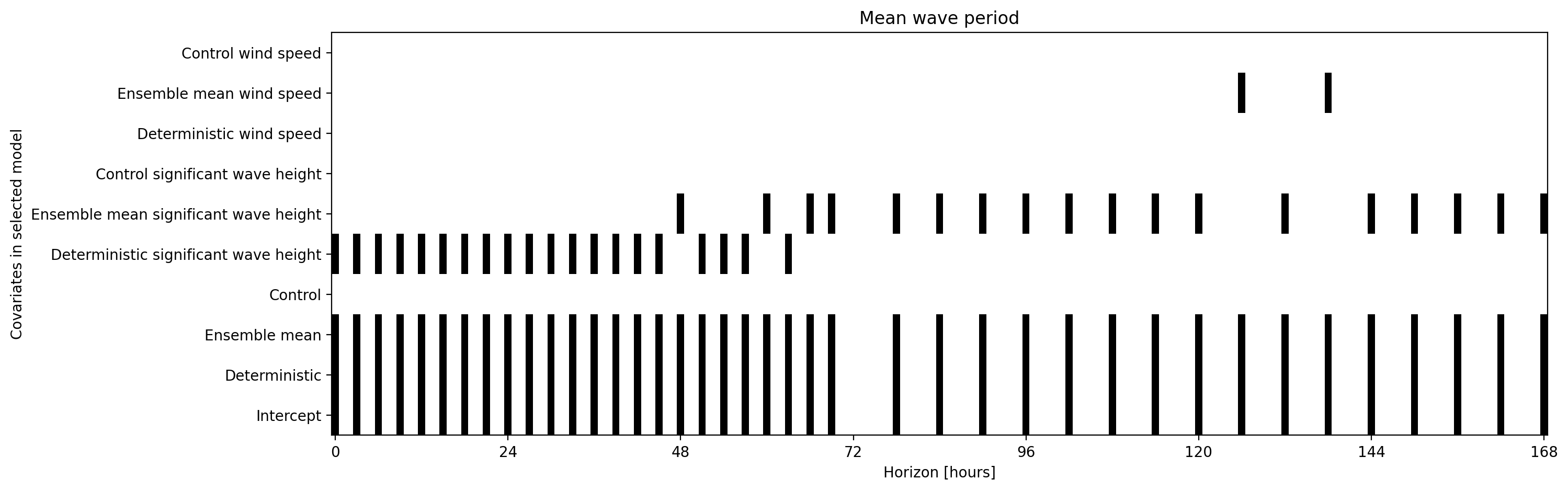}
		\caption{Mean wave period $T_M$.}
	\end{subfigure}
	\caption{Bar code plot of covariates included in optimal linear regression models for (a) significant wave height $H_S$, (b) wind speed $W$ and (c) mean wave period $T_m$ for each forecast horizon. Terms included in consistent LR calibration model for each response are given in Table~\ref{Tbl:LR}.}
	\label{Fgr:LR:CvrBar}
\end{figure}

For each combination of metocean response variable and forecast horizon, the optimal LR model was identified as that which minimises AIC, with the constraint that $n_X+n_Z \le 3$. Bar code plots of the covariates present in optimal LR models for each forecast horizon are shown in Figure~\ref{Fgr:LR:CvrBar}.
Inspection of the figure suggests that the ensemble mean and deterministic forecasts (for the same physical quantity as the response) occur almost always in calibration models; the regression intercept is always imposed. A ``consistent'' model (with the same model form over all forecast horizons for a given variable) was then selected as that which minimises AIC most often over all forecast horizons. The chosen consistent model forms are listed in Table~\ref{Tbl:LR}.
\begin{table} [h!]
	\centering
	\begin{tabular} { | c | c | c | }
		\hline
		$H_S$ & $W$ & $T_m$ \\
		\hline
		Intercept & Intercept & Intercept \\
		Deterministic $H_S$ & Deterministic $W$ & Deterministic $T_m$\\
		Ensemble mean $H_S$ & Ensemble mean $W$ & Ensemble mean $T_m$\\
		Ensemble mean $W$ & Ensemble mean $H_S$ & Deterministic $H_S$ \\
		\hline
	\end{tabular}
	\caption{Terms included in the consistent LR calibration models for significant wave height $H_S$ (left), wind speed $W$ (centre) and mean wave period $T_m$ (right).}
	\label{Tbl:LR}
\end{table}

\begin{figure}[!ht]
	\centering
	\includegraphics[width=0.32\textwidth]{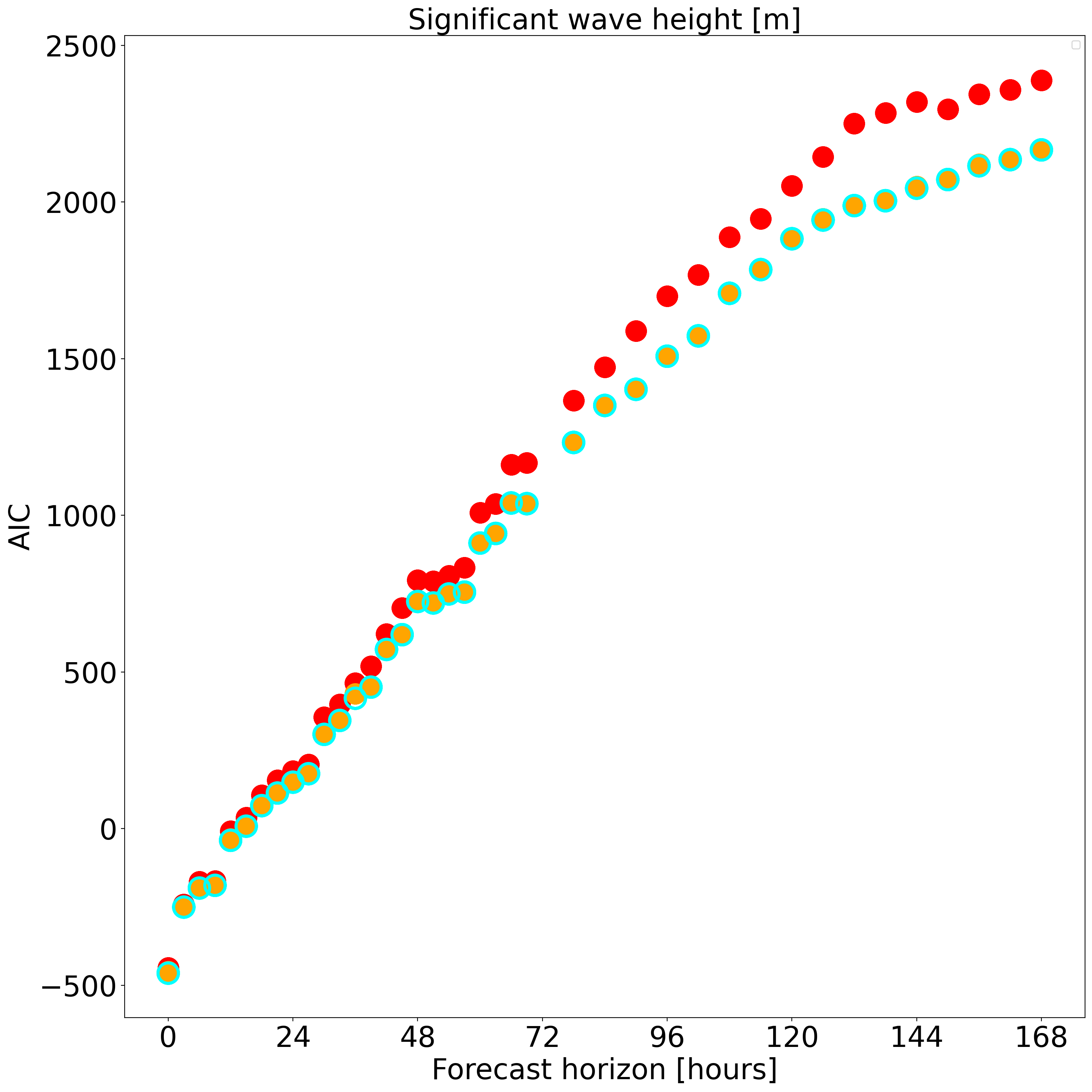}
	\includegraphics[width=0.32\textwidth]{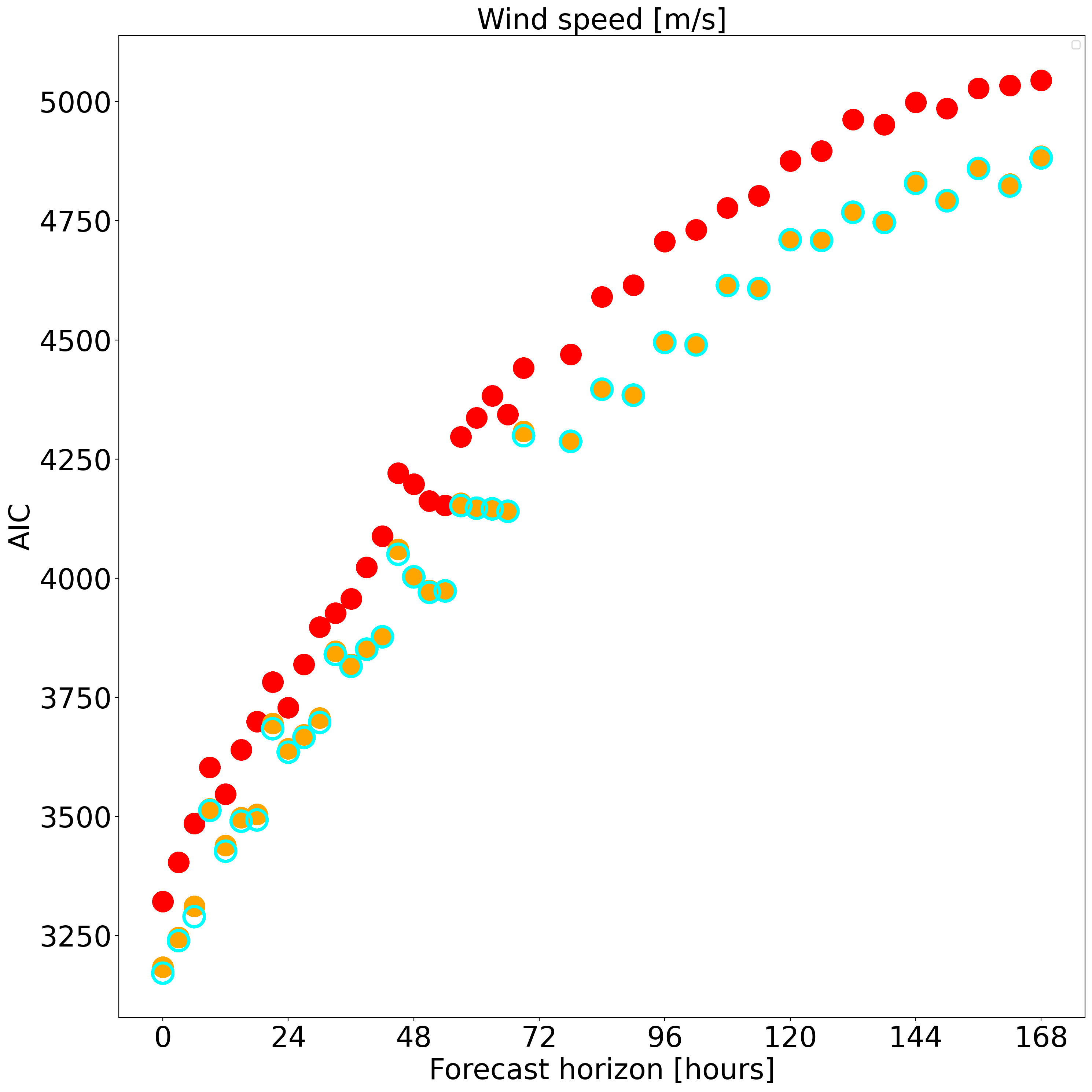}
	\includegraphics[width=0.32\textwidth]{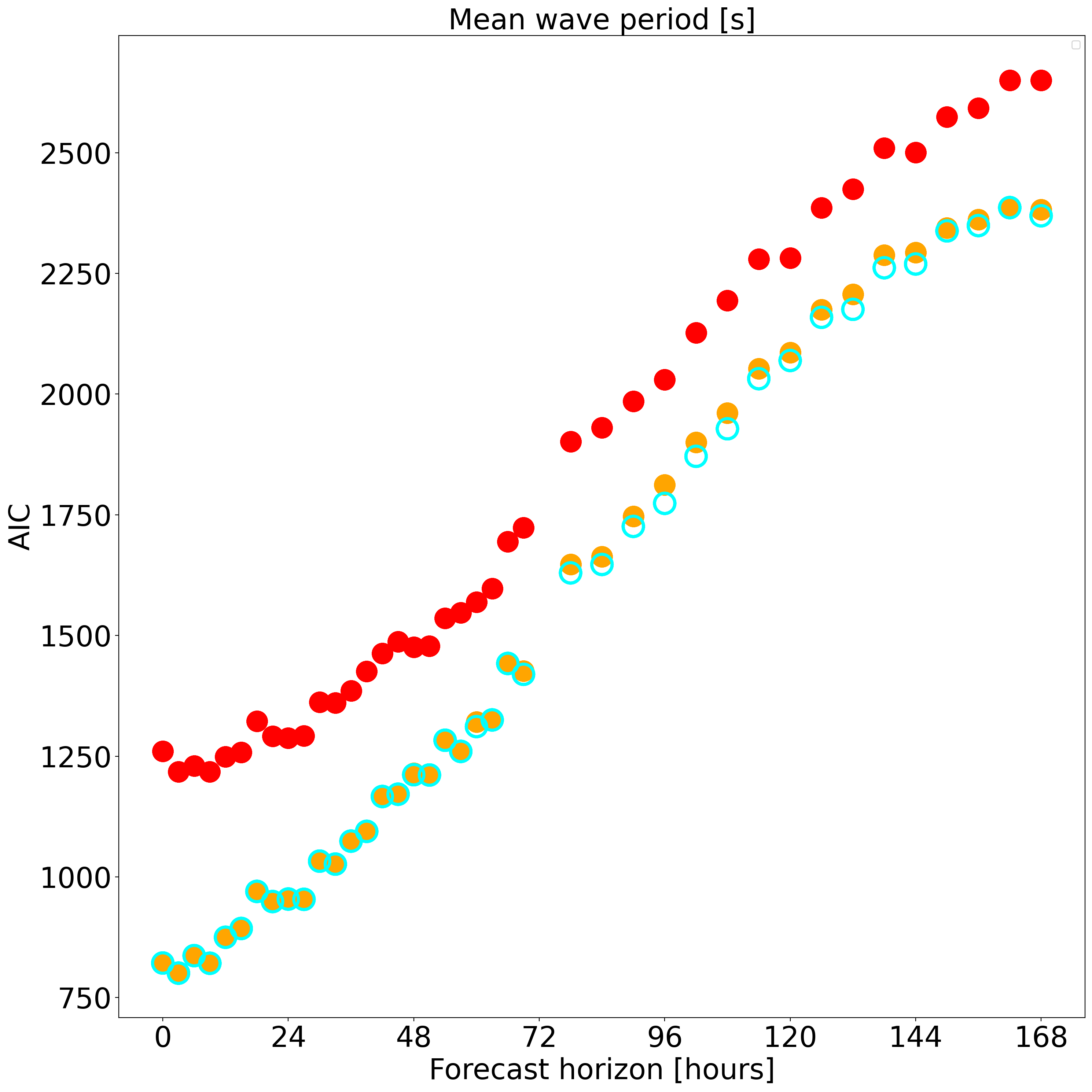}
	\caption{Variation of AIC with forecast horizon for significant wave height ($H_S$, left), wind speed ($W$, centre) and mean wave period ($T_M$, right). Each panel shows the growth of AIC with forecast horizon, for three calibration models: based on the deterministic forecast only (red disc), based on the consistent LR calibration model (cyan circle) and based on the optimal LR calibration model choice for that forecast horizon (orange disc). The consistent model form for each of $H_S$, $W$ and $T_M$ is given in Table~\ref{Tbl:LR}.} 
	\label{Fgr:LR:AIC}
\end{figure}

Figure~\ref{Fgr:LR:AIC} then illustrates the growth of AIC with forecast horizon of LR calibration models for each of $H_S$, $W$ and $T_M$. To provide a scale for comparisons, each panel of the figure gives AIC values for a calibration model using the deterministic forecast only (red), using the consistent calibration model for each response (cyan, see Table~\ref{Tbl:LR}), and using the optimal calibration model for the combination of response and forecast horizon (orange, see Figure~\ref{Fgr:LR:CvrBar}). We see that LR calibration using additional ensemble and deterministic covariates improves performance for all responses and forecast horizons relative to calibration using the deterministic forecast (for the same physical quantity as the response) only. Only for $H_S$ and short forecast horizons is the deterministic forecast calibration competitive. We note also that the consistent calibration model is competitive with the optimal model for all variables and forecast horizons, indicating that there is little benefit in selecting different LR calibration model forms for different forecast horizons. More generally, as might be expected, calibration model performance degrades with increasing forecast horizon. 

\begin{figure}[!ht]
	\centering
	\begin{subfigure}{1\textwidth}
		\centering
		\includegraphics[width=1\textwidth]{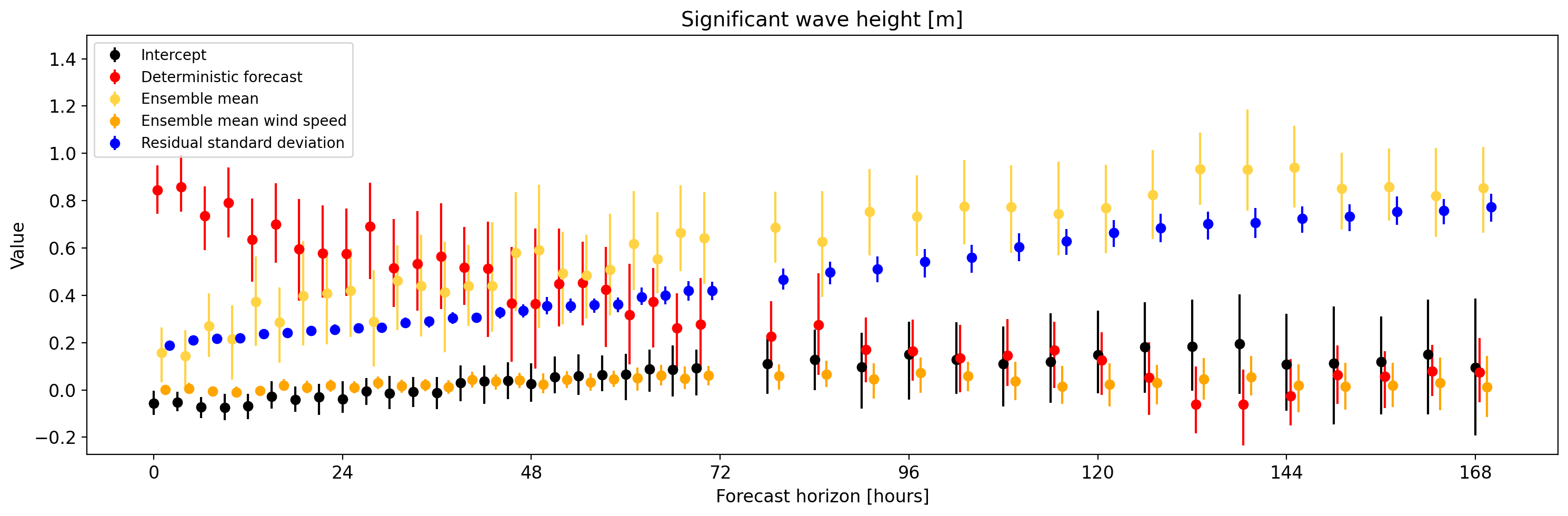}
		\caption{Significant wave height $H_S$}
	\end{subfigure}	
	\begin{subfigure}{1\textwidth}
		\centering
		\includegraphics[width=1\textwidth]{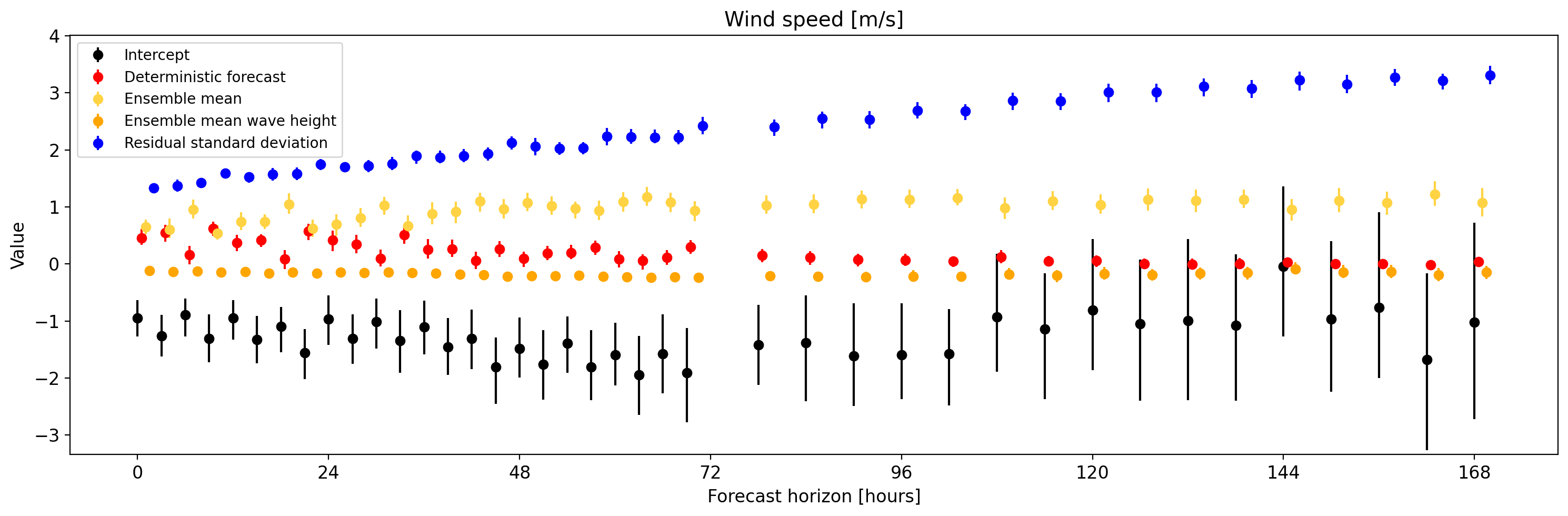}
		\caption{Wind speed $W$}
	\end{subfigure}
	\begin{subfigure}{1\textwidth}
		\centering
		\includegraphics[width=1\textwidth]{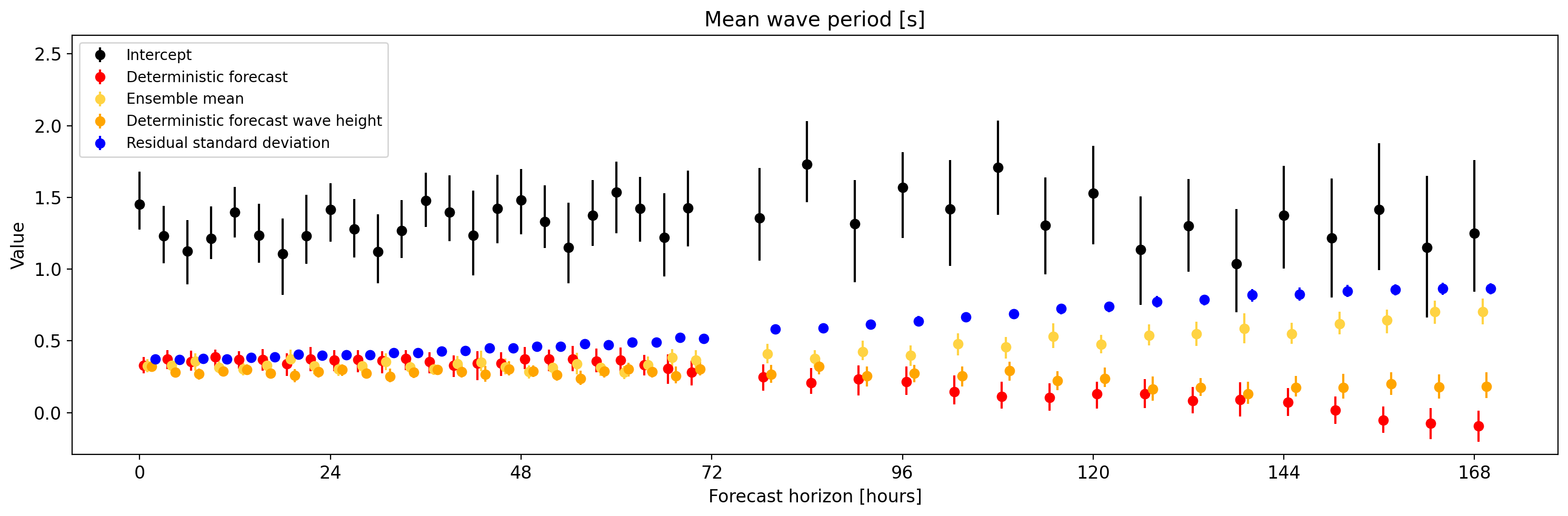}
		\caption{Mean wave period $T_M$}
	\end{subfigure}
	\caption{Variation of estimated parameters with forecast horizon from consistent LR calibration models for (a) significant wave height $H_S$, (b) wind speed $W$ and (c) mean wave period $T_M$. Consistent LR model forms are given in Table~\ref{Tbl:LR}. For each parameter, the mean parameter estimate is indicated by a disc, and the 95\% confidence interval by a vertical line. The intercept is given in black, the deterministic and ensemble mean forecasts (for the same physical quantity) in red and light orange. The remaining covariate (in dark orange) is specified in the plot legend for each panel. Also shown is the regression residual standard deviation (blue). Note that covariates are standardised as described in Section~\ref{Sct:Mth:Stn}.} \label{Fgr:LR:PrmEst}
\end{figure}

Parameter estimates for the consistent LR calibration models are illustrated in Figure~\ref{Fgr:LR:PrmEst}. Because of the standardisation of covariates employed, we expect parameter estimates for covariates to lay in the interval $(0,1)$ generally, which is the case. We also report 95\% confidence bands for the parameter estimates. For $H_S$ in Figure~\ref{Fgr:LR:PrmEst}(a), the deterministic forecast (for $H_S$, in red) dominates for short forecast horizons, but the ensemble mean forecast (for $H_S$, in light orange) dominates for long forecast horizons. The roles of the intercept (black) and ensemble mean $W$ (dark orange) are minor. The residual standard deviation (blue) grows from 0.2 m for shortest forecast horizons to approximately 0.7 m at 168 hours forecast horizon. For $W$ in Figure~\ref{Fgr:LR:PrmEst}(b), the ensemble mean forecast $W$ plays a more important role at all forecast horizons. The intercept provides a correction of around -1 m/s on average over all forecast horizons. Again the effect of the covariate corresponding to a Z-type physical quantity different to the response is small. For $T_m$ in Figure~\ref{Fgr:LR:PrmEst}(c), the contributions of all covariates are similar for short forecast horizons, but the ensemble mean forecast for $T_m$ dominates at larger horizons. The intercept provides a correction of around 1.2 m/s on average over all forecast horizons. 
These trends are further summarised in Figures~SM1 and SM3 in terms of plots of variable contributions. 

\begin{figure}[!ht]
	\centering
	\begin{subfigure}{1\textwidth}
		\includegraphics[width=1\textwidth]{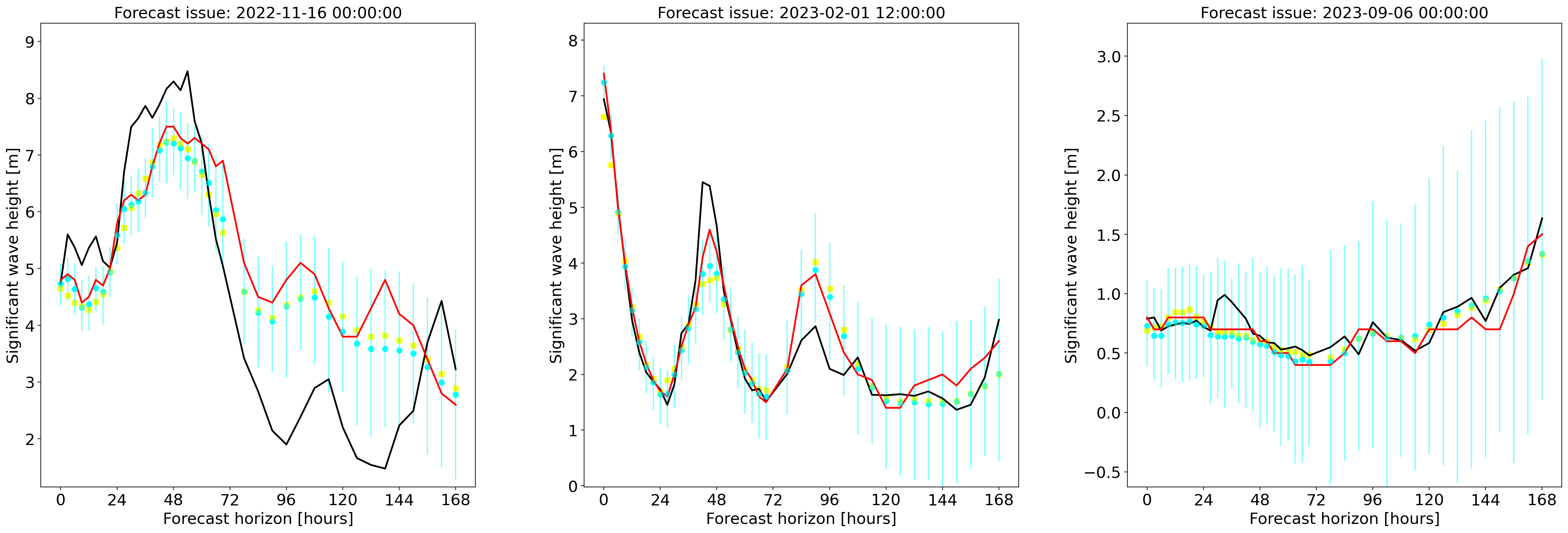}
		\caption{Significant wave height $H_S$.}
	\end{subfigure}
	\begin{subfigure}{1\textwidth}
		\includegraphics[width=1\textwidth]{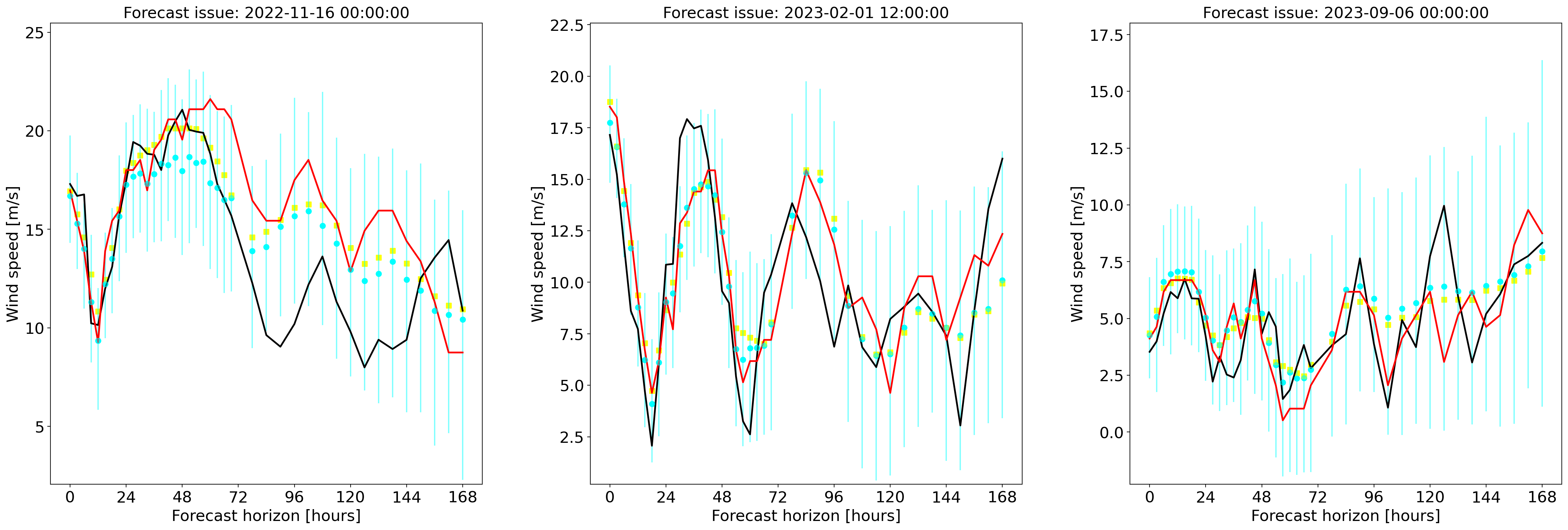}
		\caption{Wind speed $W$.}
	\end{subfigure}
	\begin{subfigure}{1\textwidth}
		\includegraphics[width=1\textwidth]{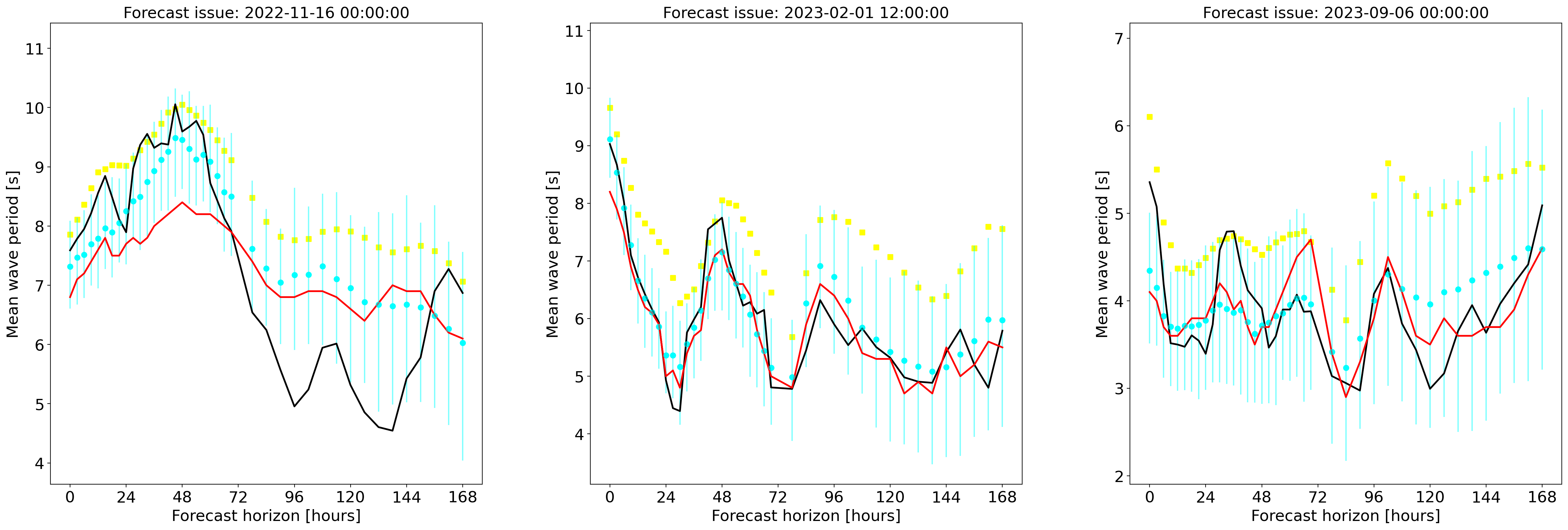}
		\caption{Mean wave period $T_M$.}
	\end{subfigure}
	\caption{Consistent LR-calibrated forecasts and (future) reality for given variable at given forecast time. Three examples (columns) of forecasts on horizons $\in [1,168]$ for (a) significant wave height $H_S$, (b) wind speed $W$ and (c) mean wave period $T_M$. The title of each column gives the time of forecast issue. (Future) reality illustrated using black line. The (uncalibrated) deterministic forecast illustrated using a red line. The (uncalibrated) ensemble mean forecast is shown as yellow discs. Optimal calibrated forecast given in cyan, in terms of the mean (disc) and 95\% Gaussian forecast uncertainty band (vertical line). Note that (future) reality and deterministic forecast are shown as lines (rather than as discrete time points) for ease of interpretation. The forms for the consistent LR-calibrated models are given in Table~\ref{Tbl:LR}.}
	\label{Fgr:LR:ClbFrcExm}
\end{figure}

Figure~\ref{Fgr:LR:ClbFrcExm} provides LR calibrated forecasts and (future) reality for each metocean response variable at three given forecast issue times. The actual measured response is shown in black, and the corresponding deterministic forecast in red. The ensemble mean forecast is in yellow. Note that both the deterministic and ensemble mean forecasts are uncalibrated. The consistent LR calibrated forecasts are shown as the mean (cyan disc) and estimated 95\% Gaussian forecast uncertainty band (calculated using the estimated model error standard deviation, see Equation~\ref{Eqn:LR}; cyan vertical line). For both $H_S$ and $W$, there is little evidence that either the ensemble mean forecast or the mean calibrated LR forecast provide much improvement over the deterministic forecast. The previously-noted bias in the ensemble mean forecast for $T_M$ is clear. Visual inspection suggests that the prediction band, increasing in width with forecast horizon, generally does a relatively good job of including actual (future) reality.

\FloatBarrier
\subsection{NHGR}  \label{Sct:Rsl:NHGR}
A model selection procedure similar to that described in Section~\ref{Sct:Rsl:LR} was used to identify optimal NHGR calibration models for each combination of metocean response variable and forecast horizon. The resulting optimal models are illustrated in the bar plot in Figure~SM2, and the consistent model forms given in Table~\ref{Tbl:NHGR}. As for optimal LR calibration, inspection of Figure~SM2 shows that both the deterministic and ensemble mean forecast (type-$X$ covariates, for the same metocean quantity, see Section~\ref{Sct:Mth:LR}) are again included in the model, but that the type-$Z$ covariate differs for NHGR calibration; e.g. for $H_S$, LR calibration favours ensemble mean $W$, whereas NHGR calibration favours ensemble mean $T_m$. Note also that for all forecast horizons, a term in the ensemble standard deviation $s_E$ (for the same metocean quantity) is included.
\begin{table} [h!]
	\centering
	\begin{tabular} { | c | c | c | }
		\hline
		$H_S$ & $W$ & $T_m$ \\
		\hline
		Intercept & Intercept & Intercept \\
		Deterministic $H_S$ & Deterministic $W$ & Deterministic $T_m$\\
		Ensemble mean $H_S$ & Ensemble mean $W$ & Ensemble mean $T_m$\\
		Ensemble mean $T_m$ & Deterministic $H_S$ & Deterministic $H_S$ \\
		\hline
	\end{tabular}
	\caption{Terms included in the consistent NHGR calibration models for significant wave height $H_S$ (left), wind speed $W$ (centre) and mean wave period $T_m$ (right).}
	\label{Tbl:NHGR}
\end{table}

\begin{figure}[!ht]
	\centering
	\includegraphics[width=0.32\textwidth]{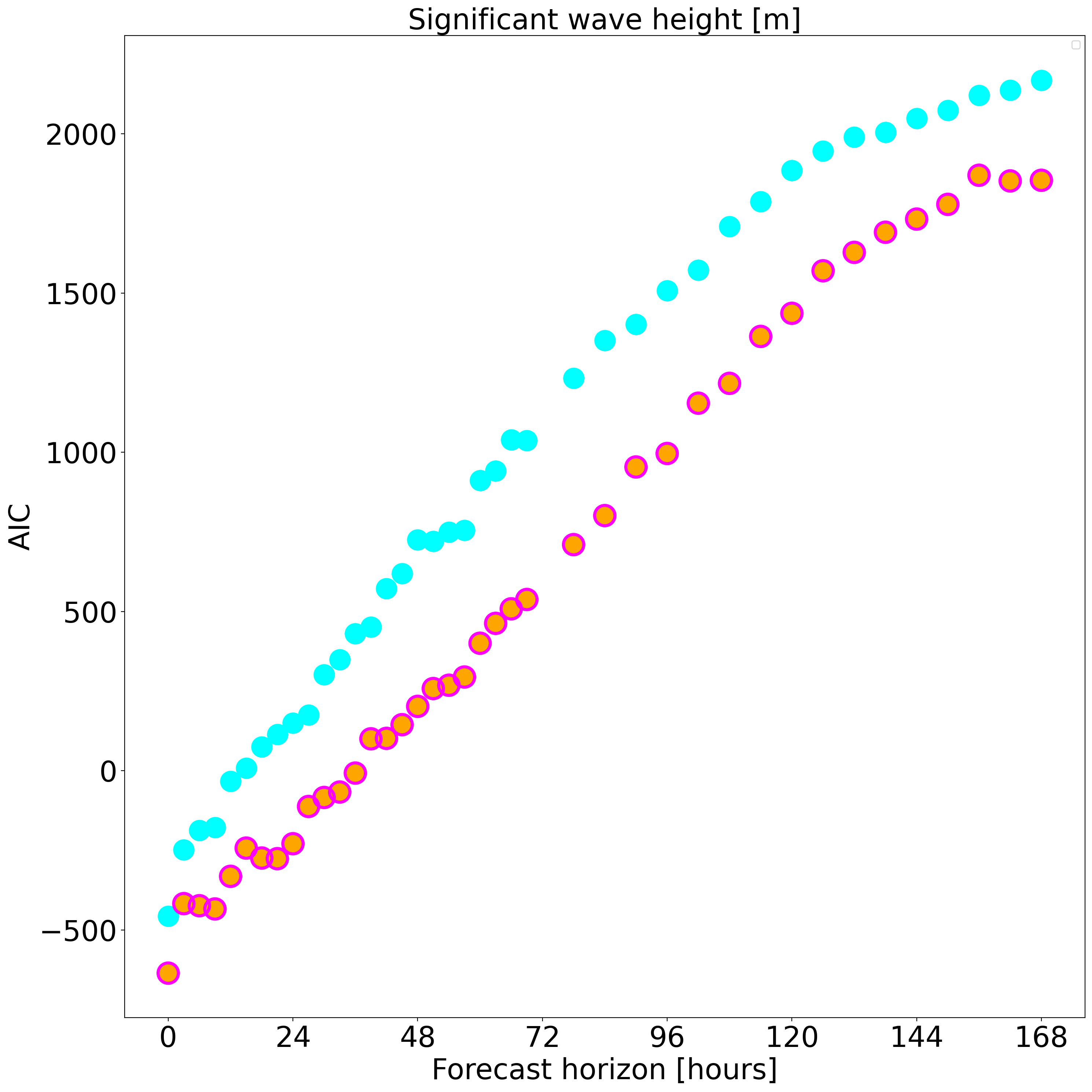}
	\includegraphics[width=0.32\textwidth]{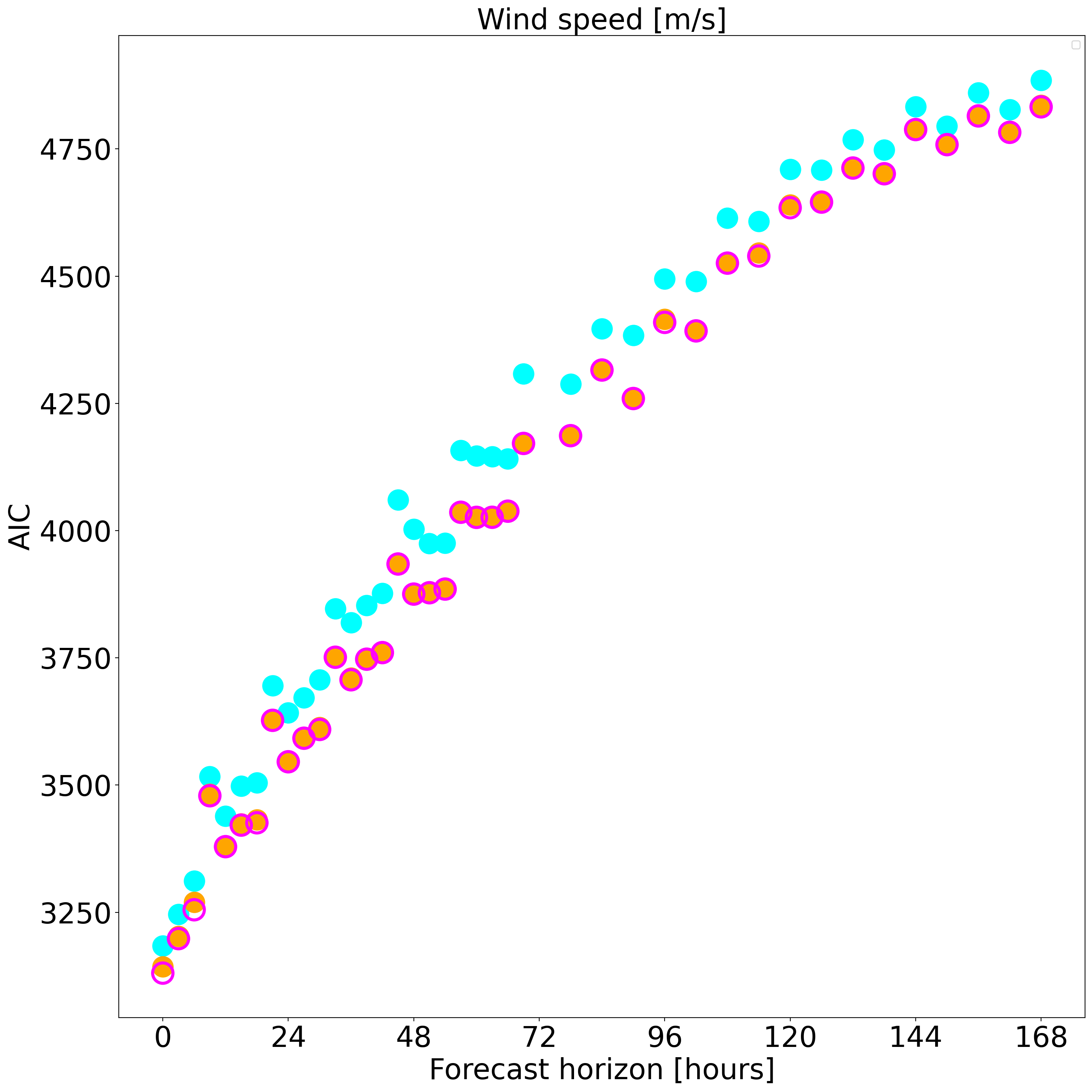}
	\includegraphics[width=0.32\textwidth]{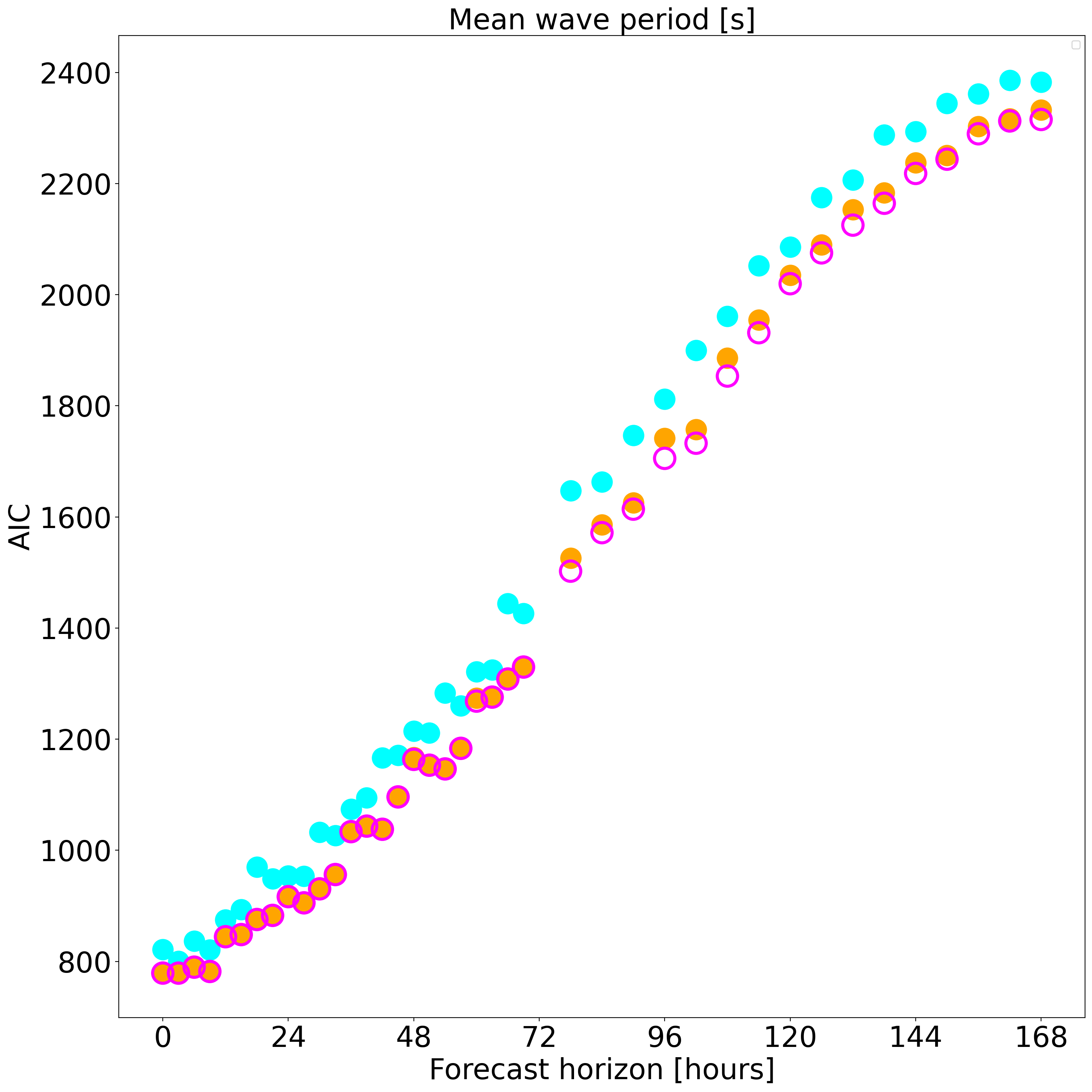}
	\caption{Variation of AIC with forecast horizon for significant wave height ($H_S$, left), wind speed ($W$, centre) and mean wave period ($T_M$, right). Each panel shows the growth of AIC with forecast horizon, for three calibration models: based on the consistent model for linear regression (cyan disc), based on the consistent model for NHGR (orange disc) and based on the optimal NHGR model choice (red circle). The consistent NHGR model form for each of $H_S$, $W$ and $T_M$ is identified in Table~\ref{Tbl:NHGR}.}
	\label{Fgr:NHGR:AIC}
\end{figure}

Model performance for NHGR calibration is again assessed using AIC, and illustrated in Figure~\ref{Fgr:NHGR:AIC}. As a reference, AIC from consistent LR calibration is provided (as cyan discs), together with AIC for consistent NHGR calibration (orange discs) and for optimal NHGR calibration per forecast horizon (red circles). We see from the three panels that (i) consistent NHGR calibration is an improvement on consistent LR calibration, and (ii) there is minimal difference between the performance of consistent and optimal NHGR calibration. There is some evidence however that a different choice of type-$Z$ covariate (specifically, ensemble mean $H_S$ forecast instead of deterministic $H_S$ forecast; see Figure~SM2) would improve performance for $T_m$ at long forecast horizons.

\begin{figure}[!ht]
	\centering
	\begin{subfigure}{1\textwidth}
		\centering
		\includegraphics[width=1\textwidth]{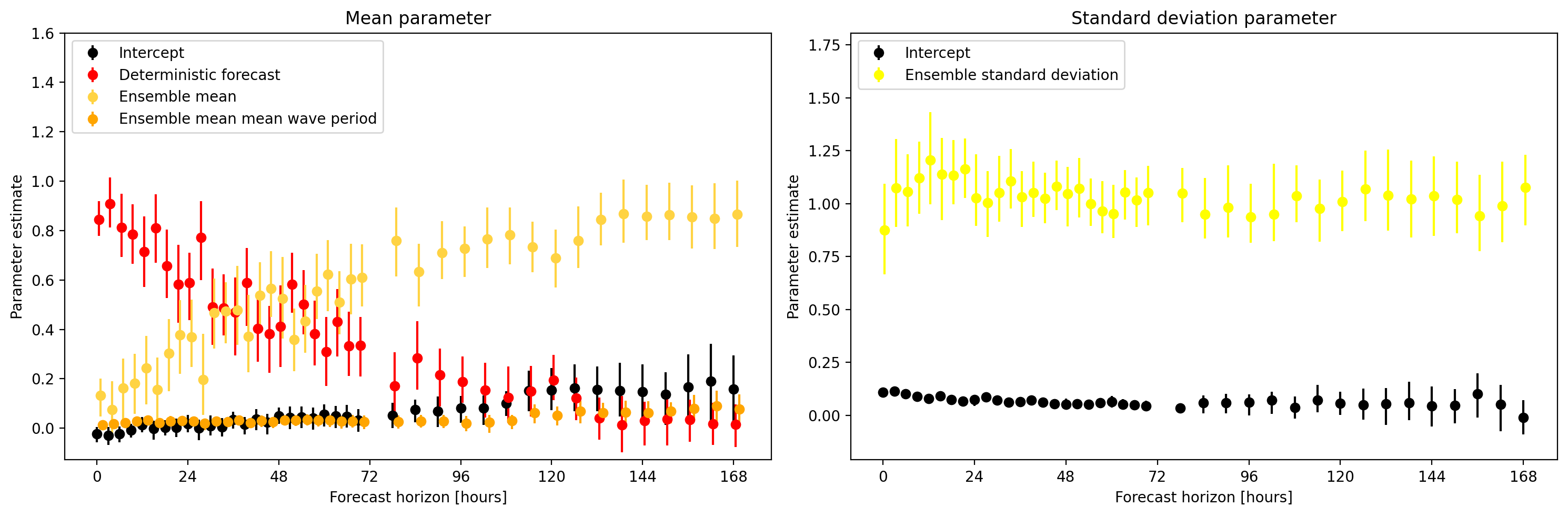}
		\caption{Significant wave height $H_S$.}
	\end{subfigure}	
	\begin{subfigure}{1\textwidth}
		\centering
		\includegraphics[width=1\textwidth]{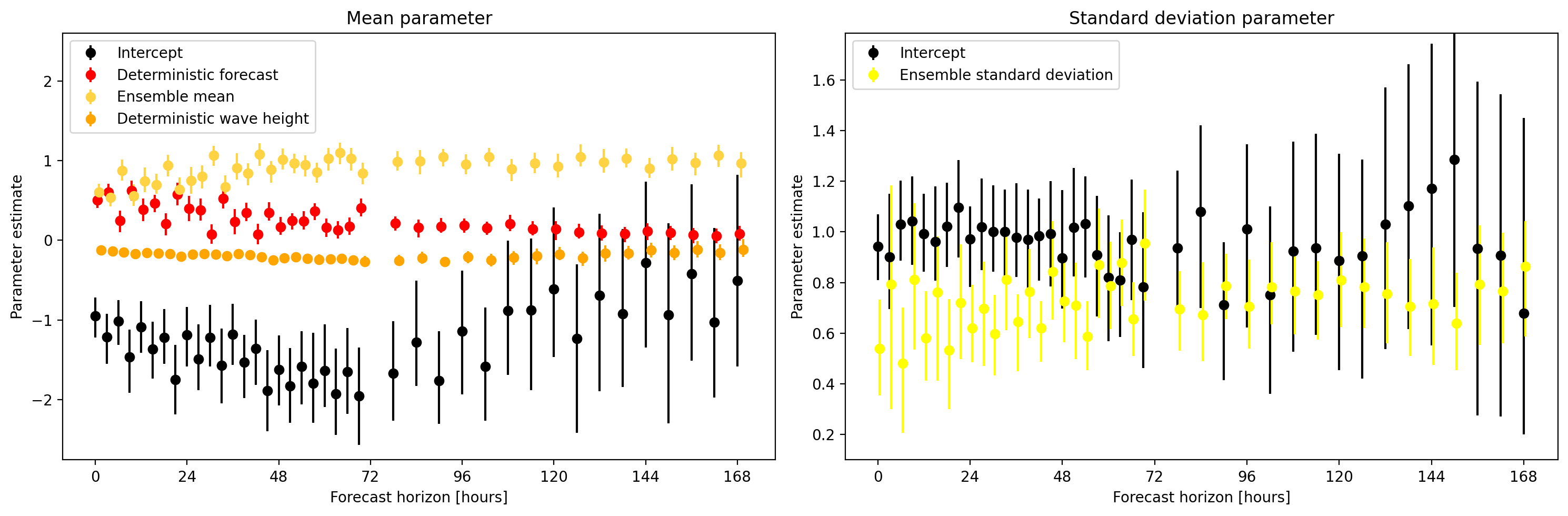}
		\caption{Wind speed $W$.}
	\end{subfigure}
	\begin{subfigure}{1\textwidth}
		\centering
		\includegraphics[width=1\textwidth]{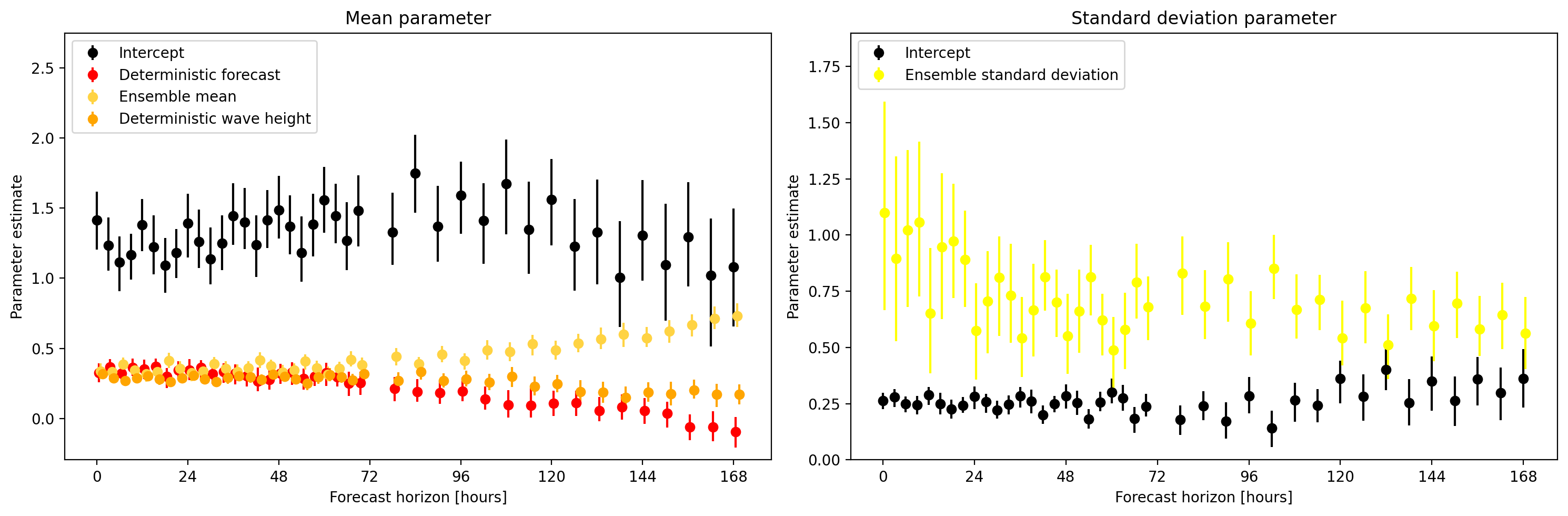}
		\caption{Mean wave period $T_M$.}
	\end{subfigure}
	\caption{Variation of estimated parameters with forecast horizon from consistent NHGR models for variables (a) $H_S$, (b) $W$ and (c) $T_M$. Form of consistent NHGR calibration model given in Table~\ref{Tbl:NHGR}. Central 95\% uncertainty bands for parameter estimates are calculated using bootstrap resampling. Other details are given in Figure~\ref{Fgr:LR:PrmEst}.}
	\label{Fgr:NHGR:PrmEst}
\end{figure}

Parameter estimates for NHGR calibration, again using the standardisation scheme for covariates of the distributional mean described in Section~\ref{Sct:Mth:Stn}, are illustrated in Figure~\ref{Fgr:NHGR:PrmEst}. For the distributional mean parameters in the left hand panels, estimates are similar to those found using LR calibration for all of $H_S$, $W$ and $T_m$. For the distributional standard deviation parameters, it is striking that the ensemble standard deviation is very informative for $H_S$, and for $T_m$ at short forecast horizons. Central 95\% uncertainty bands for parameter estimates are calculated using bootstrap resampling. Figure~SM3 quantifies benefit of incorporating ensemble variability (for the same physical quantity) to explain forecast variability in terms of percentage covariate contributions; covariate contributions are similar to those for LR. The importance of the ensemble standard deviation in estimating the distributional standard deviation is clear, especially for $H_S$.

\begin{figure}[!ht]
	\centering
	\begin{subfigure}{1\textwidth}
		\includegraphics[width=1\textwidth]{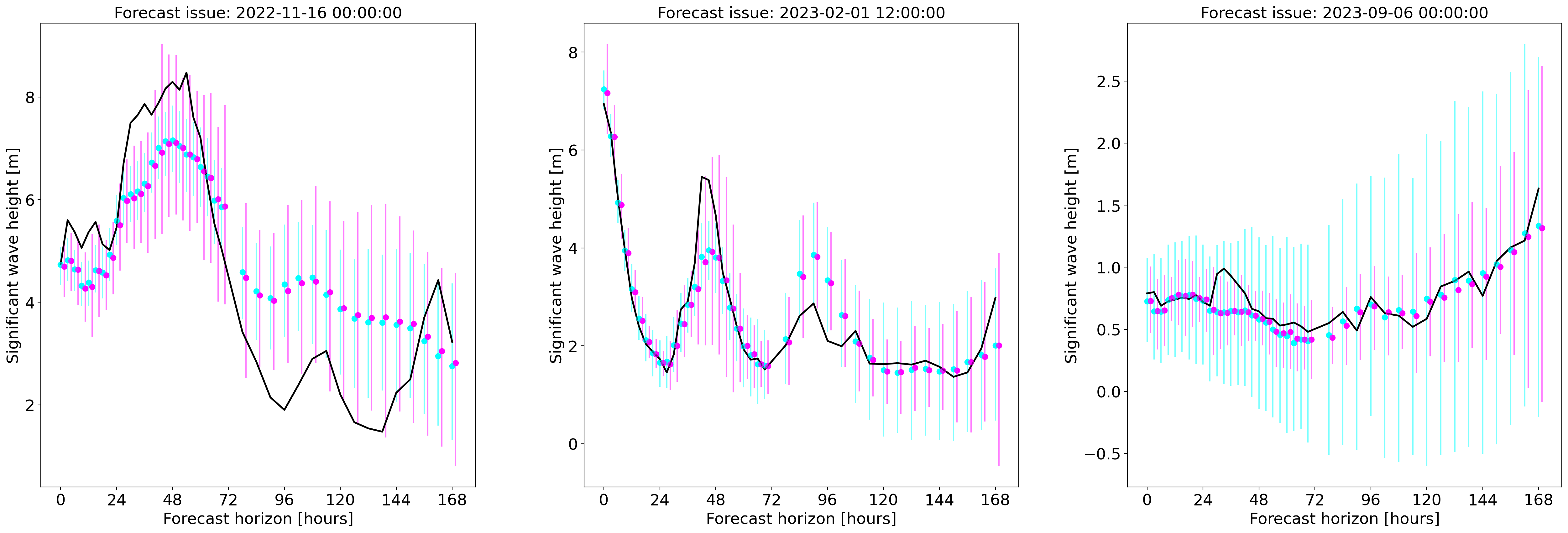}
		\caption{Significant wave height $H_S$. }
	\end{subfigure}
	\begin{subfigure}{1\textwidth}
		\includegraphics[width=1\textwidth]{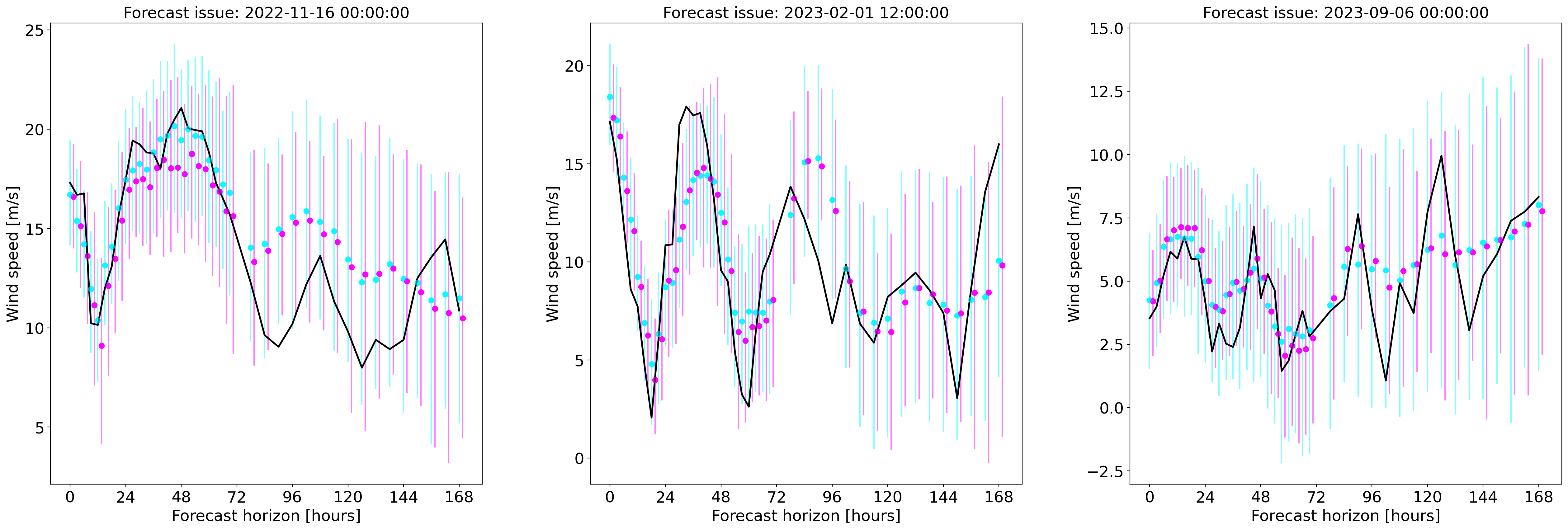}
		\caption{Wind speed $W$. }
	\end{subfigure}
	\begin{subfigure}{1\textwidth}
		\includegraphics[width=1\textwidth]{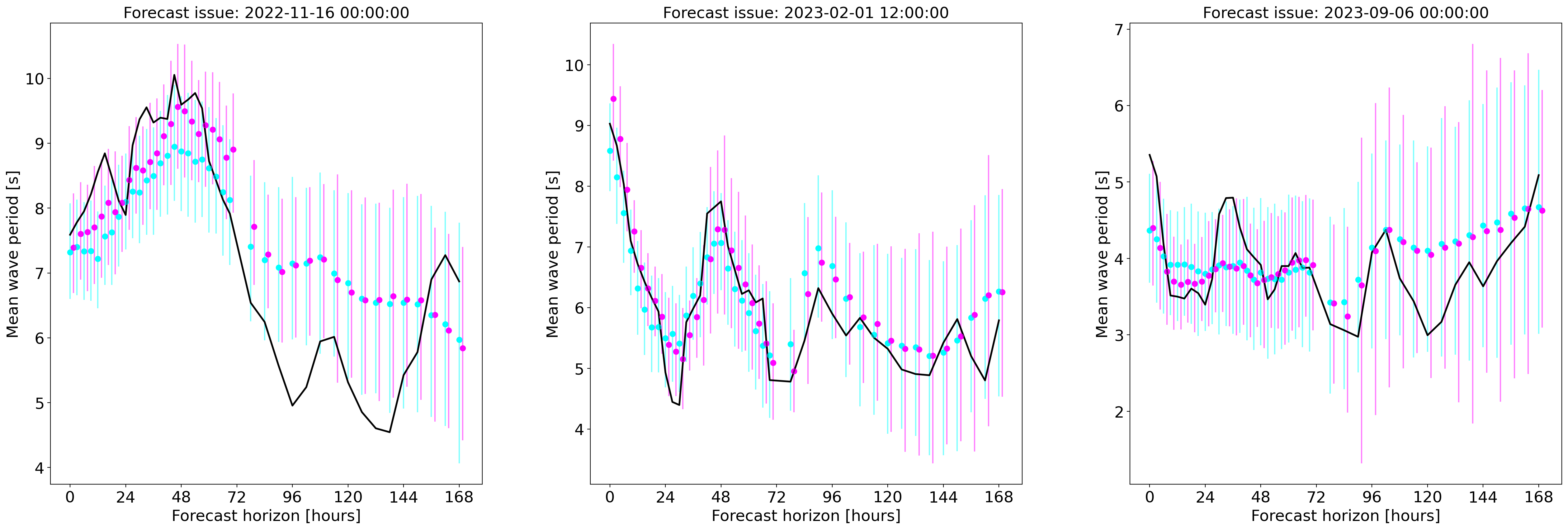}
		\caption{Mean wave period $T_M$. }
	\end{subfigure}
	\caption{NHGR-calibrated forecasts and (future) reality for given variable at given forecast time. Three examples (columns) of forecasts on horizons $\in [1,168]$ for significant wave height (top), wind speed (middle) and mean wave period (bottom). Title of each column gives the time of forecast issue. (Future) reality illustrated using black line. Optimal calibrated forecast given in magenta, in terms of the mean (disc) and 95\% Gaussian forecast uncertainty band (vertical line). Also shown for comparison are the corresponding optimal calibrated LR-forecasts (cyan) from Figure~\ref{Fgr:LR:ClbFrcExm}. Form of consistent LR and NHGR calibration models given in Tables~\ref{Tbl:LR} and \ref{Tbl:NHGR}. Box-whiskers have been translated horizontally by a small amount for clarity to avoid them being superimposed.}
	\label{Fgr:NHGR:ClbFrcExm}
\end{figure}

Figure~\ref{Fgr:NHGR:ClbFrcExm} compares forecasts and their variability from LR (cyan) and NHGR (magenta) models. Given observations already made, LR and NHGR are similar in terms of mean forecasts (discs in figure). The difference between model performance is noticeable in the width of Gaussian forecast uncertainty bands (calculated using the estimated parameter values and error standard deviations, see Equations~\ref{Eqn:LR} and \ref{Eqn:NHGR}), particularly for $H_S$.

\FloatBarrier
\subsection{Comparing deterministic forecast, LR and NHGR forecast calibration model performance: \ed{within-sample assessment}}
\begin{figure}[!ht]
	\centering
	\begin{subfigure}{0.97\textwidth}
		\centering
		\includegraphics[width=1\textwidth]{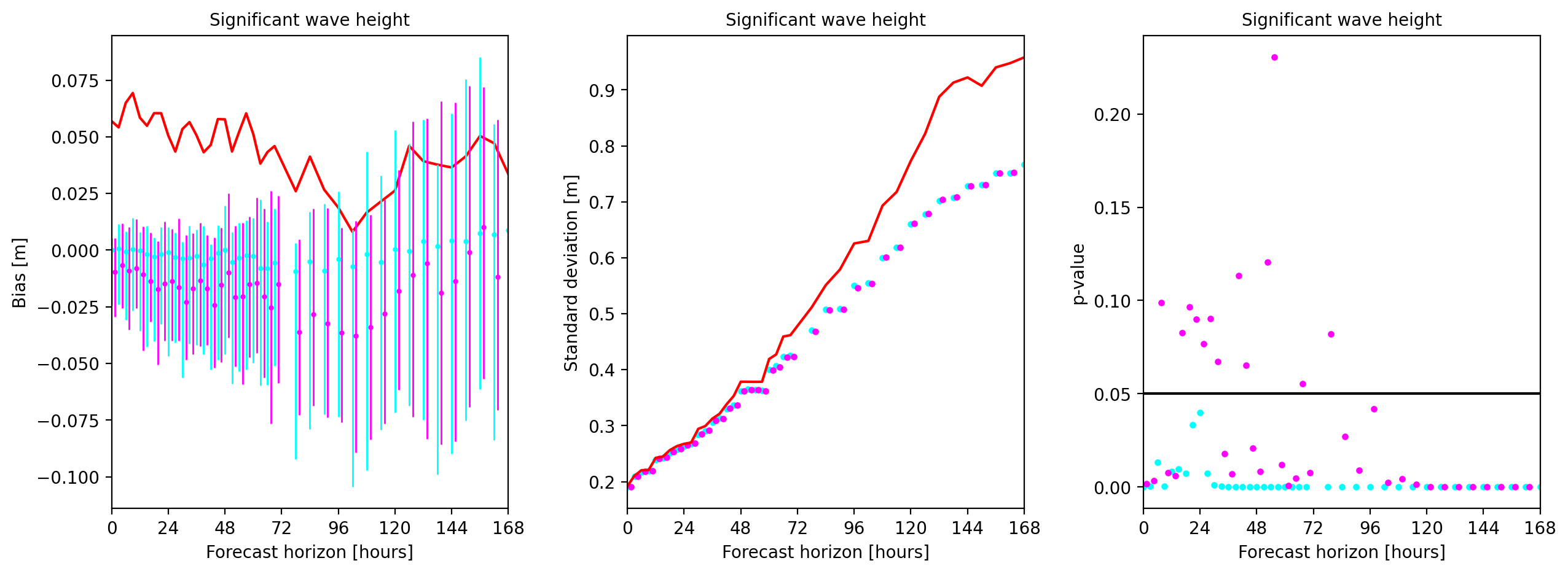}
		\caption{Significant wave height $H_S$.}
	\end{subfigure}
	\begin{subfigure}{0.97\textwidth}
		\centering
		\includegraphics[width=1\textwidth]{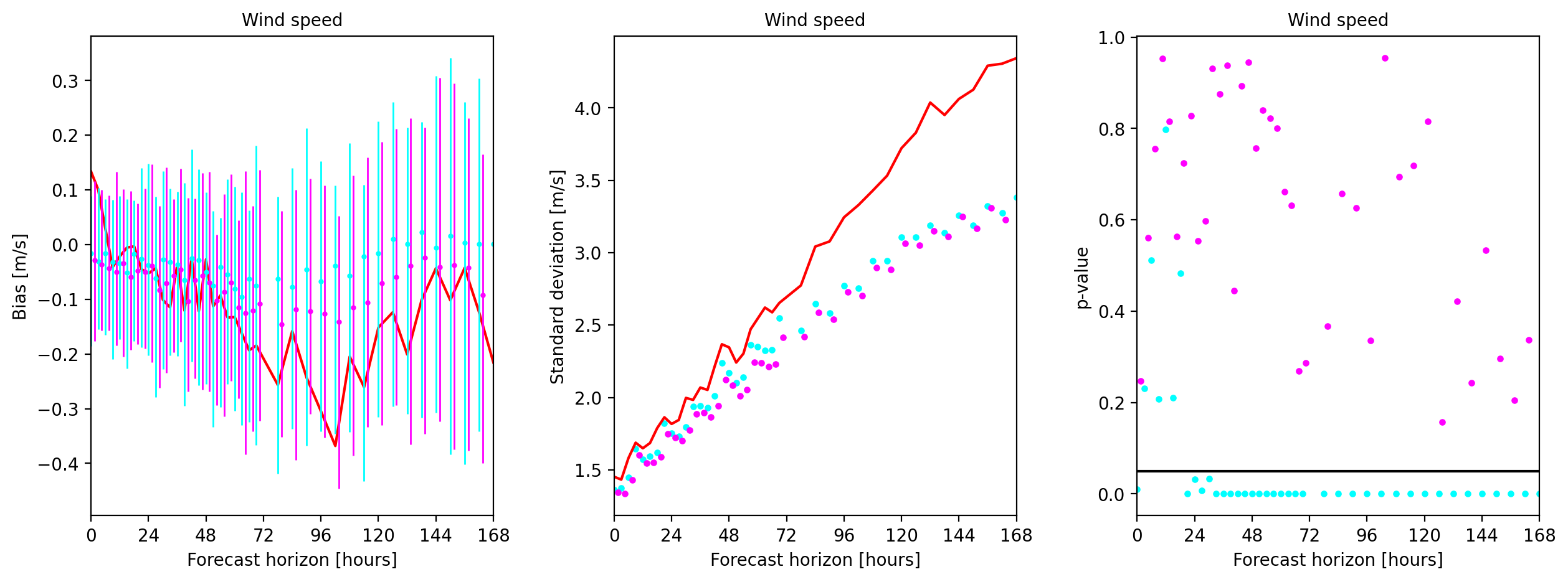}
		\caption{Wind speed $W$.}
	\end{subfigure}
	\begin{subfigure}{0.97\textwidth}
		\centering
		\includegraphics[width=1\textwidth]{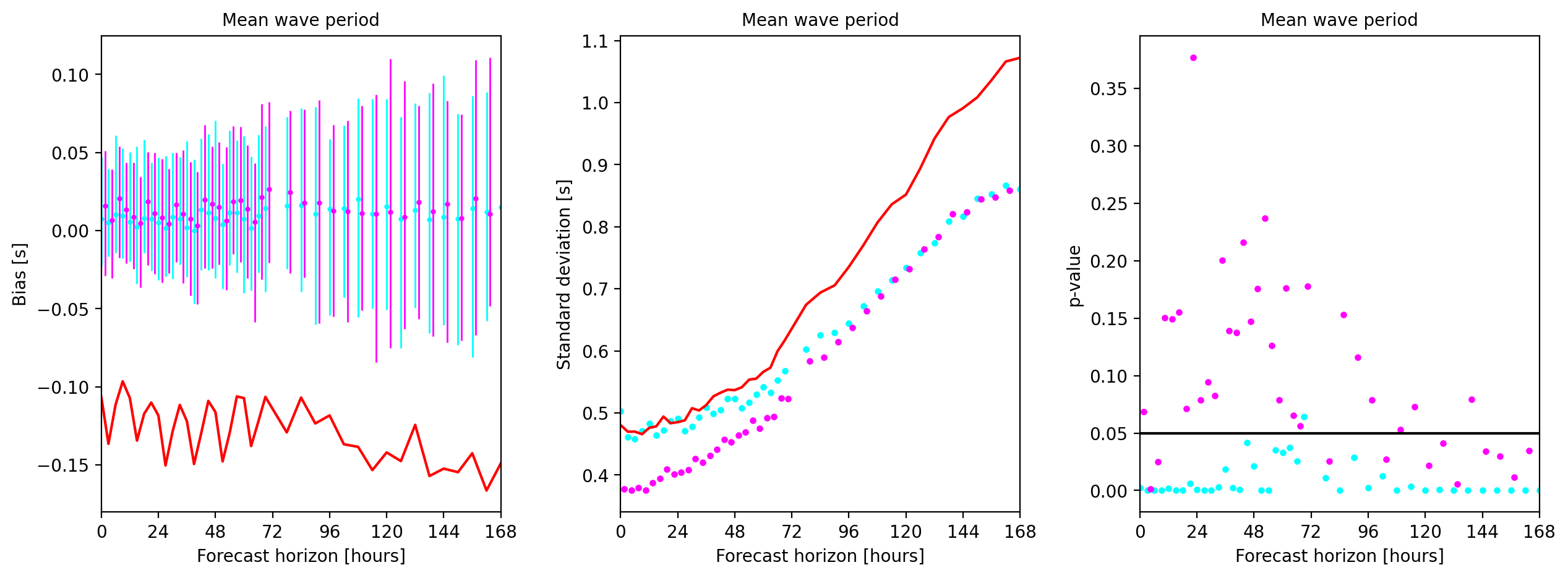}
		\caption{Mean wave period $T_M$.}
	\end{subfigure}
	\caption{Estimated mean (left) and standard deviation (centre) of forecast errors as a function of horizon, for (a) significant wave height $H_S$, (b) wind speed $W$ and (c) mean wave period $T_M$. The right column gives p-values of a KS test, with null hypothesis that standardised forecast residuals are standard Gaussian distributed, as a function of horizon. Red lines refer to the (uncalibrated) deterministic forecast; cyan (magenta) to LR (NHGR) calibration models. Box-whiskers (left panels) and discs (centre and right panels) have been translated horizontally by a small amount for clarity to avoid them being superimposed. 95\% uncertainty bands for bias (left) are calculated using bootstrap resampling.}
	\label{Fgr:BisVrnLRNHGR}
\end{figure}
Here we summarise the difference in performance of LR and NHGR calibration models by characterising the distribution of their forecast error as a function of horizon, specifically by estimating the bias and standard deviation of mean forecasts for the full period of data available. We also assess the similarity of the distribution of standardised residuals from the fitted models to the assumed Gaussian distribution, using a Kolmogorov-Smirnov (KS) test. 

With $Y(t+\tau)$ representing the in-situ measured metocean variable $Y$ at time $t+\tau$, we seek to assess the performance of its calibrated forecasts $\hat{Y}_S(\tau|t)$ at forecast issue time $t$ and forecast horizon $\tau \ge 0$, where $S$ indicates the source of the forecast. Specifically, we will consider three cases $S \in \{\text{D},\text{LR},\text{NHGR}\}$, corresponding to the (uncalibrated) deterministic forecast, the LR calibration model and the NHGR calibration model. The deterministic forecast is included as a benchmark which with to assess the improvement in forecast performance offered by the LR and NHGR models. Each of the calibrated forecasts can be written in the form
\begin{eqnarray}
	\hat{Y}_S(\tau|t) = \mu_S(\tau|t) + \epsilon \ \sigma_S(\tau|t), \text{ for } S \in \{\text{D},\text{LR},\text{NHGR}\}
	\label{Eqn:GnrGssFrm}
\end{eqnarray}
since both LR and NHGR have a Gaussian error structure, where $\mu_S(\tau|t)$ is the mean forecast and $\sigma_S(\tau|t)$ the corresponding error standard deviation. See Equations~\ref{Eqn:LR} and \ref{Eqn:NHGR} for the functional forms of these parameters for LR and NHGR; for deterministic forecast D, $\mu_D(\tau|t)$ is simply the deterministic forecast itself, direct from the forecast provider, and $\sigma_D(\tau|t)=0$. Further $\epsilon$ a standard Gaussian random variate. 

We then characterise model performance for D, LR and NHGR in terms of three summary statistics per forecast horizon, including sample estimates for the mean $\mathbb{E}(Y(t+\tau)-\mu_S(\tau|t))$ and standard deviation $\text{sd}(Y(t+\tau)-\mu_S(\tau|t))$ of the difference between reality and mean forecast. For \text{LR} and \text{NHGR} models, we also estimate the p-value associated with the \ed{KS} test with null hypothesis that the standardised residuals $(Y(t+\tau)-\mu_S(\tau|t))/\sigma_S(\tau|t)$ are drawn from a standard Gaussian distribution; p-values less than 0.05 indicate evidence that the null hypothesis can be rejected. 

Results are given in Figure~\ref{Fgr:BisVrnLRNHGR}. In each row of the figure, the left hand panel illustrates the estimated bias $\mathbb{E}(Y(t+\tau)-\mu_S(\tau|t))$ for LR (cyan) and NHGR (magenta) models, with the (uncalibrated) deterministic forecasts used as a benchmark. We also provide central 95\% uncertainty intervals for bias estimates using a non-parametric bootstrap analysis. The centre panel illustrates the corresponding standard deviation $\text{sd}(Y(t+\tau)-\mu_S(\tau|t))$, again using the (uncalibrated) deterministic forecast standard deviation as a benchmark. The right hand panel illustrates p-values for the KS test for standardised residuals. 

From the figure we see that the LR and NHGR calibration models provide reduced bias, especially for forecasting of $T_m$. We see considerable asymmetry in the bootstrap uncertainty intervals for bias of LR models for $H_S$, but note that the uncertainty intervals for bias of both LR and NHGR are very narrow relative to the typical range of values of $H_S$. We also see that forecast error standard deviation is smaller for LR and NHGR than for the uncalibrated deterministic forecast, particularly for longer forecast horizons. There is also evidence that the NHGR model outperforms LR for $T_m$ for shorter horizons. Results of the KS test suggest that there is evidence to reject the null hypothesis of standard Gaussian standardised residuals from LR models (cyan) for all variables, and that standardised residuals from NHGR model fits (magenta) are more consistent with modelling assumptions, particularly for shorter horizons. {Biases in forecasts from Figure~\ref{Fgr:BisVrnLRNHGR} are unlikely to be of material concern to the metocean engineer, and may well be at the level of measurement error in practice. However, uncertainties in forecasts grow to levels which are likely to be of practical concern for horizons of 3 days and longer, for which 95\% uncertainty bands are at least $\pm 1$ m for $H_S$, $\pm 5$ m/s for $W$ and $\pm 1$ s for $T_M$.} 

\FloatBarrier
\subsection{\ed{Comparing deterministic forecast, LR and NHGR forecast calibration model performance: out-of-sample assessment}}

Provided that data for the calibration model training period is representative of future observations of the environment, we can be confident that future model performance will be similar to that reported in Figure~11. We can also directly evaluate forecast performance for a time period following that used to estimate the calibration models. Figures SM4 and SM5 in the Supplementary Material illustrate forecast performance of models over out-of-sample Periods 1 (1 September 2023 - 30 April 2024) and Period 2 (1 September 2024 - 31 December 2024). Regrettably, for Period 1, comparable wave period data were not available, and for Period 2, comparable wind speed data were not available. \ed{(More specifically, for Period 1, for reasons beyond the authors' control, the ensemble forecast data from the forecast provider were not retained and hence not available for analysis. Further, just prior to Period 2, a recalibration of the wind sensor was performed, making fair assessment of forecast performance unviable.)} Nevertheless, for $H_S$ and $W$ (Period 1) and $H_S$ and $T_m$ (Period 2), we see that the general characteristics of Figures SM4 and SM5 are similar to those of Figure~11. Forecast bias is small relative to standard deviation for the uncalibrated deterministic, LR-calibrated and NHGR-calibrated forecasts. Reduction in forecast standard deviation at longer horizons is clear for both LR and NHGR relative to the deterministic forecast. The distribution of residuals from the NHGR model is generally more similar to the assumed Gaussian form.

Bias and variance are fundamental quantities used to characterise the performance of an estimator, favoured by us in the current work. Similarly, the KS statistic is a generic measure of the dissimilarity between distributions. However there are additional performance scoring rules and diagnostics, particularly interesting when evaluating ensemble forecasts, which could also be used for the current work. These include the continuous ranked probability score (CRPS;  e.g. \citealt{GntEA05}) and the probability integral transform (PIT) histogram (e.g. \citealt{Dwd84}). More generally, the work of \cite{HrnEA18a} provides a review of performance, skill and accuracy assessment in operational oceanography, and \cite{MssEA20} reviews forecast verification tools, with a focus on wind power applications.

\begin{figure}[!ht]
	\centering
	\begin{subfigure}[t]{0.49\textwidth}
		\includegraphics[width=1\textwidth]{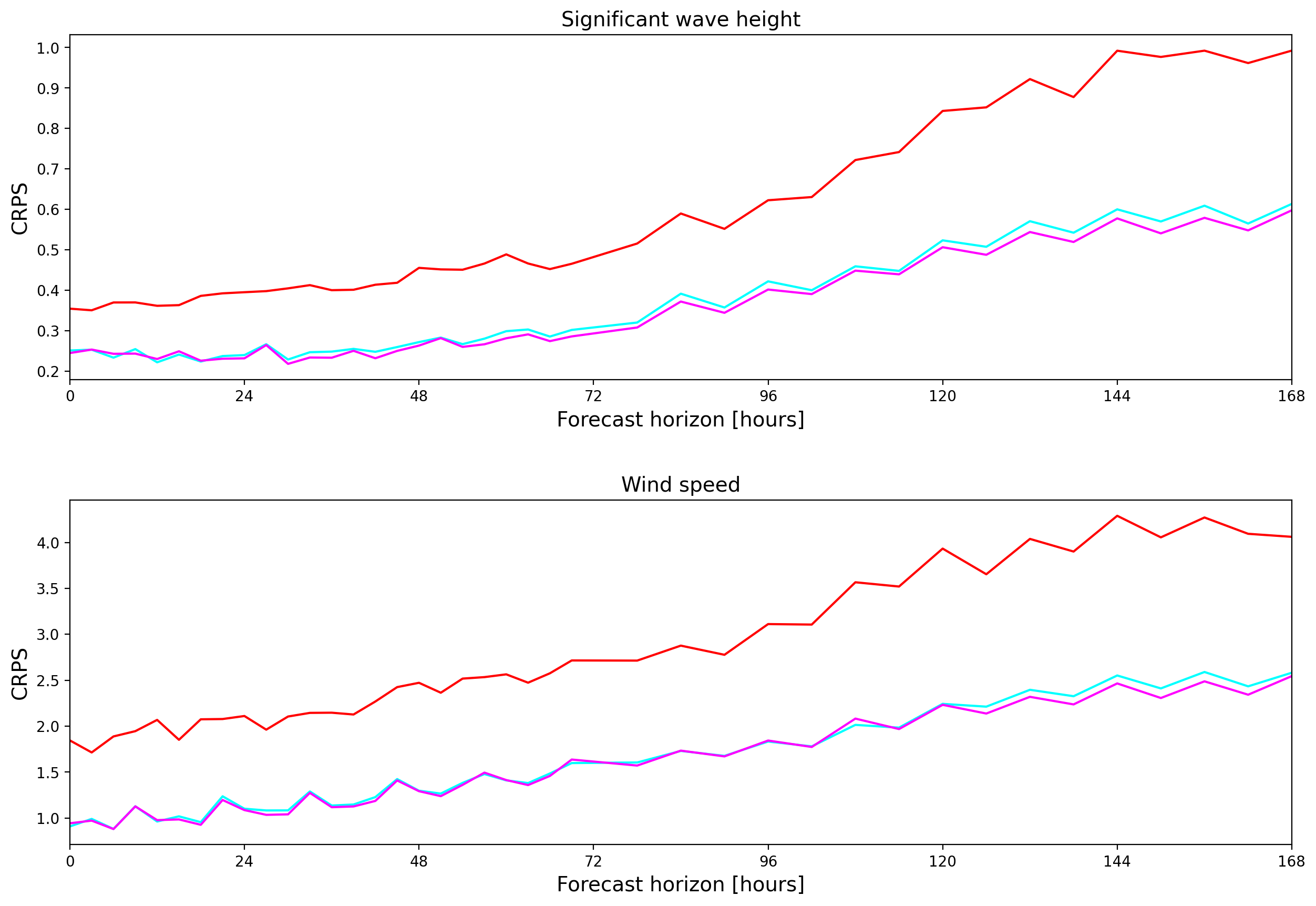}{a}
		\caption{{Period 1.}}
	\end{subfigure}
	\begin{subfigure}[t]{0.49\textwidth}
		\includegraphics[width=1\textwidth]{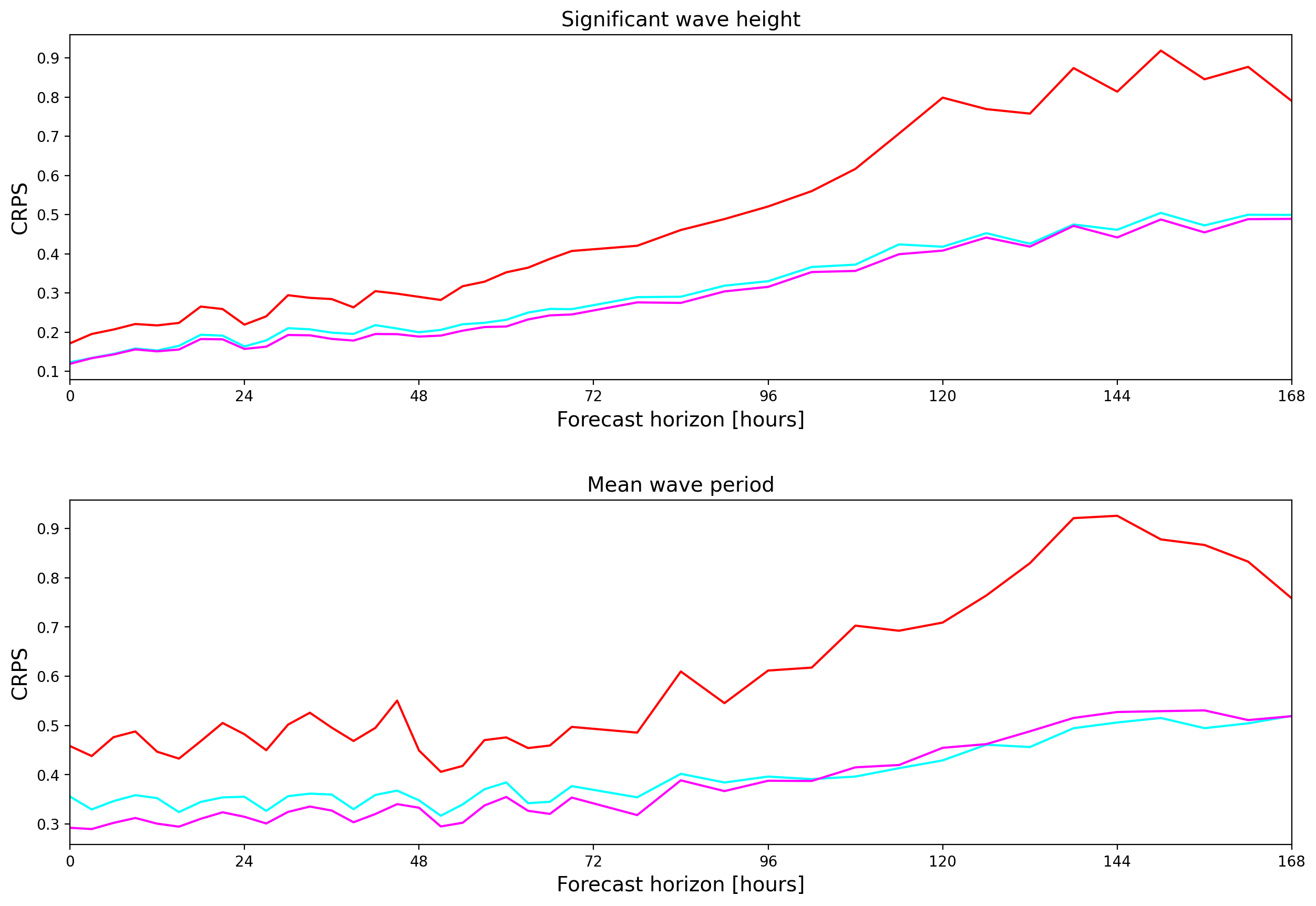}{b}
		\caption{{Period 2.}}
	\end{subfigure}
	\caption{{Variation of CRPS with forecast horizon for (a) out-of-sample Period 1 (1 September 2023 - 30 April 2024) and (b) out-of-sample Period 2 (1 September 2024 - 31 December 2024) for the uncalibrated deterministic forecast (red), the forecast calibrated using LR (cyan) and the forecast calibrated using NHGR (magenta).}}
\label{Fgr:CrpsSmm}
\end{figure}

As an illustration, Figure~\ref{Fgr:CrpsSmm} shows the variation of CRPS with forecast horizon for the two out-of-sample test Periods 1 and 2.
{For value $y$ of the response at forecast horizon $\tau$, CRPS is calculated from the distribution $F_{\hat{Y}}(\hat{y}; \tau)$ of the forecast response $\hat{Y}$, using the expression}
\begin{eqnarray*}
	{\text{CRPS}(y,\tau) = \int_{\hat{y}} \left( F_{\hat{Y}}(\hat{y}; \tau) - I(\hat{y} \ge y) \right)^2 d\hat{y}}
\end{eqnarray*} 
{for indicator function $I$ (with $I(x)=1$ if $x$ is true and $=0$ otherwise), which can then be averaged over $y$ to provide a function of $\tau$ only (e.g. \citealt{GntEA05}), or inspected as a function of $y$ for given $\tau$. CRPS $=0$ indicates perfect agreement between forecast and truth. Figure~\ref{Fgr:CrpsSmm} shows that the average CRPS performance of calibrated forecasts betters that of the uncalibrated deterministic forecast, with an approximate reduction of 50\% in value. Moreover, the NHGR calibrated forecast (magenta) is generally (but not always) slightly better than that calibrated using LR (cyan). Further, Figures SM5 and SM6 of the Supplementary Material show variation of CRPS with the value of response for four representative forecast horizons, for Periods 1 and 2.}

\begin{figure}[!ht]
	\centering
	\begin{subfigure}[t]{0.49\textwidth}
		\includegraphics[width=1\textwidth]{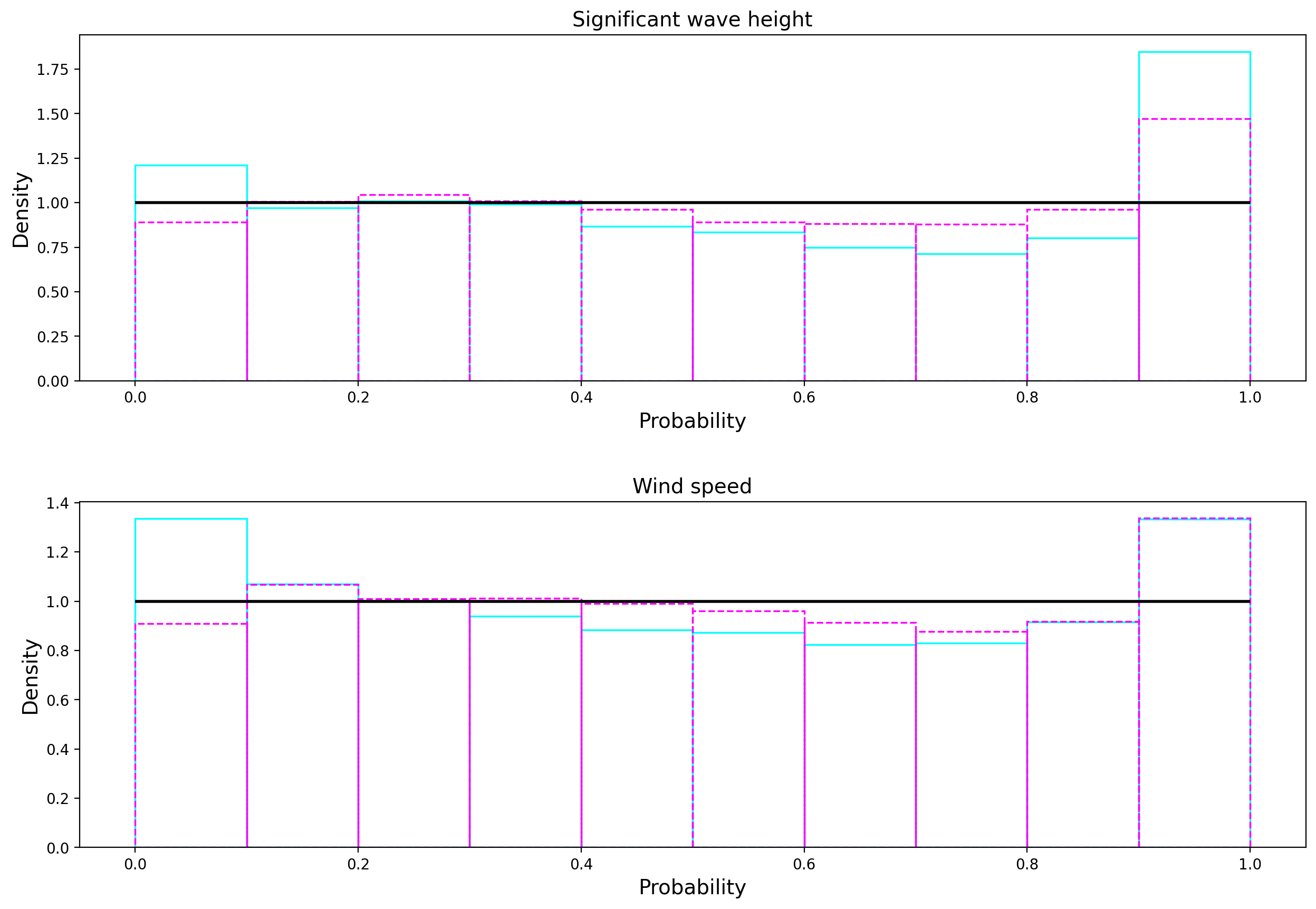}{a}
		\caption{\ed{Period 1.}}
	\end{subfigure}
	\begin{subfigure}[t]{0.49\textwidth}
		\includegraphics[width=1\textwidth]{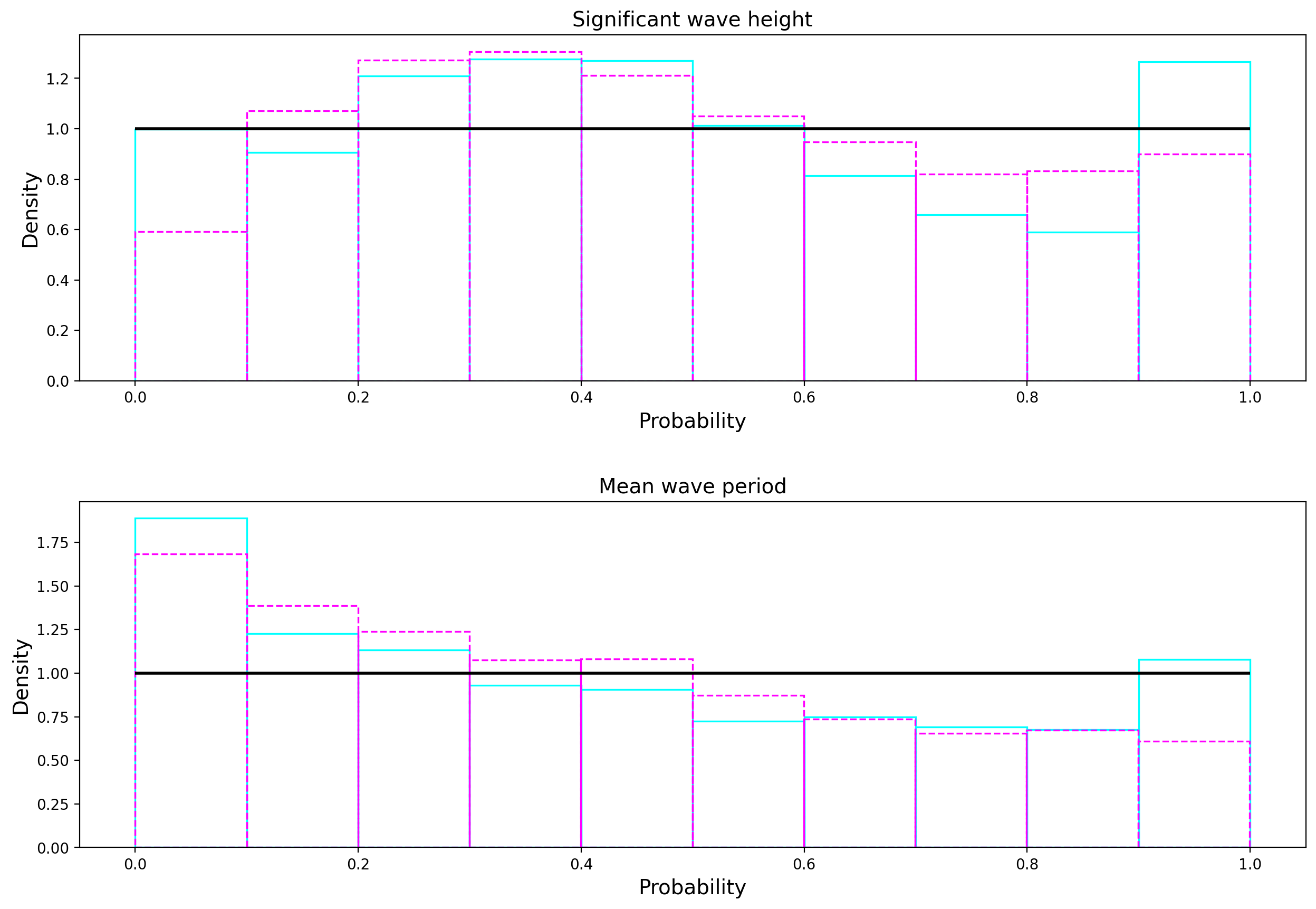}{b}
		\caption{\ed{Period 2.}}
	\end{subfigure}
	\caption{\ed{Rank histogram plots over all forecast horizons for (a) out-of-sample Period 1 (1 September 2023 - 30 April 2024) and (b) out-of-sample Period 2 (1 September 2024 - 31 December 2024) for the the forecast calibrated using LR (cyan) and the forecast calibrated using NHGR (magenta). The horizontal line (black) indicates the model-assumed histogram in each case.}}
	\label{Fgr:RnkHstSmm}
\end{figure}

\ed{Further, Figure~\ref{Fgr:RnkHstSmm} shows corresponding rank histograms which provide a visual interpretation of the distribution of observed values of response relative to the distribution of probabilistic forecast. As shown by Equation~\ref{Eqn:GnrGssFrm}, both LR- and NHGR-calibration models yield a probabilistic forecast with Gaussian error structure. Therefore, the standardised residual $r(\tau|t)$ for observation $y(t+\tau)$ given the forecast at time $t$ and forecast horizon $\tau$, defined by}
\begin{eqnarray*}
	\ed{r(\tau|t) = \frac{y(t+\tau)-\mu_S(\tau|t)}{\sigma_S(\tau|t)}}
\end{eqnarray*} 
\ed{is expected to follow the standard Gaussian distribution with zero mean and unit variance (and cumulative distribution function $\Phi(\bullet)$), with $\mu_S$ and $\sigma_S$ defined in Equation~\ref{Eqn:GnrGssFrm}. Therefore, we expect the cumulative distribution function $\Phi(r(\tau|t))$ evaluated at $r(\tau|t)$ to be a uniform random number on $[0,1]$. We can hence inspect the empirical density of $\Phi(r(\tau|t))$ over all $t$ and $\tau$, as shown in Figure~\ref{Fgr:RnkHstSmm}, to assess its correspondence to a constant uniform density on $[0,1]$.}
\ed{The figure suggests that there is reasonable overall agreement between residuals from the LR- and NHGR-calibrated forecasts and the model-assumed Gaussian distributional forms; departures from the horizontal black line in each panel indicate lack of agreement between model and reality. It appears that the NHGR-calibrated forecast provides somewhat better general agreement. We note that results of the KS testing visualised in Figures~SM4 and SM5 for the out-of-sample periods, provide rather similar diagnostic information.}

\section{Discussion and conclusions}  \label{Sct:DscCnc}

In this article, we assess the value of calibrating forecast models for significant wave height $H_S$, wind speed $W$ and mean spectral wave period $T_m$ for forecast horizons between zero and 168 hours from a commercial forecast provider, to improve forecast performance for a location in the central North Sea. We consider two straightforward calibration models, linear regression (LR) and non-homogeneous Gaussian regression (NHGR), incorporating deterministic, control and ensemble mean forecast covariates. We show that relatively simple calibration models (with at most three covariates) provide good calibration; addition of further covariates cannot be justified. Optimal calibration models (for the forecast mean of a physical quantity) always make use of the deterministic forecast and ensemble mean forecast for the same quantity, together with a covariate associated with a different physical quantity. The selection of optimal covariates is performed independently per forecast horizon, and the set of optimal covariates shows a large degree of consistency across forecast horizons. As a result, it is possible to specify a consistent model to calibrate a given physical quantity, incorporating a common set of three covariates for all horizons. For NHGR models of a given physical quantity, the ensemble forecast standard deviation for that quantity is skilful in predicting forecast error standard deviation, especially for $H_S$. We show that the consistent LR and NHGR calibration models facilitate reduction in forecast bias to near zero for all of $H_S$, $W$ and $T_m$, and that there is little difference between LR and NHGR calibration for the mean. Both LR and NHGR models facilitate reduction in forecast error standard deviation relative to naive adoption of the (uncalibrated) deterministic forecast, with NHGR providing somewhat better performance. Distributions of standardised residuals from NHGR models are generally considerably more like standard Gaussians. 

Direct adoption of (uncalibrated) ensemble forecasts is not recommended, since there is evidence of large bias for $T_m$. For short horizons, the contribution of the deterministic forecast to the calibration model is highest, decaying with increasing horizon. In contrast, the importance of the ensemble mean forecast (for the same metocean quantity as that being forecast) increases. The relative contributions of deterministic and ensemble mean forecasts to the calibration varies across metocean responses. During the course of the study, we also considered different summaries of the ensemble distribution to the mean and standard deviation. We found e.g. that use of the ensemble median offered no improvement in general over the ensemble mean.

{The choices of calibration models used here represent the simplest approaches that might reasonably be adopted in practice. Consequently there are many opportunities to extend the analysis. We noted evidence in the exploratory analysis that the forecast model generally tends to underestimate the very largest values. This might be an opportunity e.g. to include quadratic and higher order terms in covariates in the parametric form for the forecast mean (and, within the NHGR framework, for the forecast standard deviation). Alternatively, given that joint largest values of measured and forecast variables can be considered extreme, it might be more appropriate to adopt extreme value models (e.g. \citealt{DvsSmt90}, \citealt{HffTwn04}, \citealt{JntEA14}, \citealt{TowEA23}, \citealt{TowEA24}) to characterise these regions more correctly, or more generally to relax the assumption of Gaussianity made by both LR and NHGR. There are opportunities also to exploit the extreme time-series structure of predictors and responses, following the Markov extremal model and related frameworks in \cite{WntTwn16}, \cite{TndEA18} and \cite{TndEA23}.} {We might also consider joint calibration of multiple metocean variables. Multivariate predictive modelling is a large field of research and applications, offering a wealth of modelling strategies to predict a multivariate response from multivariate predictors. The presence of correlated predictors can lead to inflation in estimated model parameters (which can be quantified using measures such as the variance inflation factor), and inflation of prediction uncertainty. Nevertheless, joint modelling of multiple metocean variables provides the potential for better calibrated forecasts, including extremes. In the context of weather forecasting, \cite{AllEA24} provides a discussion of methods for assessing the calibration of multivariate probabilistic forecasts. Extension to calibration for multiple locations is also possible; for some locations at least,} the calibration model might be sensitive to additional covariates, including e.g. wave and wind direction and season.

NHGR models are relatively simple to estimate. Although closed form uncertainty quantification (e.g. for parameter estimates and predictions) is not available, we find that non-parametric bootstrapping can be used successfully. Based on these findings, we recommend the adoption of NHGR models incorporating ensemble forecasts for medium-term forecasting tasks of the type described here, including for unmanning and related operations (see e.g. \citealt{TowEA21}).

\section*{Acknowledgements}
%
The authors would like to thank StormGeo for provision of the forecast data, and for useful discussions during the course of the analysis and preparation of this article. They further acknowledge the support of Shell colleagues Graham Feld, David Randell and Stan Tendijck{, and helpful comments from two reviewers.}

\bibliographystyle{elsarticle-harv}
\bibliography{phil}

\end{document}


\maketitle

\renewcommand{\thesection}{SM\arabic{section}} 
\renewcommand{\thefigure}{SM\arabic{figure}} 

\section*{Overview}
This document provides supporting plots for the article ``Calibration of medium-range metocean forecasts for the North Sea''. \\

Figures~SM1 and SM3 illustrate the contributions of different covariates to linear regression (LR, Figure~SM1) and non-homogeneous Gaussian regression (NHGR, Figure~SM3). These are calculated using estimates for LR and NHGR parameters $\beta_j(\tau)$ (see Equations~1 and 2 of main text) from model fits where all covariates have been standardised to zero mean and unit variance, as
\begin{eqnarray*}
	\frac{|\beta_j(\tau)|}{\sum_k |\beta_k(\tau)|} .
\end{eqnarray*}
Here, the set of parameters $\beta(\tau)$ corresponds to the  set of the parameters $b$ and $c$ in the expressions LR or NHGR mean, and to the set of parameters $d$ and $e$ in the expression for the NHGR standard deviation. The intercept term $a$ in LR and NHGR mean is ignored without loss of generality, but included for the NHGR standard deviation. Figures~SM1 and SM3 provide illustrations to support the discussion on Figures~6 and 9 of the main text.\\

Figure~SM2 provides the equivalent of Figure~4 of the main text for NHGR calibration.\\ 

\ed{Figures~SM4 and SM5 summarise out-of-sample forecast performance for the uncalibrated deterministic forecast, and the LR- and NHGR-calibrated forecasts, on two out-of-sample periods. Figures~SM6 and SM7 illustrate corresponding continuous rank probability scores, and Figures~SM8 and SM9 rank histograms.}

\clearpage

\begin{figure}[!ht]
	\centering
	\begin{subfigure}{1\textwidth}
		\includegraphics[width=1\textwidth]{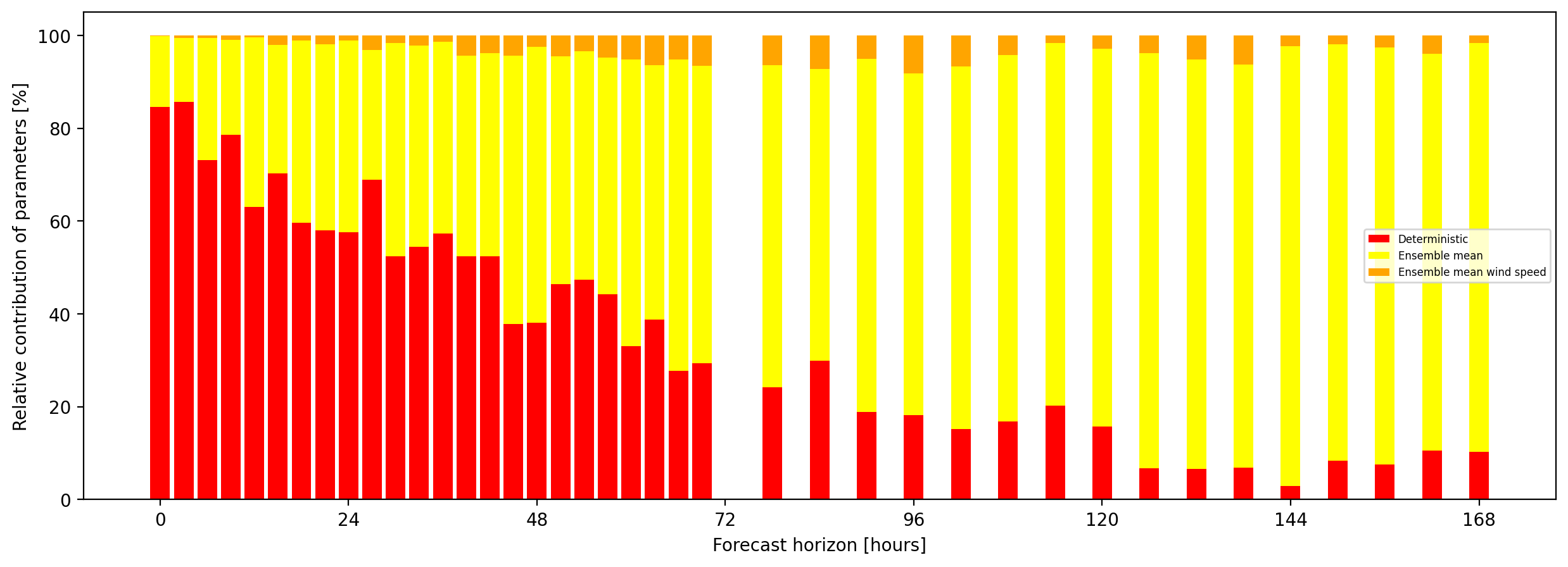}
		\caption{Significant wave height $H_S$.}
	\end{subfigure}
	\begin{subfigure}{1\textwidth}
		\includegraphics[width=1\textwidth]{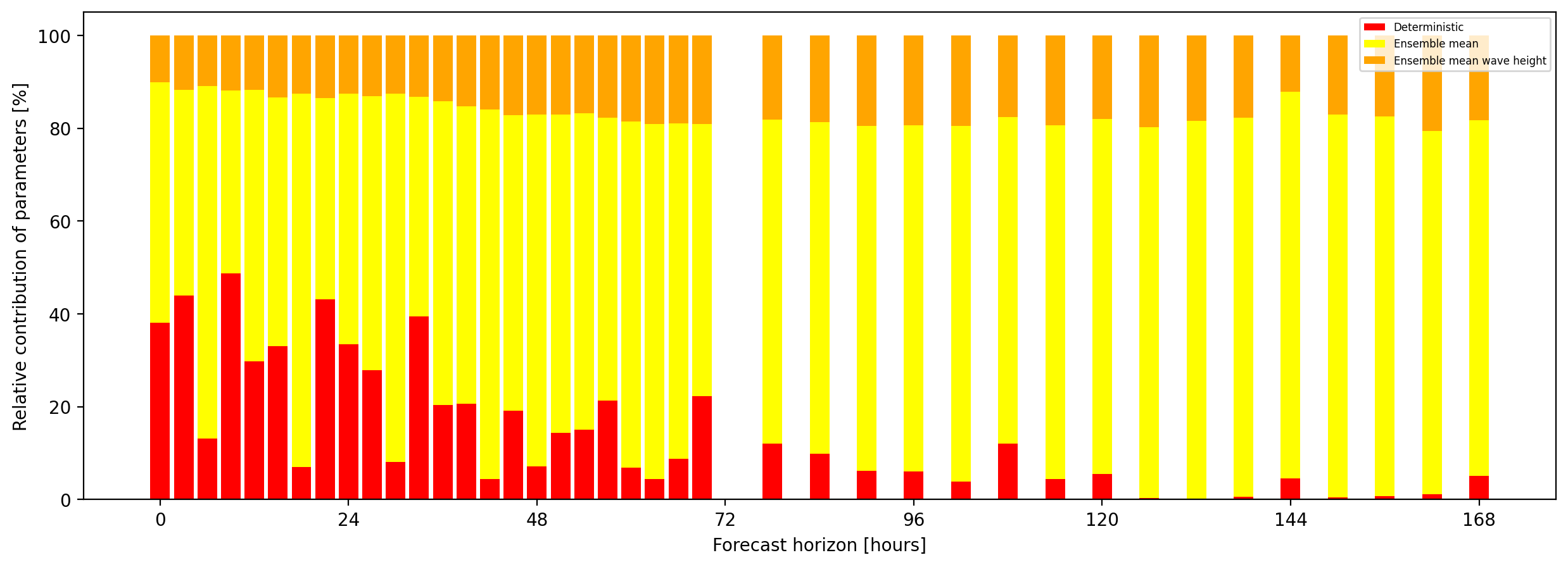}
		\caption{Wind speed $W$.}
	\end{subfigure}
	\begin{subfigure}{1\textwidth}
		\includegraphics[width=1\textwidth]{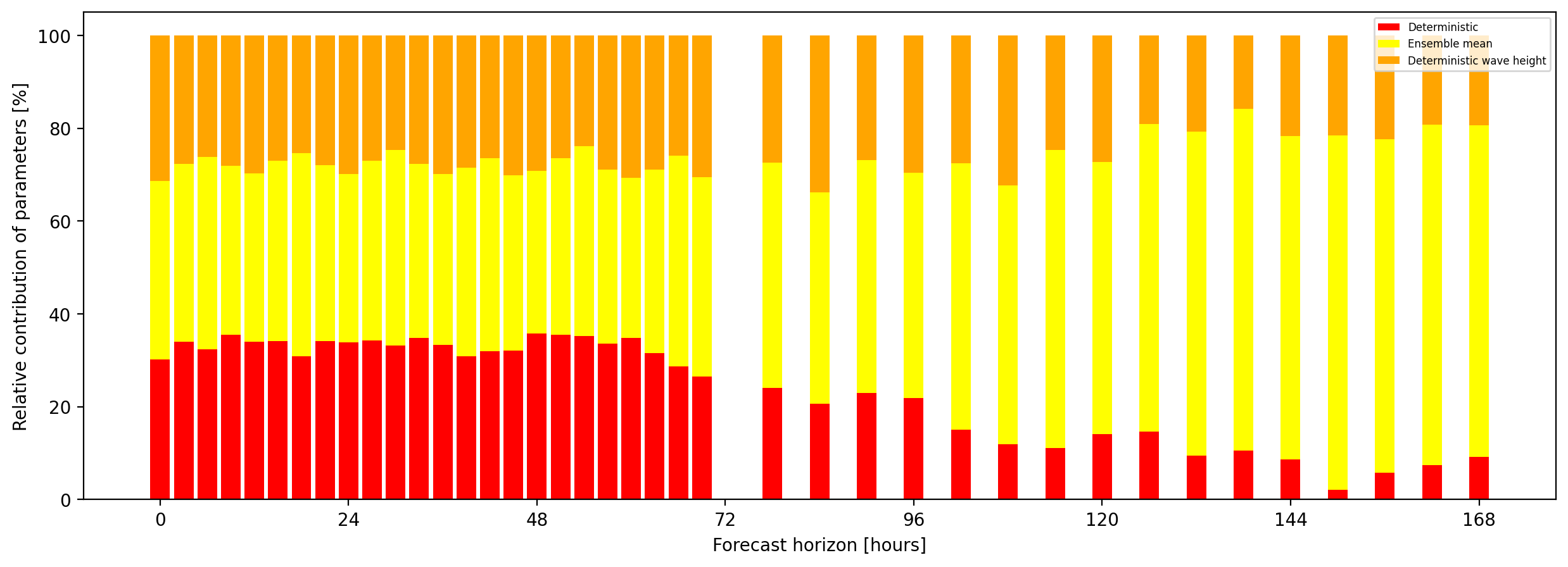}
		\caption{Mean wave period $T_M$.}
	\end{subfigure}
	\caption{Plot of linear regression covariate contributions to the calibration of forecasts for (a) significant wave height $H_S$, (b) wind speed $W$ and (c) mean wave period $T_M$, as a function of forecast horizon. All covariates are standardised to zero mean and unit standard deviation prior to regression analysis. Contributions are reported in terms of the percentage absolute value of the parameter coefficient from the regression. The intercept effect is not shown. Consistent model forms are given in Table 1 of the main text.}
	\label{Fgr:LR:CvrImp}
\end{figure}

\begin{figure}[!ht]
	\centering
	\begin{subfigure}{1\textwidth}
	\centering
		\includegraphics[width=1\textwidth]{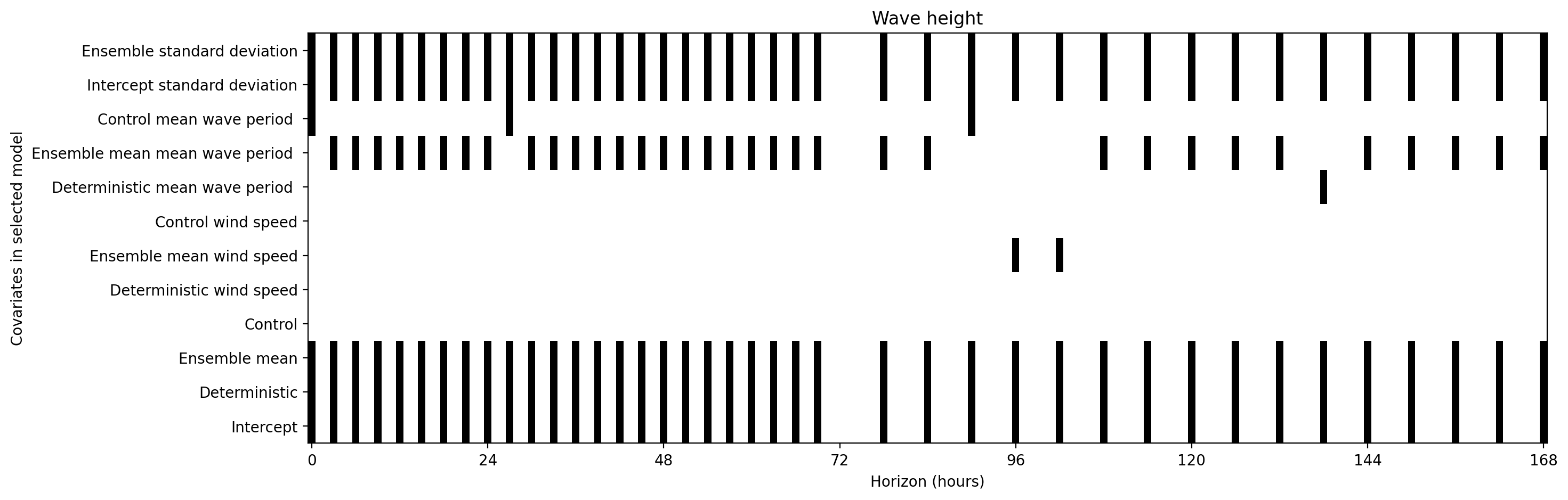}
		\caption{Significant wave height $H_S$.}
	\end{subfigure}
	\begin{subfigure}{1\textwidth}
	\centering
		\includegraphics[width=1\textwidth]{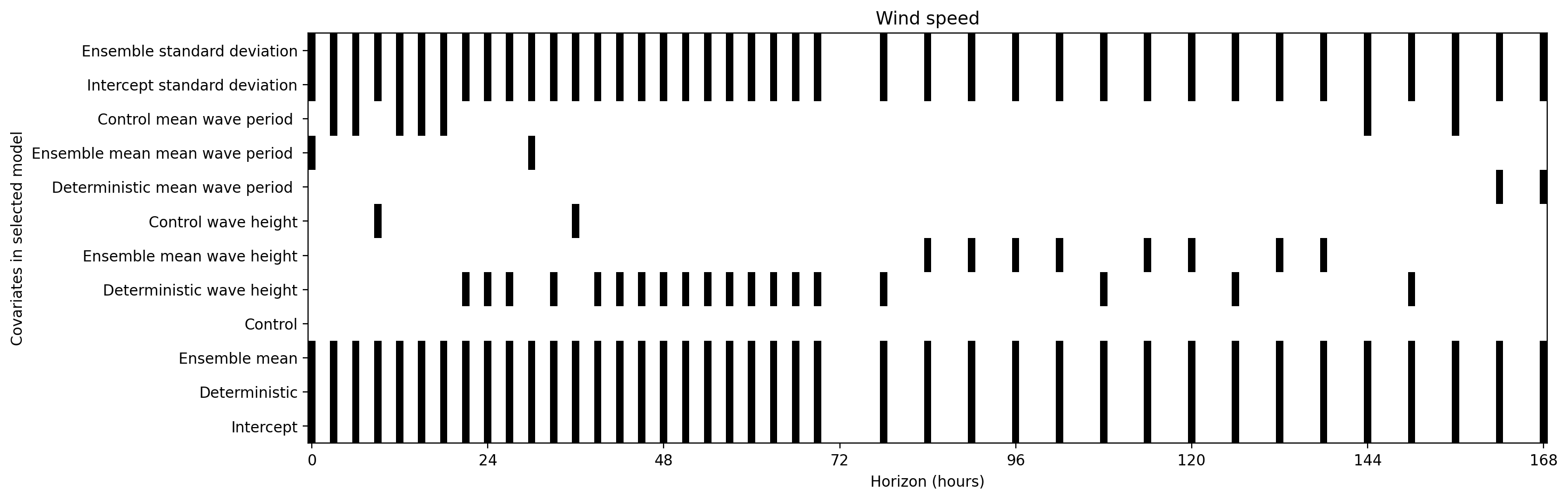}
		\caption{Wind speed $W$.}
	\end{subfigure}
	\begin{subfigure}{1\textwidth}
	\centering
		\includegraphics[width=1\textwidth]{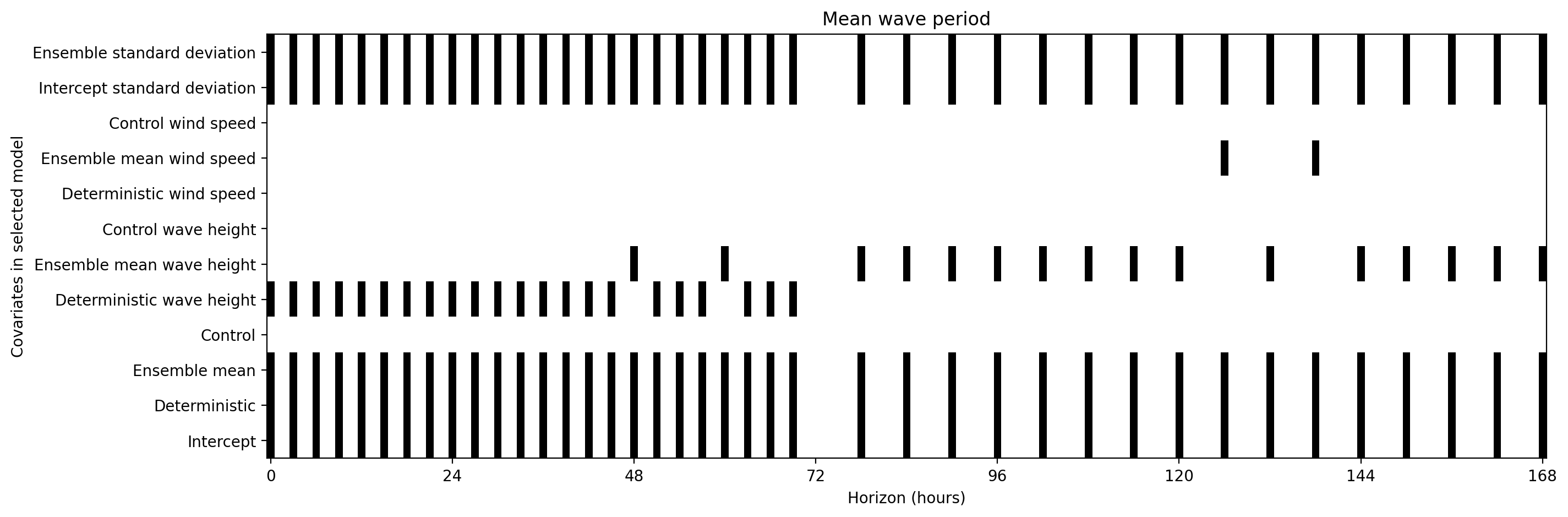}
		\caption{Mean wave period $T_M$.}
	\end{subfigure}
	\caption{Bar code plot of covariates included in optimal NHGR model for each forecast horizon. For discussion, see Section~4.2 of the main text.}
	\label{Fgr:NHGR:CvrBar}
\end{figure}

\begin{figure}[!ht]
	\centering
	\begin{subfigure}{1\textwidth}
		\includegraphics[width=1\textwidth]{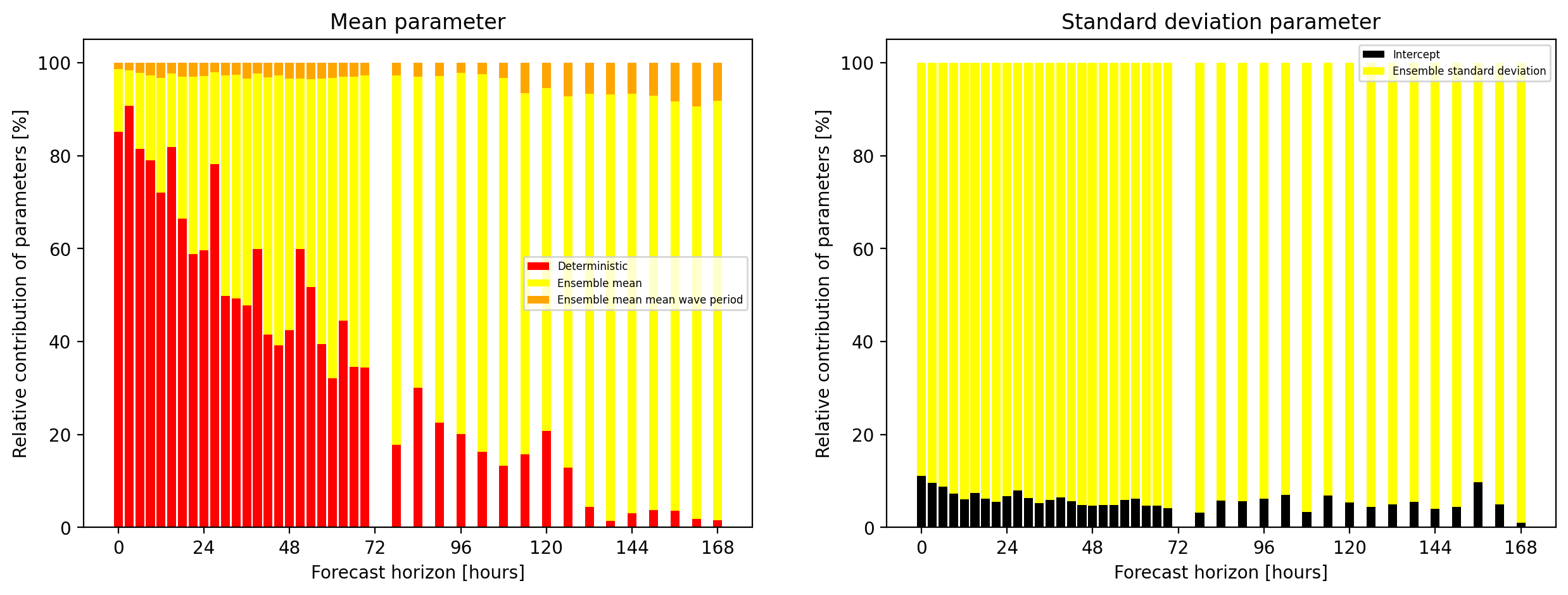}
		\caption{Significant wave height $H_S$. }
	\end{subfigure}
	\begin{subfigure}{1\textwidth}
		\includegraphics[width=1\textwidth]{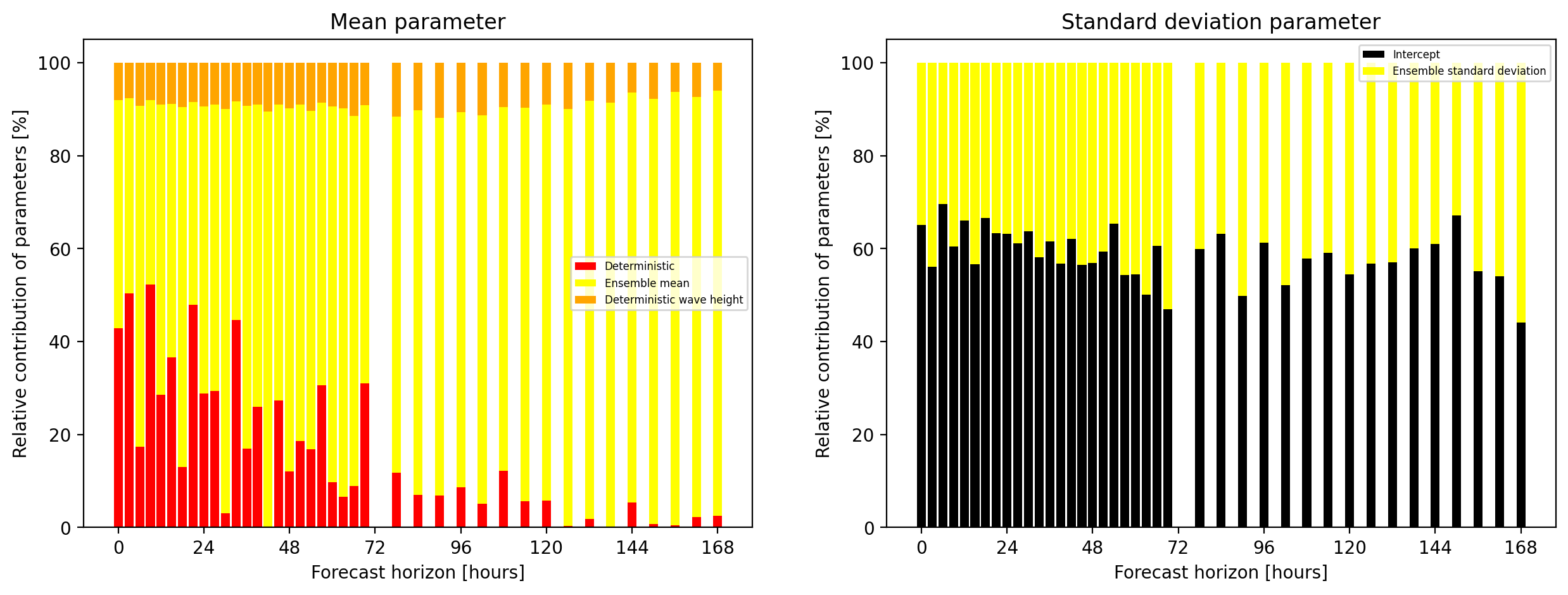}
		\caption{Wind speed $W$. }
	\end{subfigure}
	\begin{subfigure}{1\textwidth}
		\includegraphics[width=1\textwidth]{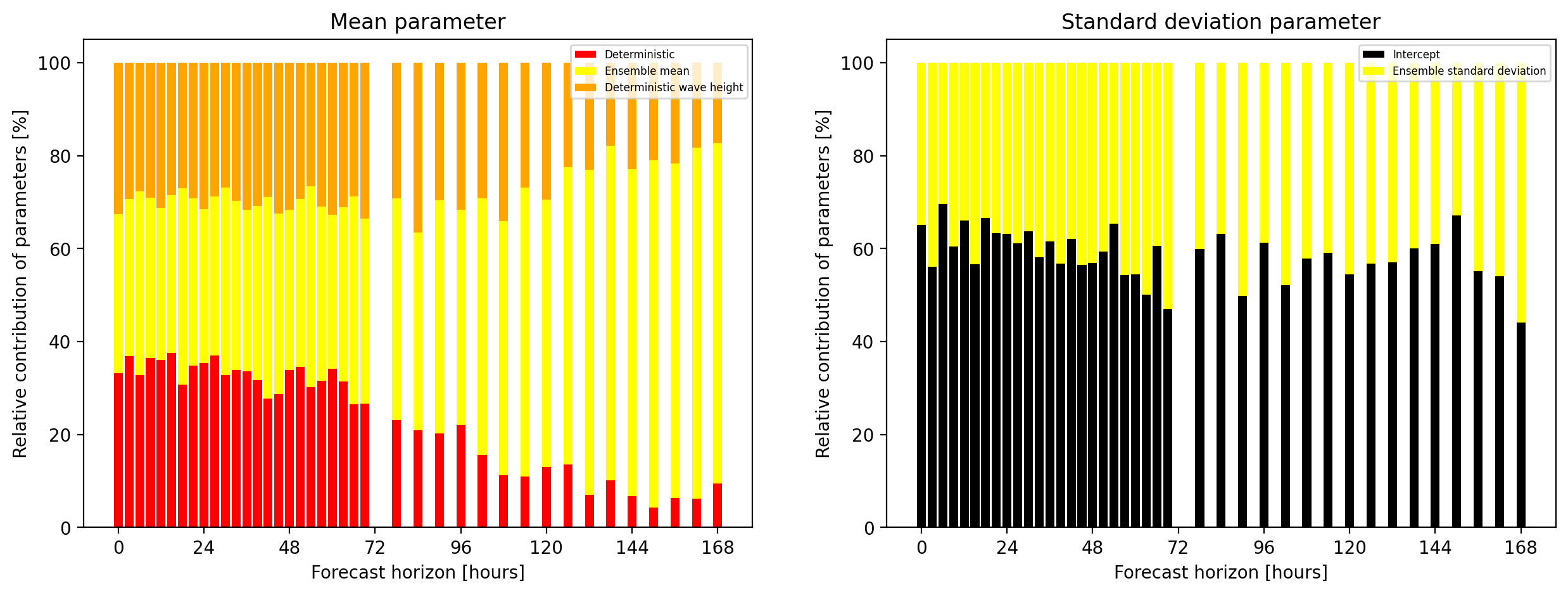}
		\caption{Mean wave period $T_M$. }
	\end{subfigure}
	\caption{Bar code plot of non-homogeneous Gaussian regression covariate contributions to the calibration of forecasts for (a) significant wave height $H_S$, (b) wind speed $W$ and (c) mean wave period $T_M$, as a function of forecast horizon. The response and all covariates are standardised to zero mean and unit standard deviation prior to regression analysis. Contributions are reported in terms of the percentage absolute value of the parameter coefficient from the regression. Consistent model forms are given in Table 2 of the main text.}
	\label{Fgr:NHGR:CvrImp}
\end{figure}

\begin{figure}[!ht]
	\centering
	\includegraphics[width=1\textwidth]{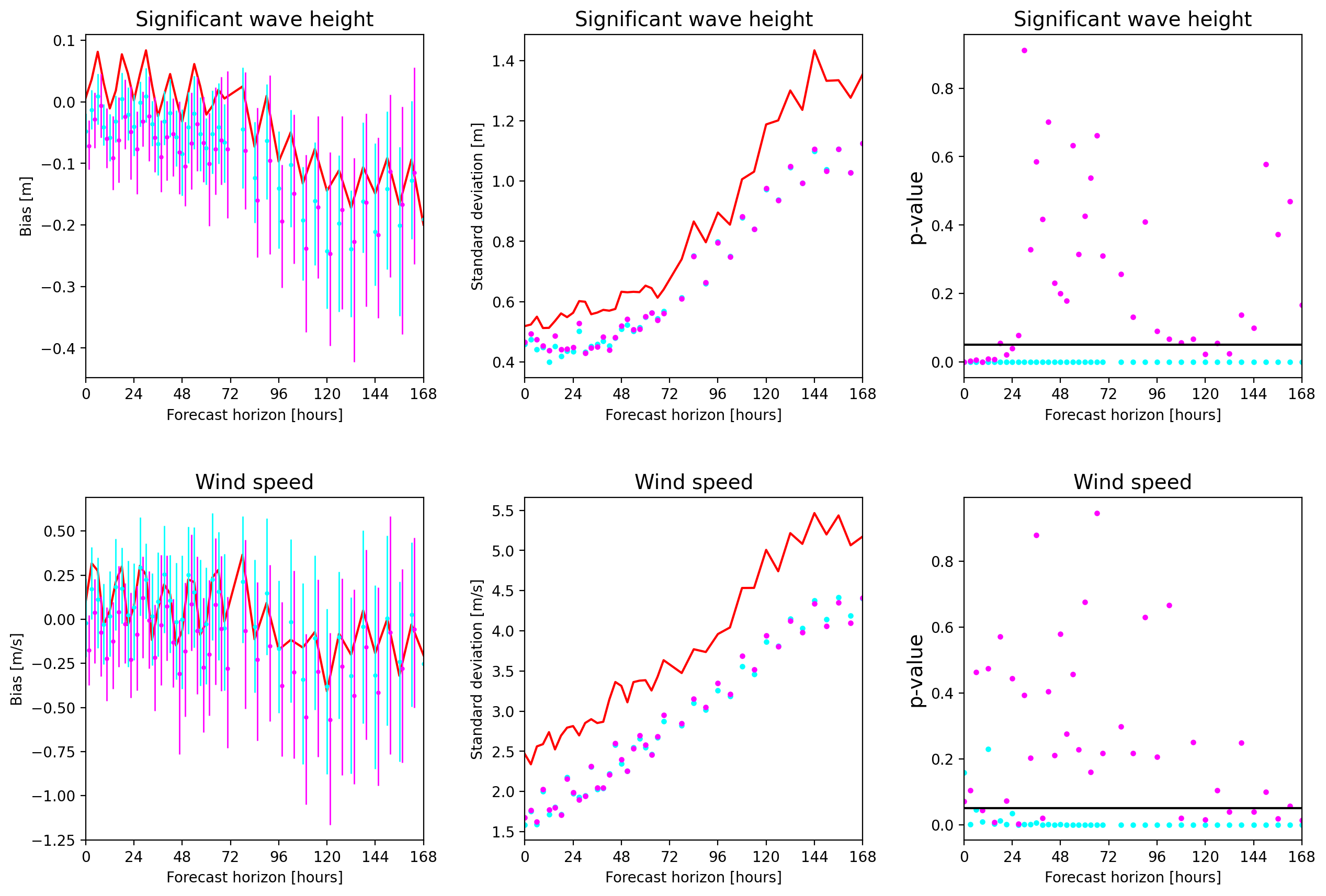}
	\caption{{Out-of-sample performance assessed for Period 1 (1 September 2023 - 30 April 2024), for calibration models estimated on data for period 17 May 2022 to 6 September 2023. Estimated mean (left) and standard deviation (centre) of forecast errors as a function of horizon, for (a) significant wave height $H_S$ and (b) wind speed $W$. The right column gives p-values of a KS test, with null hypothesis that standardised forecast residuals are standard Gaussian distributed, as a function of horizon. Red lines refer to the (uncalibrated) deterministic forecast; cyan (magenta) to LR (NHGR) calibration models. Box-whiskers (left panels) and discs (centre and right panels) have been translated horizontally by a small amount for clarity to avoid them being superimposed. 95\% uncertainty bands for bias (left) are calculated using bootstrap resampling. Results should be compared with those is Figure 11 of the main text. Bias is small relative to standard deviation. Reduction in forecast standard deviation at longer horizons is clear for both LR and NHGR relative to the deterministic forecast. The distribution of residuals from the NHGR model is generally more similar to the assumed Gaussian. Regrettably, comparable mean wave period data for the test period were not available.}}
	\label{Fgr:BisVrnLRNHGR-OOS1}
\end{figure}

\begin{figure}[!ht]
	\centering
	\includegraphics[width=1\textwidth]{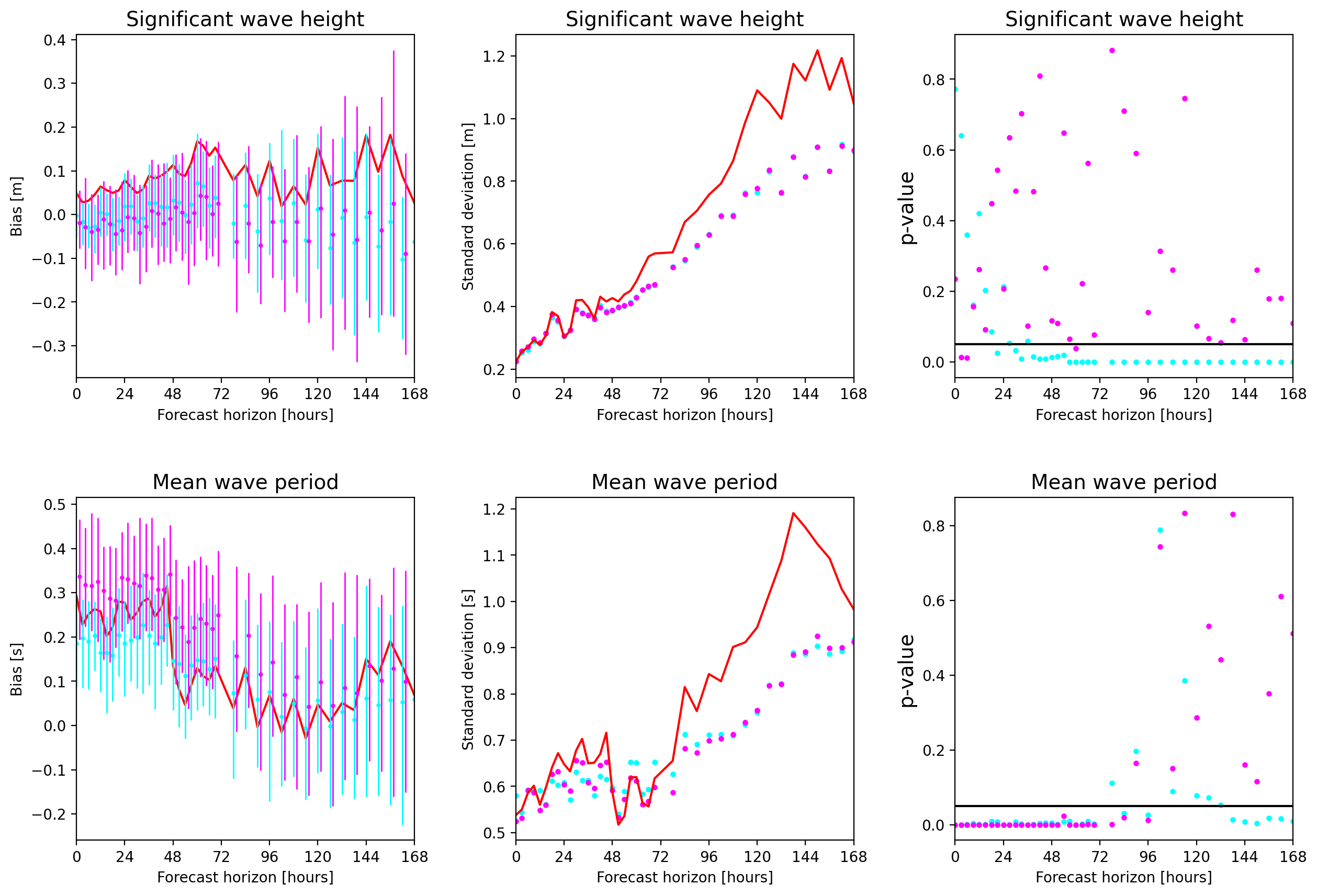}
	\caption{{Out-of-sample performance assessed for Period 2 (1 September 2024 – 31 December 2024, for calibration models estimated on data for period 17 May 2022 to 6 September 2023. Estimated mean (left) and standard deviation (centre) of forecast errors as a function of horizon, for (a) significant wave height $H_S$ and (b) mean wave period $T_M$. The right column gives p-values of a KS test, with null hypothesis that standardised forecast residuals are standard Gaussian distributed, as a function of horizon. Red lines refer to the (uncalibrated) deterministic forecast; cyan (magenta) to LR (NHGR) calibration models. Box-whiskers (left panels) and discs (centre and right panels) have been translated horizontally by a small amount for clarity to avoid them being superimposed. 95\% uncertainty bands for bias (left) are calculated using bootstrap resampling. Results should be compared with those is Figure 11 of the main text. Bias is small relative to standard deviation. Reduction in forecast standard deviation at longer horizons is clear for both LR and NHGR relative to the deterministic forecast. The distribution of residuals from the NHGR model is generally more similar to the assumed Gaussian for $H_S$. Regrettably, comparable wind speed data for the test period were not available.}}
	\label{Fgr:BisVrnLRNHGR-OOS2}
\end{figure}

\begin{figure}[!ht]
	\centering
	\begin{subfigure}{0.8\textwidth}
		\includegraphics[width=1\textwidth]{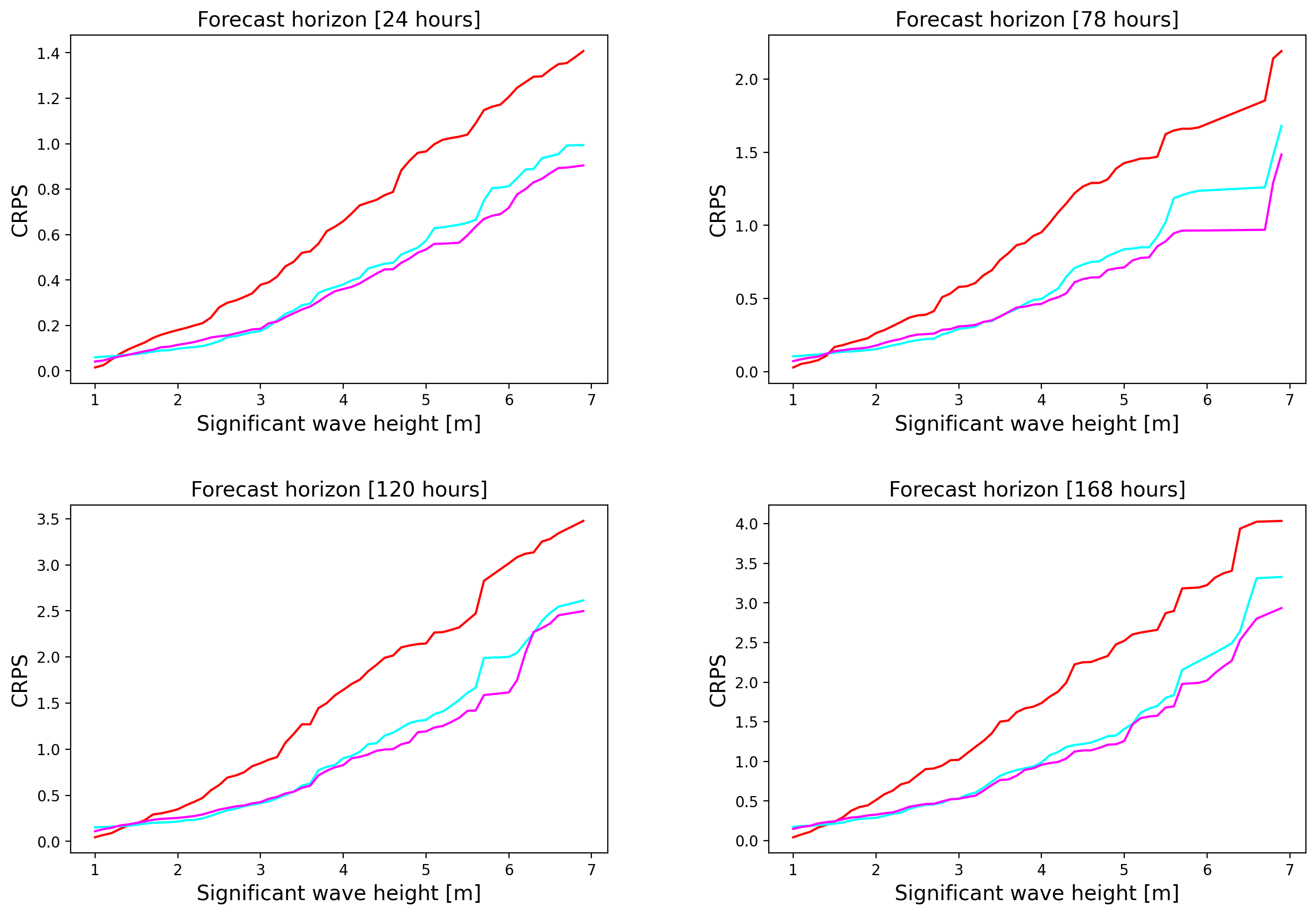}
		\caption{{Significant wave height $H_S$.} }
	\end{subfigure}
	\begin{subfigure}{0.8\textwidth}
		\includegraphics[width=1\textwidth]{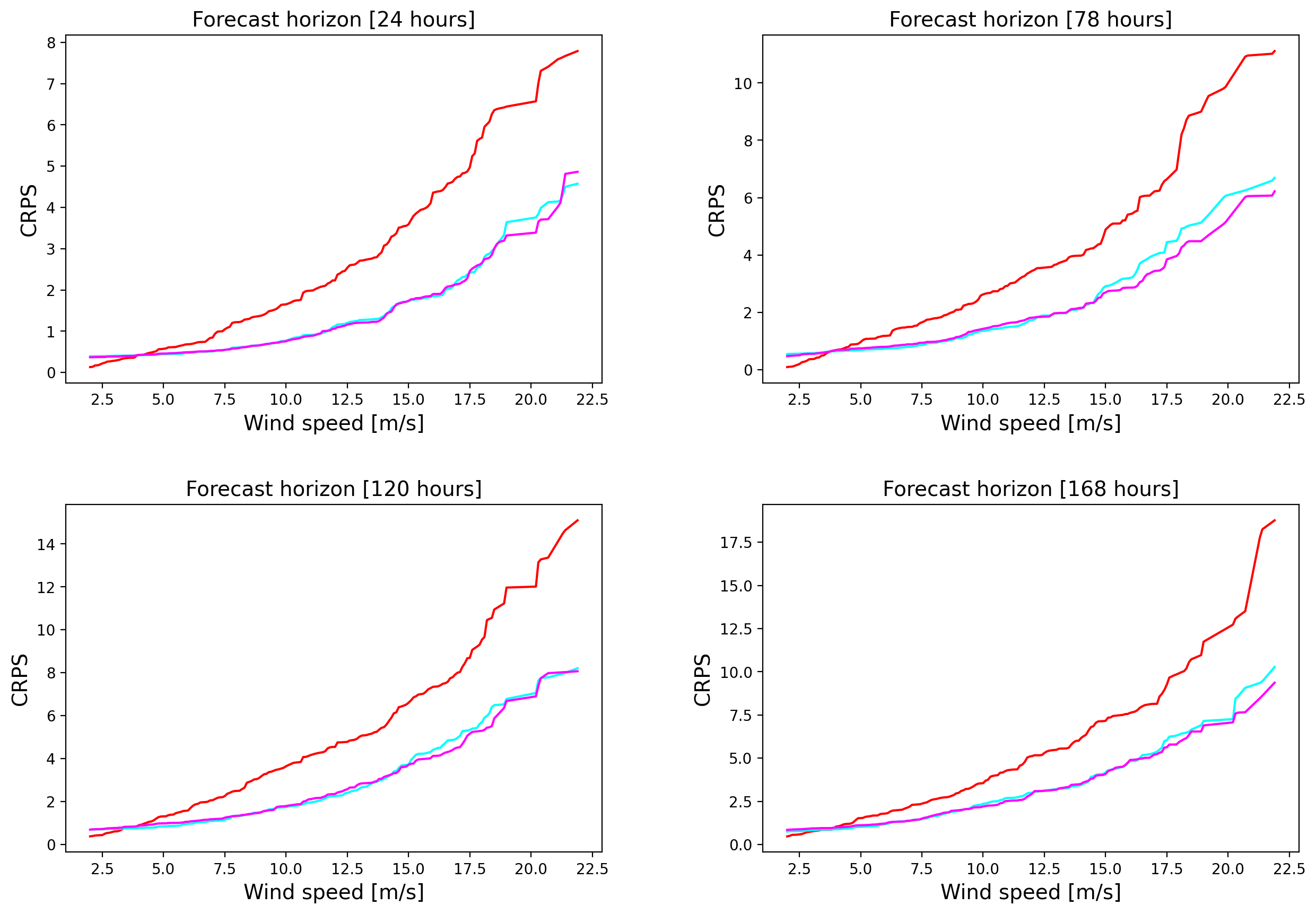}
		\caption{{Wind speed $W$.} }
	\end{subfigure}
	\caption{{Variation of CRPS with value of response $y$ for four forecast horizons $\tau$ on out-of-sample Period~1 (1 September 2023 - 30 April 2024) for (a) significant wave height $H_S$ and (b) wind speed $W$. CRPS shown for the uncalibrated deterministic forecast (red), the forecast calibrated using LR (cyan) and the forecast calibrated using NHGR (magenta). CRPS performance of calibrated forecasts betters that of the uncalibrated forecast, with an approximate reduction of 50\% in value. Moreover, the NHGR calibrated forecast is generally slightly better than that calibrated using LR.}}
	\label{Fgr:CrpsSmm_P1}
\end{figure}

\begin{figure}[!ht]
	\centering
	\begin{subfigure}{0.8\textwidth}
		\includegraphics[width=1\textwidth]{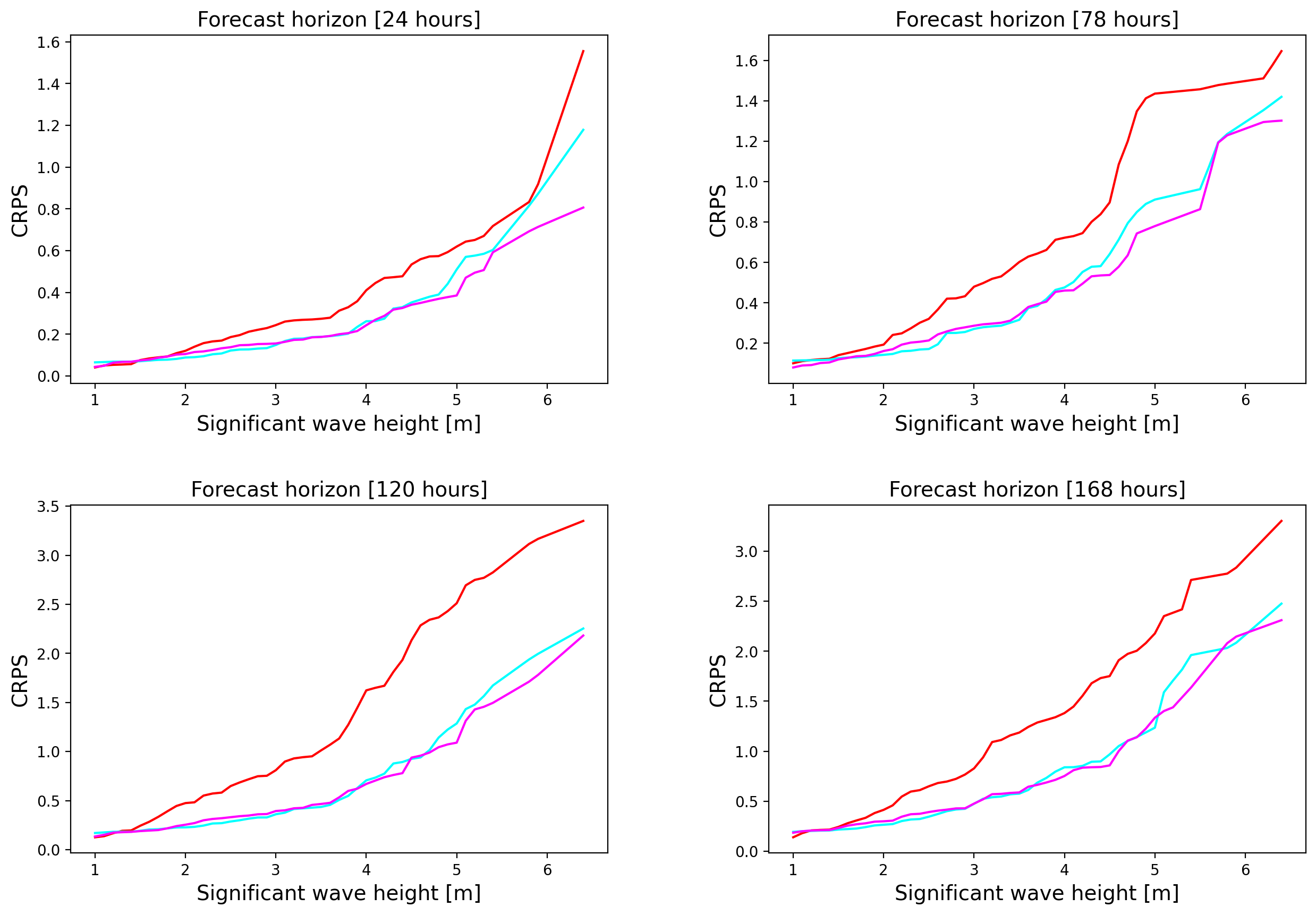}
		\caption{{Significant wave height $H_S$.} }
	\end{subfigure}
	\begin{subfigure}{0.8\textwidth}
		\includegraphics[width=1\textwidth]{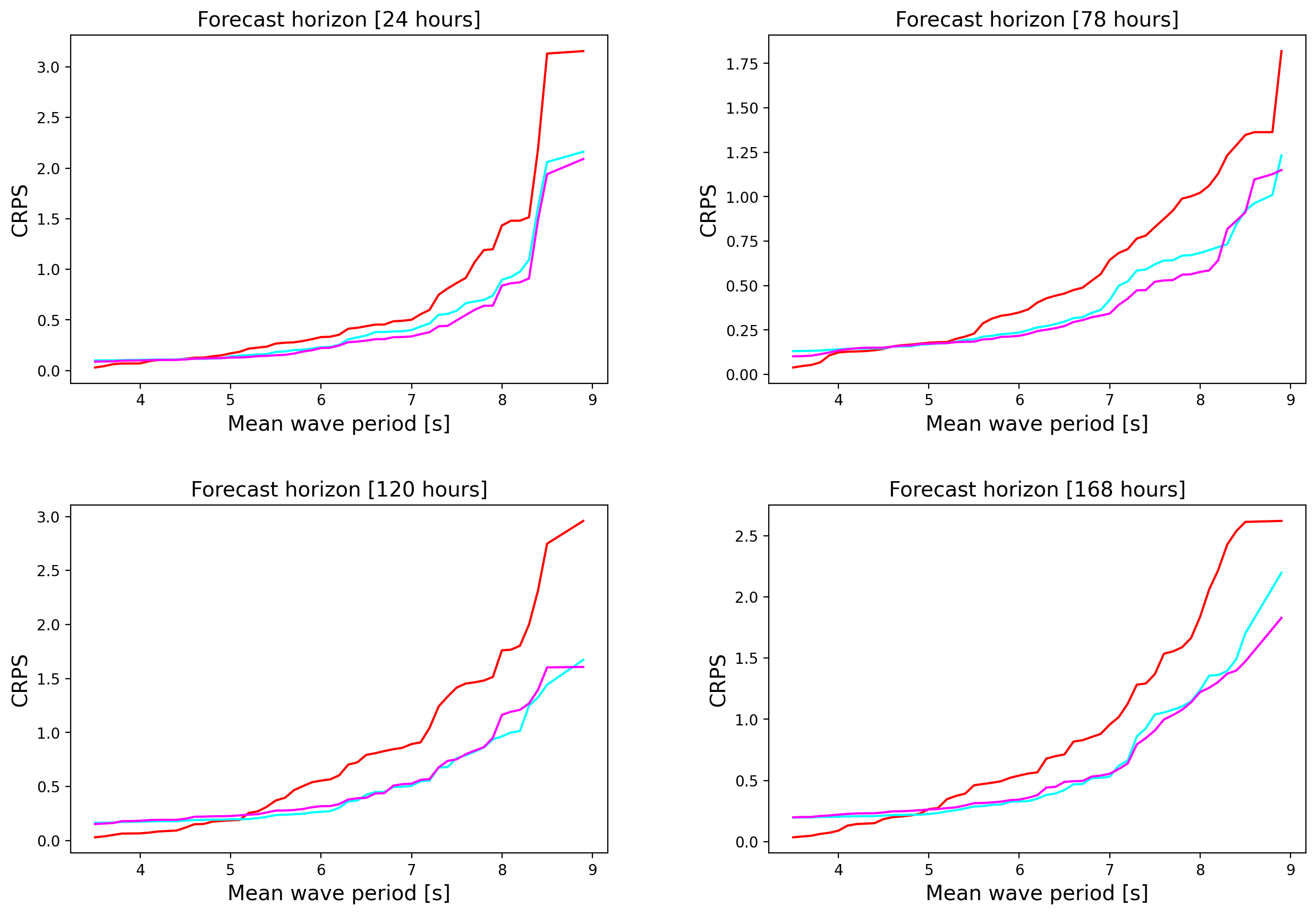}
		\caption{{Wind speed $W$. }}
	\end{subfigure}
	\caption{{Variation of CRPS with value of response $y$ for four forecast horizons $\tau$ on out-of-sample Period~2 (1 September 2024 - 31 December 2024) for (a) significant wave height $H_S$ and (b) mean wave period $T_M$. CRPS shown for the uncalibrated deterministic forecast (red), the forecast calibrated using LR (cyan) and the forecast calibrated using NHGR (magenta). CRPS performance of calibrated forecasts betters that of the uncalibrated forecast, with an approximate reduction of 50\% in value. Moreover, the NHGR calibrated forecast is generally (but not always) slightly better than that calibrated using LR.}}
	\label{Fgr:CrpsSmm_P2}
\end{figure}

\begin{figure}[!ht]
	\centering
	\begin{subfigure}{0.8\textwidth}
		\includegraphics[width=1\textwidth]{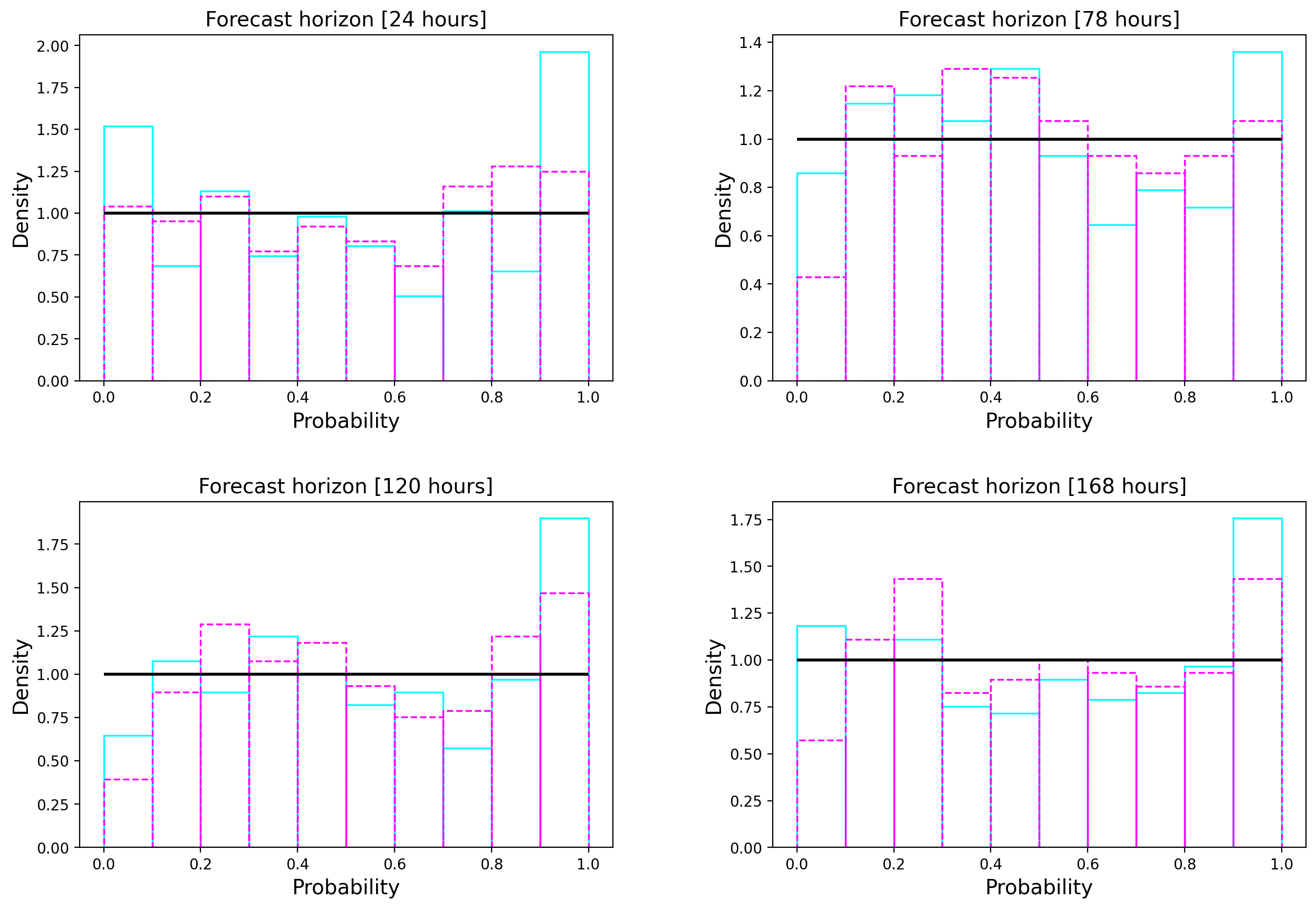}
		\caption{\ed{Significant wave height $H_S$.} }
	\end{subfigure}
	\begin{subfigure}{0.8\textwidth}
		\includegraphics[width=1\textwidth]{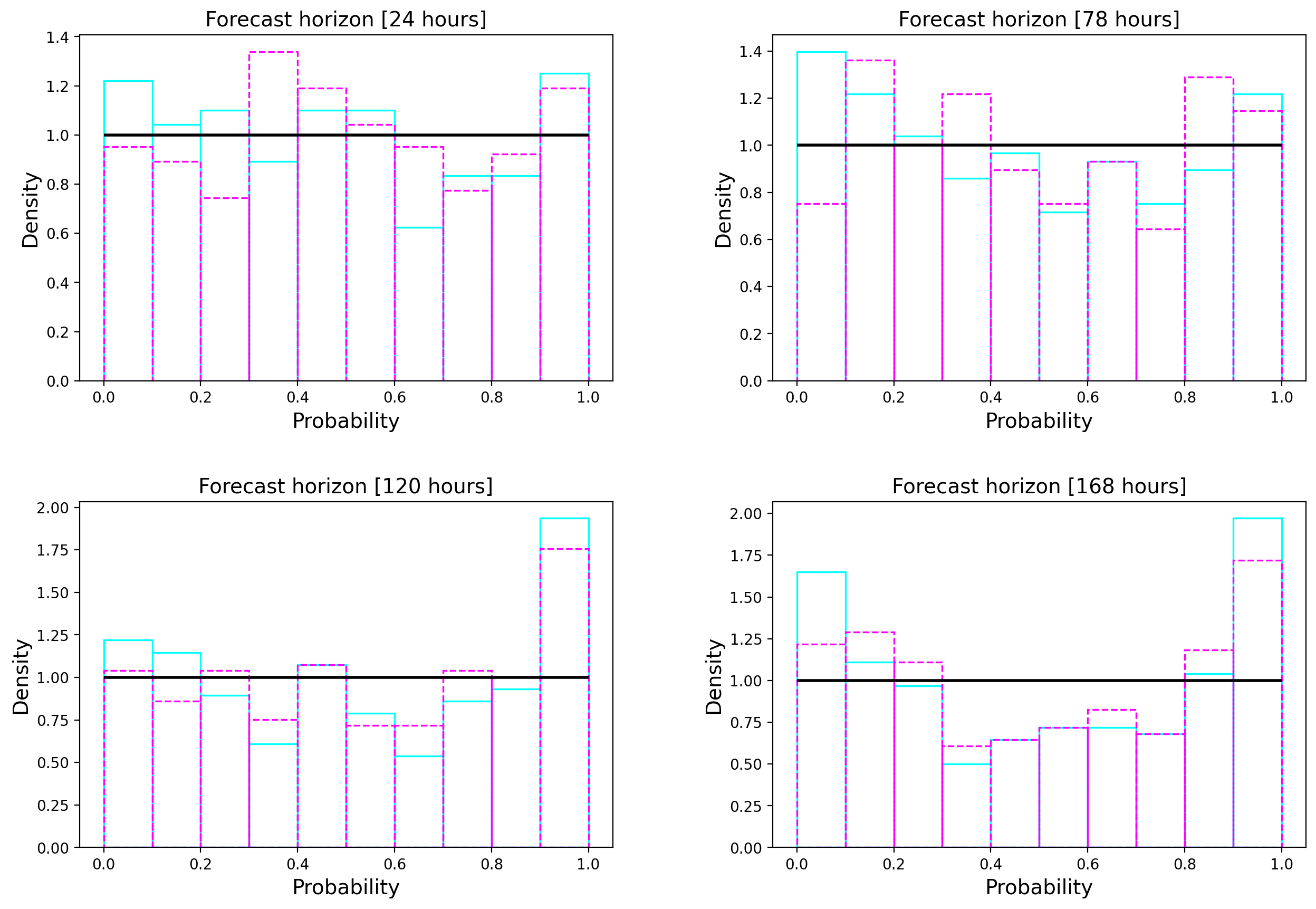}
		\caption{\ed{Wind speed $W$.} }
	\end{subfigure}
	\caption{\ed{Rank histogram plots for four forecast horizons $\tau$ on out-of-sample Period~1 (1 September 2023 - 30 April 2024) for (a) significant wave height $H_S$ and (b) wind speed $W$. Histograms are shown for the forecast calibrated using LR (cyan) and the forecast calibrated using NHGR (magenta). The horizontal line (black) indicates the model-assumed histogram, and departures from it indicate lack of agreement between model and reality. Model performance for both LR- and NHGR-calibrated forecasts is seen to be reasonable, with NHGR calibration to be preferred.}}
	\label{Fgr:RnkHstSmm_P1}
\end{figure}

\begin{figure}[!ht]
	\centering
	\begin{subfigure}{0.8\textwidth}
		\includegraphics[width=1\textwidth]{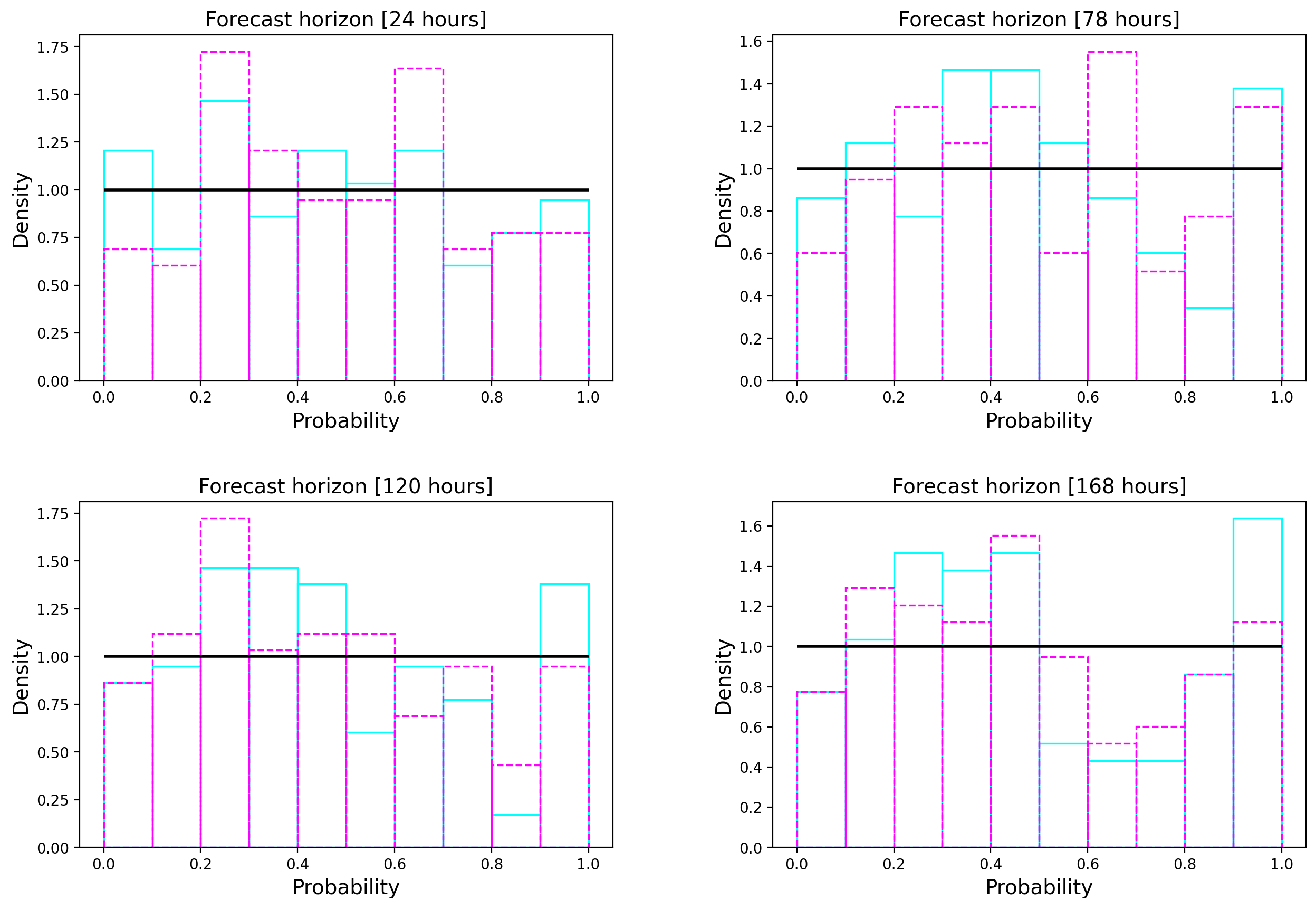}
		\caption{\ed{Significant wave height $H_S$.} }
	\end{subfigure}
	\begin{subfigure}{0.8\textwidth}
		\includegraphics[width=1\textwidth]{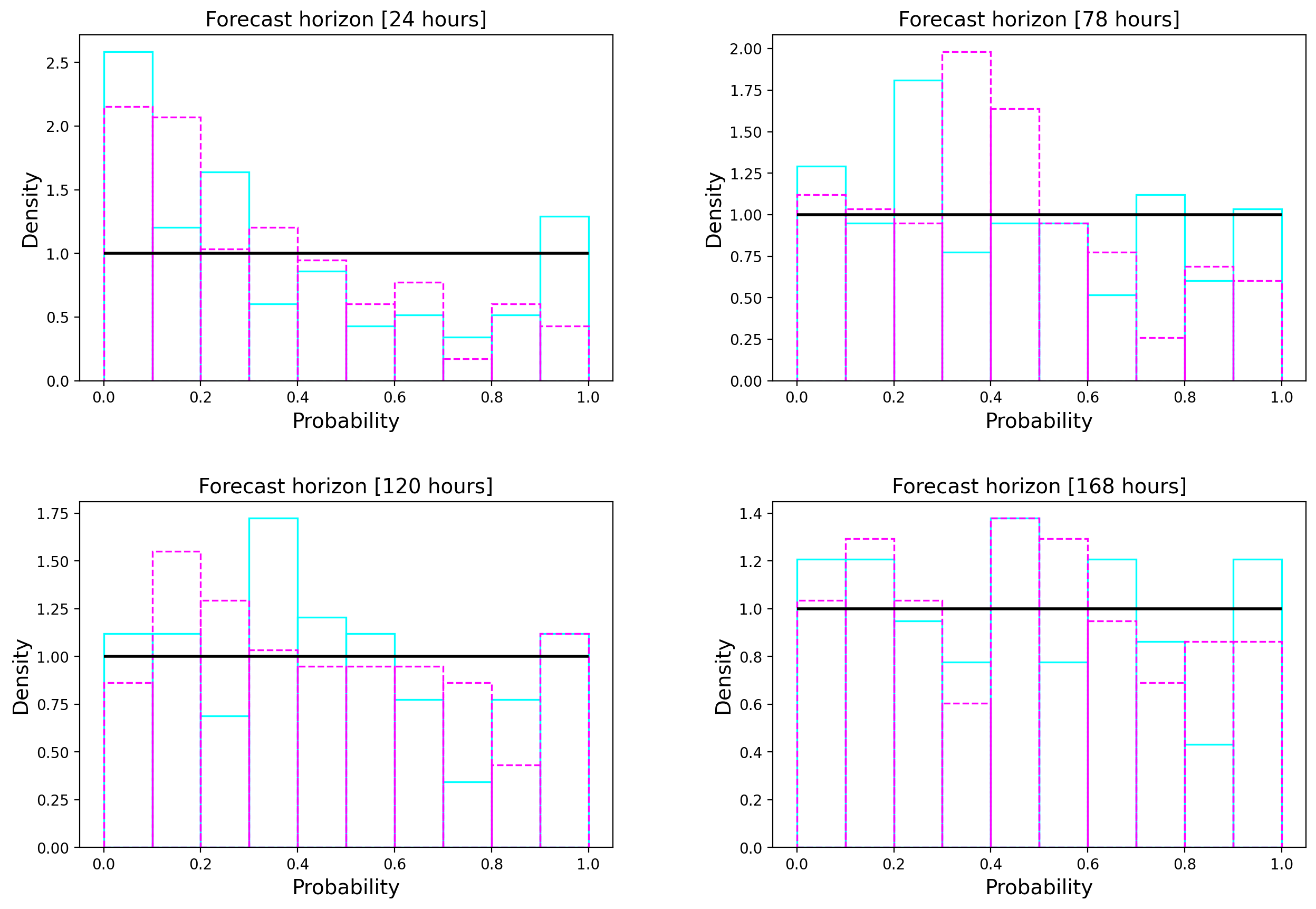}
		\caption{\ed{Mean wave period $T_M$.} }
	\end{subfigure}
	\caption{\ed{Rank histogram plots for four forecast horizons $\tau$ on out-of-sample Period~2 (1 September 2024 - 31 December 2024) for (a) significant wave height $H_S$ and (b) mean wave period $T_M$. Histograms are shown for the forecast calibrated using LR (cyan) and the forecast calibrated using NHGR (magenta). The horizontal line (black) indicates the model-assumed histogram, and departures from it indicate lack of agreement between model and reality. Model performance for both LR- and NHGR-calibrated forecasts is seen to be reasonable, with NHGR calibration to be preferred generally.}}
	\label{Fgr:RnkHstSmm_P2}
\end{figure}
